\documentclass[zpreprint,zbstepj,british]{./LaTeX/zeus/zeus_Kpaper}

\usepackage{./LaTeX/zeus/zeusdet_units}
\usepackage{subcaption}
\usepackage{caption}
\usepackage{rotating}
\usepackage{./LaTeX/styles/lineno}
\usepackage{amsmath}
\usepackage{verbatim}
\usepackage{stackengine}
\usepackage{graphicx}
\usepackage{adjustbox}
\usepackage{pdfpages}
\usepackage{multirow}
\usepackage{hyperref}
\usepackage[title]{appendix}
\usepackage{orcidlink}

\chardef\usc=95
\chardef\til=126
\catcode`\@=11 % @ signs are now treated as letters
\DeclareRobustCommand\xdotspace{\futurelet\@let@token\@xdotspace}
\def\@xdotspace{%
  \ifx\@let@token.\else
  \ifx\@let@token\bgroup.\else
  \ifx\@let@token\egroup.\else
  \ifx\@let@token\/.\else
  \ifx\@let@token\ .\else
  \ifx\@let@token~.\else
  \ifx\@let@token!.\else
  \ifx\@let@token,.\else
  \ifx\@let@token:.\else
  \ifx\@let@token;.\else
  \ifx\@let@token?.\else
  \ifx\@let@token/.\else
  \ifx\@let@token'.\else
  \ifx\@let@token).\else
  \ifx\@let@token-.\else
  \ifx\@let@token\@xobeysp.\else
  \ifx\@let@token\space.\else
  \ifx\@let@token\@sptoken.\else
   .\space
   \fi\fi\fi\fi\fi\fi\fi\fi\fi\fi\fi\fi\fi\fi\fi\fi\fi\fi}
\catcode`\@=12 % @ signs are no longer letters

\newcommand{\stru}[2]{%
   \relax\ifmmode\hbox{\vrule height#1 depth#2 width0pt}%
   \else\vrule height#1 depth#2 width0pt\fi}
\newcommand{\Ronum}[1]{\uppercase\expandafter{\romannumeral#1}}
\newcommand{\ronum}[1]{\expandafter{\romannumeral#1}}
\DeclareRobustCommand{\LaTeXZ}{%
  \LaTeX\kern-.05em4\kern-.1em
  {\raisebox{-0.2ex}{$\scriptstyle\text{ZEUS}$}}\xspace}
\newcommand{\slashfrac}[2]{%
  \raisebox{0.5ex}{\ensuremath #1}\kern-0.12em/\kern-0.08em
  \raisebox{-.8ex}{\ensuremath #2}}
\newcommand{\sqr}[3]{%
    {\vcenter{\hrule height.#3ex\hbox{\vrule width.#2ex height#1ex
     \kern#1ex\vrule width.#3ex}\hrule height.#2ex}}}

\catcode`\@=11 % @ signs are now treated as letters
\newcommand{\parenbar}{\mathpalette\p@renb@r}
\def\p@renb@r#1#2{\vbox{%
  \ifx#1\scriptscriptstyle \dimen@.7em\dimen@ii.2em\else
  \ifx#1\scriptstyle \dimen@.8em\dimen@ii.25em\else
  \dimen@1em\dimen@ii.4em\fi\fi \offinterlineskip
  \ialign{\hfill##\hfill\cr
    \vbox{\hrule width\dimen@ii}\cr
    \noalign{\vskip-.3ex}%
    \hbox to\dimen@{$\mathchar300\hfil\mathchar301$}\cr
    \noalign{\vskip-.3ex}%
    $#1#2$\cr}}}
\catcode`\@=12 % @ signs are no longer letters

\ifzeusunit
  
\else
  
\fi

\newcommand{\IP}{{\rm I$\kern-0.01667em$P}\xspace}

\mathchardef\qsm=63
\mathchardef\pls=43
\mathchardef\mns=512
\mathchardef\plm=518
\mathchardef\eql=61
\mathchardef\smallleft=300
\mathchardef\smallright=301
\mathchardef\les=316
\mathchardef\gre=318
\mathchardef\leq=532
\mathchardef\grq=533
\catcode`\@=11 % @ signs are now treated as letters
\newcounter{pict@width}
\newcounter{pict@height}
\newlength{\pict@scale}
\newcommand{\psfigadd}[4]{%
\setcounter{pict@width}{1*\ratio{#2+\pict@scale/2}{\pict@scale}}
\setcounter{pict@height}{1*\ratio{#3+\pict@scale/2}{\pict@scale}}
\setlength{\unitlength}{\pict@scale}
\hbox to #2{\hspace{-\fill}\begin{picture}(\thepict@width,\thepict@height)
\put(0,0){\psfig{figure=#1,width=#2,height=#3,clip=}}
\SetScale{0.283466457}
\SetWidth{1.763889}
{#4}
\end{picture}}
}
\newcounter{pict@widthfst}
\newcounter{pict@widthscd}
\newcounter{pict@widthtot}
\newcommand{\psfigaddtwo}[7]{%
\setcounter{pict@widthfst}{1*\ratio{#2+\pict@scale/2}{\pict@scale}}
\setcounter{pict@widthscd}{1*\ratio{#2+#4+\pict@scale/2}{\pict@scale}}
\setcounter{pict@widthtot}{1*\ratio{#2+#4+#6+\pict@scale/2}{\pict@scale}}
\setcounter{pict@height}{1*\ratio{#3+\pict@scale/2}{\pict@scale}}
\setlength{\unitlength}{\pict@scale}
\hbox{\hspace{-\fill}\begin{picture}(\thepict@widthtot,\thepict@height)
\put(0,0){\psfig{figure=#1,width=#2,height=#3,clip=}}
\put(\thepict@widthscd,0){\psfig{figure=#5,width=#6,height=#3,clip=}}
\SetScale{0.283466457}
\SetWidth{1.763889}
{#7}
\end{picture}}
}
\newcommand{\psfigror}[4]{%
\setcounter{pict@width}{1*\ratio{#2+\pict@scale/2}{\pict@scale}}
\setcounter{pict@height}{1*\ratio{#3+\pict@scale/2}{\pict@scale}}
\setlength{\unitlength}{\pict@scale}
\hbox{\begin{picture}(\thepict@width,\thepict@height)
\put(0,\thepict@height){\psfig{figure=#1,width=#3,height=#2,clip=,angle=270}}
\SetScale{0.283466457}
\SetWidth{1.763889}
{#4}
\end{picture}}
}
\newcommand{\psfigrol}[4]{%
\setcounter{pict@width}{1*\ratio{#2+\pict@scale/2}{\pict@scale}}
\setcounter{pict@height}{1*\ratio{#3+\pict@scale/2}{\pict@scale}}
\setlength{\unitlength}{\pict@scale}
\hbox{\begin{picture}(\thepict@width,\thepict@height)
\put(0,0){\psfig{figure=#1,width=#3,height=#2,clip=,angle=90}}
\SetScale{0.283466457}
\SetWidth{1.763889}
{#4}
\end{picture}}
}
\catcode`\@=12 % @ signs are no longer letters
\newlength\listtextwidth

\catcode`\@=11 % @ signs are now treated as letters
\newlength{\@tabfninsert}
\newlength{\@tabfnwidth}
\newcommand{\tabfootnote}[2]{%
  \setlength{\@tabfninsert}{0.8em}
  \setlength{\@tabfnwidth}{\textwidth}
  \addtolength{\@tabfnwidth}{-\@tabfninsert}
  \addtolength{\@tabfnwidth}{-0.4em}
  \noindent\makebox[\@tabfninsert][r]{\footnotesize$^{#1}$\hfil}\hfill%
  \parbox[t]{\@tabfnwidth}{\footnotesize #2\hfill}}
\catcode`\@=12 % @ signs are no longer letters

\def\citeCTD{{\cite{%
nim:a279:290,*npps:b32:181,*nim:a338:254%
}}\xspace}
\def\citeMVD{{\cite{%
nim:a581:656%
}}\xspace}
\def\citeSTT{{\cite{%
nim:a535:191%
}}\xspace}
\def\citeCAL{{\cite{%
nim:a309:77,*nim:a309:101,*nim:a321:356,*nim:a336:23%
}}\xspace}
\def\citeHES{{\cite{%
nim:a277:176%
}}\xspace}
\def\citeSixmT{{\cite{%
thesis:gosau:2007%
}}\xspace}
\def\citeBAC{{\cite{%
nim:a300:480%
}}\xspace}
\def\citePCAL{{\cite{%
desy-92-066,*zfp:c63:391,*acpp:b32:2025%
}}\xspace}
\def\citeSPECTRO{{\cite{%
nim:a565:572%
}}\xspace}

\begin{document}
%------------------------------------------------------------------------------
%       Title sheet
%------------------------------------------------------------------------------
\prepnum{DESY-24-070}
\prepdate{May 2024}

\draftversion{3.5}

\zeustitle{
The azimuthal correlation between the leading jet and the scattered lepton in deep inelastic scattering at HERA
}

\zeusauthor{ZEUS Collaboration}

\date{}

\maketitle

%\vspace*{-2cm}

\begin{abstract}
\noindent
The azimuthal correlation angle, $\Delta\phi$, between the scattered lepton and the leading jet in deep inelastic $e^{\pm}p$ scattering at HERA has been studied using data collected with the ZEUS detector at a centre-of-mass energy of $\sqrt{s} = 318 \gev$, corresponding to an integrated luminosity of $326 \pb^{-1}$.
A measurement of jet cross sections in the laboratory frame was made in a fiducial region corresponding to photon virtuality $10 \gev^2 < Q^2 < 350 \gev^2$, inelasticity $0.04 < y < 0.7$, outgoing lepton energy $E_e > 10 \gev$, lepton polar angle $140^\circ < \theta_e < 180^\circ$, jet transverse momentum $2.5 \gev < p_\mathrm{T,jet} < 30 \gev$, and jet pseudorapidity $-1.5 < \eta_\mathrm{jet} < 1.8$.
Jets were reconstructed using the $k_\mathrm{T}$ algorithm with the radius parameter $R = 1$.
The leading jet in an event is defined as the jet that carries the highest $p_\mathrm{T,jet}$.
Differential cross sections, $d\sigma/d\Delta\phi$, were measured as a function of the azimuthal correlation angle in various ranges of leading-jet transverse momentum, photon virtuality and jet multiplicity.
Perturbative calculations at $\mathcal{O}(\alpha_{s}^2)$ accuracy successfully describe the data within the fiducial region, although a lower level of agreement is observed near $\Delta\phi \rightarrow \pi$ for events with high jet multiplicity, due to limitations of the perturbative approach in describing soft phenomena in QCD.
The data are equally well described by Monte Carlo predictions that supplement leading-order matrix elements with parton showering.

\end{abstract}

\thispagestyle{empty}
\cleardoublepage

\topmargin-1.cm
\evensidemargin-0.3cm
\oddsidemargin-0.3cm
\textwidth 16.cm
\textheight 680pt
\parindent0.cm
\parskip0.3cm plus0.05cm minus0.05cm

\pagenumbering{Roman}

\newcommand{\authorinfo}[3][\textit]{%                                                                                                                        
\makebox[3ex]{$^{#2}$}
\begin{minipage}[t]{14cm}\raggedright
#1{#3}
\end{minipage}\par
}

\begin{center}
{\Large The ZEUS Collaboration}
\end{center}

{\small\raggedright

% members                                                                                                                                                     
I.~Abt$^{1}$,
R. Aggarwal$^{2}$,
V.~Aushev$^{3}$,
O.~Behnke$^{4}$,
A.~Bertolin$^{5}$,
I.~Bloch$^{6}$,
I.~Brock$^{7}$,
N.H.~Brook$^{8, a}$,
R.~Brugnera$^{9}$,
A.~Bruni$^{10}$,
P.J.~Bussey$^{11}$,
A.~Caldwell$^{1}$,
C.D.~Catterall$^{12}$,
J.~Chwastowski$^{13}$,
J.~Ciborowski$^{14, b}$,
R.~Ciesielski$^{4, c}$,
A.M.~Cooper-Sarkar$^{15}$,
M.~Corradi$^{10, d}$,
R.K.~Dementiev$^{16}$,
S.~Dusini$^{5}$,
J.~Ferrando$^{17}$,
B.~Foster$^{15, e}$,
E.~Gallo$^{18, f}$,
D.~Gangadharan$^{19, g}$,
A.~Garfagnini$^{9}$,
A.~Geiser$^{4}$,
G.~Grzelak$^{14}$,
C.~Gwenlan$^{15}$,
D.~Hochman$^{20}$,
N.Z.~Jomhari$^{4}$,
I.~Kadenko$^{3}$,
U.~Karshon$^{20}$,
P.~Kaur$^{21}$,
R.~Klanner$^{18}$,
U.~Klein$^{4, h}$,
I.A.~Korzhavina$^{16}$,
N.~Kovalchuk$^{18}$,
M.~Kuze$^{22}$,
B.B.~Levchenko$^{16}$,
A.~Levy$^{23}$,
B.~L\"ohr$^{4}$,
E.~Lohrmann$^{18}$,
A.~Longhin$^{9}$,
F.~Lorkowski$^{24}$\orcidlink{0000-0003-2677-3805},
E.~Lunghi$^{25}$,
I.~Makarenko$^{4}$,
J.~Malka$^{4, i}$,
S.~Masciocchi$^{26, j}$,
K.~Nagano$^{27}$,
J.D.~Nam$^{28}$,
Yu.~Onishchuk$^{3}$,
E.~Paul$^{7}$,
I.~Pidhurskyi$^{29}$,
A.~Polini$^{10}$,
M.~Przybycie\'n$^{30}$,
A.~Quintero$^{28}$,
M.~Ruspa$^{31}$,
U.~Schneekloth$^{4}$,
T.~Sch\"orner-Sadenius$^{4}$,
I.~Selyuzhenkov$^{26}$,
M.~Shchedrolosiev$^{4}$,
L.M.~Shcheglova$^{16}$,
N.~Sherrill$^{32}$,
I.O.~Skillicorn$^{11}$,
W.~S{\l}omi\'nski$^{33}$,
A.~Solano$^{34}$,
L.~Stanco$^{5}$,
N.~Stefaniuk$^{4}$,
B.~Surrow$^{28}$,
K.~Tokushuku$^{27}$,
O.~Turkot$^{4, i}$,
T.~Tymieniecka$^{35}$,
A.~Verbytskyi$^{1}$,
W.A.T.~Wan Abdullah$^{36}$,
K.~Wichmann$^{4}$,
M.~Wing$^{8, k}$,
S.~Yamada$^{27}$,
Y.~Yamazaki$^{37}$,
A.F.~\.Zarnecki$^{14}$,
O.~Zenaiev$^{4, l}$
\newpage

% institutes                                                                                                                                                  
{\setlength{\parskip}{0.4em}
\authorinfo{1}{Max-Planck-Institut f\"ur Physik, M\"unchen, Germany}
\authorinfo{2}{DST-Inspire Faculty, Department of Technology, SPPU, India}
\authorinfo{3}{Department of Nuclear Physics, National Taras Shevchenko University of Kyiv, Kyiv, Ukraine}
\authorinfo{4}{Deutsches Elektronen-Synchrotron DESY, Hamburg, Germany}
\authorinfo{5}{INFN Padova, Padova, Italy~$^{A}$}
\authorinfo{6}{Deutsches Elektronen-Synchrotron DESY, Zeuthen, Germany}
\authorinfo{7}{Physikalisches Institut der Universit\"at Bonn, Bonn, Germany~$^{B}$}
\authorinfo{8}{Physics and Astronomy Department, University College London, London, United Kingdom~$^{C}$}
\authorinfo{9}{Dipartimento di Fisica e Astronomia dell' Universit\`a and INFN, Padova, Italy~$^{A}$}
\authorinfo{10}{INFN Bologna, Bologna, Italy~$^{A}$}
\authorinfo{11}{School of Physics and Astronomy, University of Glasgow, Glasgow, United Kingdom~$^{C}$}
\authorinfo{12}{Department of Physics, York University, Ontario, Canada M3J 1P3~$^{D}$}
\authorinfo{13}{The Henryk Niewodniczanski Institute of Nuclear Physics, Polish Academy of Sciences, Krakow, Poland}
\authorinfo{14}{Faculty of Physics, University of Warsaw, Warsaw, Poland}
\authorinfo{15}{Department of Physics, University of Oxford, Oxford, United Kingdom~$^{C}$}
\authorinfo{16}{Affiliated with an institute covered by a current or former collaboration agreement with DESY}
\authorinfo{17}{Physics Department, Lancaster University, Lancaster, United Kingdom}
\authorinfo{18}{Hamburg University, Institute of Experimental Physics, Hamburg, Germany~$^{E}$}
\authorinfo{19}{Physikalisches Institut of the University of Heidelberg, Heidelberg, Germany}
\authorinfo{20}{Department of Particle Physics and Astrophysics, Weizmann Institute, Rehovot, Israel}
\authorinfo{21}{Sant Longowal Institute of Engineering and Technology, Longowal, Punjab, India}
\authorinfo{22}{Department of Physics, Tokyo Institute of Technology, Tokyo, Japan~$^{F}$}
\authorinfo{23}{Raymond and Beverly Sackler Faculty of Exact Sciences, School of Physics, Tel Aviv University, Tel Aviv, Israel~$^{G}$}
\authorinfo{24}{Physik-Institut, University of Zurich, Zurich, Switzerland}
\authorinfo{25}{Department of Physics, Indiana University Bloomington, Bloomington, IN 47405, USA}
\authorinfo{26}{GSI Helmholtzzentrum f\"{u}r Schwerionenforschung GmbH, Darmstadt, Germany}
\authorinfo{27}{Institute of Particle and Nuclear Studies, KEK, Tsukuba, Japan~$^{F}$}
\authorinfo{28}{Department of Physics, Temple University, Philadelphia, PA 19122, USA~$^{H}$}
\authorinfo{29}{Institut f\"ur Kernphysik, Goethe Universit\"at, Frankfurt am Main, Germany}
\authorinfo{30}{AGH University of Science and Technology, Faculty of Physics and Applied Computer Science, Krakow, Poland}
\authorinfo{31}{Universit\`a del Piemonte Orientale, Novara, and INFN, Torino, Italy~$^{A}$}
\authorinfo{32}{Department of Physics and Astronomy, University of Sussex, Brighton, BN1 9QH, United Kingdom~$^{I}$}
\authorinfo{33}{Department of Physics, Jagellonian University, Krakow, Poland~$^{J}$}
\authorinfo{34}{Universit\`a di Torino and INFN, Torino, Italy~$^{A}$}
\authorinfo{35}{National Centre for Nuclear Research, Warsaw, Poland}
\authorinfo{36}{National Centre for Particle Physics, Universiti Malaya, 50603 Kuala Lumpur, Malaysia~$^{K}$}
\authorinfo{37}{Department of Physics, Kobe University, Kobe, Japan~$^{F}$}
}
\vspace{3em}

% references concerning institutes                                                                                                                            
{\setlength{\parskip}{0.4em}
\authorinfo[]{ A}{ supported by the Italian National Institute for Nuclear Physics (INFN)}
\authorinfo[]{ B}{ supported by the German Federal Ministry for Education and Research (BMBF), under contract No.\ 05 H09PDF}
\authorinfo[]{ C}{ supported by the Science and Technology Facilities Council, UK}
\authorinfo[]{ D}{ supported by the Natural Sciences and Engineering Research Council of Canada (NSERC)}
\authorinfo[]{ E}{ supported by the German Federal Ministry for Education and Research (BMBF), under contract No.\ 05h09GUF, and the SFB 676 of the Deutsche \
Forschungsgemeinschaft (DFG)}
\authorinfo[]{ F}{ supported by the Japanese Ministry of Education, Culture, Sports, Science and Technology (MEXT) and its grants for Scientific Research}
\authorinfo[]{ G}{ supported by the Israel Science Foundation}
\authorinfo[]{ H}{ supported in part by the Office of Nuclear Physics within the U.S.\ DOE Office of Science}
\authorinfo[]{ I}{supported in part by the Science and Technology Facilities Council grant number ST/T006048/1}
\authorinfo[]{ J}{supported by the Polish National Science Centre (NCN) grant no.\ DEC-2014/13/B/ST2/02486}
\authorinfo[]{ K}{ supported by HIR grant UM.C/625/1/HIR/149 and UMRG grants RU006-2013, RP012A-13AFR and RP012B-13AFR from Universiti Malaya, and ERGS grant\
 ER004-2012A from the Ministry of Education, Malaysia}
}
\pagebreak[4]

% references concerning members                                                                                                                               
{\setlength{\parskip}{0.4em}
\authorinfo[]{ a}{now at University of Bath, United Kingdom}
\authorinfo[]{ b}{also at Lodz University, Poland}
\authorinfo[]{ c}{now at Rockefeller University, New York, NY 10065, USA}
\authorinfo[]{ d}{now at INFN Roma, Italy}
\authorinfo[]{ e}{also at DESY and University of Hamburg, Hamburg, Germany and supported by a Leverhulme Trust Emeritus Fellowship}
\authorinfo[]{ f}{also at DESY, Hamburg, Germany}
\authorinfo[]{ g}{now at University of Houston, Houston, TX 77004, USA}
\authorinfo[]{ h}{now at University of Liverpool, United Kingdom}
\authorinfo[]{ i}{now at European X-ray Free-Electron Laser facility GmbH, Hamburg, Germany}
\authorinfo[]{ j}{also at Physikalisches Institut of the University of Heidelberg, Heidelberg,  Germany}
\authorinfo[]{ k}{also supported by DESY, Hamburg, Germany}
\authorinfo[]{ l}{now at Hamburg University, II. Institute for Theoretical Physics, Hamburg, Germany }
}
}

\pagenumbering{arabic}

\pagestyle{scrheadings}

 %----------------------------------------------------------------------------
%       Introduction
% %----------------------------------------------------------------------------
\section{Introduction}
\label{sec-int}

% P1
The HERA collider provided $e^{\pm}p$ events\footnote{In this paper, both electrons and positrons are referred to as electrons.} that are a unique basis for tests of a wide range of predictions based on perturbative Quantum Chromodynamics (pQCD).
Jet production at HERA continues to be used for rigorous tests of the validity of pQCD~\cite{H1:2021xxi,ZEUS:2023zie}.
The azimuthal distribution of jets with respect to the outgoing lepton in deep inelastic scattering (DIS) provides an interesting means of investigating both soft and hard phenomena in QCD, and is the subject of the present paper.
In neutral current (NC) $ep$ DIS mediated by a virtual boson, a final-state jet can be produced at the Born limit ($\mathcal{O}(\alpha_s^0)$) of DIS via the following process:

\begin{equation}
\label{eq1}
e + p \rightarrow e + \mathrm{jet} + X.
\end{equation}

The azimuthal correlation angle, $\Delta\phi = |\phi_{e} - \phi_\mathrm{jet}|$, is defined as the difference in the azimuthal angle between the scattered lepton, $\phi_{e}$, and the final-state jet, $\phi_\mathrm{jet}$, where all quantities are specified in the laboratory frame.
The lepton--jet pairs in reaction (\ref{eq1}) are produced in a back-to-back topology, $\Delta\phi = \pi$.
Small deviations from the back-to-back topology arise if soft gluons are emitted and/or if the struck parton carries a non-zero transverse momentum~\cite{Feng:2019,Liu:2020dct}.
Larger deviations from $\Delta\phi = \pi$ are expected when additional jets are produced through hard gluon radiation.
This sensitivity to various QCD phenomena, including both soft and hard processes, allows evaluation of theoretical models without explicitly describing the additional jets arising from higher-order ($\mathcal{O}(\alpha_s^k)$, $k > 0$) processes.

Azimuthal correlations in photoproduction have been studied by the ZEUS collaboration for various final-state systems~\cite{ZEUS:2001lmk,ZEUS:2005qyx,ZEUS:2007njl} to test the validity of perturbative QCD predictions.
Measurements of azimuthal correlations in multijet systems in hadron collisions have been performed by the D$\emptyset$ experiment at the Tevatron~\cite{D0:2005}, as well as by the CMS~\cite{CMS:2011, CMS:2016, CMS:2018} and ATLAS~\cite{ATLAS:2011, ATLAS:2018} experiments at the LHC, in order to investigate the effects of soft and hard QCD radiation in the high-energy regime.
The H1 collaboration recently published~\cite{H1:2021wkz} a measurement of the azimuthal correlation between the DIS scattered lepton and jets in the event.
The azimuthal correlation in dijet production in transversely polarised hadron collisions has been measured by the STAR experiment at RHIC~\cite{STAR:2023xvk}.

A study of $\Delta\phi$ between the scattered lepton and the jet of highest transverse momentum\footnote{From this point, these jets are referred to as the "leading jets".} in inclusive jet production in NC DIS at HERA is presented in this paper.
Differential cross sections of the pairs of lepton and leading jet were measured as a function of the azimuthal correlation angle using data collected with the ZEUS detector, representing an integrated luminosity of $326 \pb^{-1}$.
Jets were reconstructed with the $k_\mathrm{T}$ algorithm in the laboratory frame.
The measurement was performed for photon virtuality $10 \gev^2 < Q^2 < 350 \gev^2$, inelasticity\footnote{The inelasticity, $y$, quantifies the energy transfer from the electron to the hadronic system~\cite{H1:2015dma}.} $0.04 < y < 0.7$, and jet transverse momentum $2.5 \gev < p_\mathrm{T,jet} < 30 \gev$.
Calculations based on perturbative QCD~\cite{deFlorian:2020,deFlorian:2021} and predictions obtained from Monte Carlo (MC) simulations based on the ARIADNE colour-dipole model~\cite{Ariadne} for parton showering are compared to the extracted cross section.
The performance of these calculations in describing both soft and hard QCD processes and their evolution are evaluated.

% ----------------------------------------------------------------------------
%       Experimental set-up %This is mostly from the prompt photon paper-->
% ----------------------------------------------------------------------------
\section{Experimental set-up}
\label{sec-exp}

\Zdetdesc

%\Zctdmvdsttdesc{\ZcoosysfnBEeta}

In the kinematic range of the analysis, charged particles were tracked
in the central tracking detector (CTD)~\citeCTD, the microvertex
detector (MVD)~\citeMVD and the straw-tube tracker (STT)~\citeSTT. The CTD and the MVD
operated in a magnetic field of $1.43\Tesla$ provided by a thin
superconducting solenoid. The CTD drift chamber covered the
polar-angle\ZcoosysfnBEeta \; region \mbox{$15^\circ<\theta<164^\circ$}. The MVD
silicon tracker consisted of a barrel (BMVD) and a forward (FMVD)
section. The BMVD provided polar angle coverage for tracks with three
measurements from $30^\circ$ to $150^\circ$. The FMVD extended the
polar-angle coverage in the forward region to $7^\circ$. The STT
covered the polar-angle region \mbox{$5^\circ<\theta<25^\circ$}.

\Zcaldesc

%\Zlumidesc
%$\;$The fractional systematic uncertainty on the measured luminosity was $2{\%}$.

The luminosity was measured using the Bethe--Heitler reaction
$ep\,\rightarrow\, e\gamma p$ by a luminosity detector which consisted of independent lead-scintillator calorimeter\citePCAL and magnetic spectrometer\citeSPECTRO systems.
The fractional systematic
uncertainty on the measured luminosity was $1.9{\%}$.

% ----------------------------------------------------------------------------
%       Simulation
% ----------------------------------------------------------------------------
\section{Data sample and Monte Carlo simulation}
\label{sec-MC}

This analysis was performed using $ep$ collision data collected with the ZEUS detector in the years 2004--2007, comprising both $e^-p$ and $e^+p$ collisions.
The incoming energies of the leptons and protons were $27.5 \gev$ and $920 \gev$, respectively, corresponding to a centre-of-mass energy of $\sqrt{s} = 318 \gev$.
The integrated luminosity was $188 \pb^{-1}$ for $e^-p$ collisions and $138 \pb^{-1}$ for $e^+p$ collisions.
No significant dependence on the incoming lepton charge was observed in control distributions of the resulting DIS events and jets in the considered $Q^{2}$ range.

Monte Carlo samples were generated in the leading order (LO) plus parton showering (PS) approach.
Inclusive NC DIS samples were generated using DJANGOH 1.6 \cite{Djangoh} with the CTEQ5D PDF sets \cite{CTEQ5} for $Q^2 > 4\;\mathrm{GeV}^2$.
Hard parton scattering was simulated using LO matrix elements supplemented with the ARIADNE 4.12 parton-showering algorithm based on the colour-dipole model~\cite{Ariadne} to account for higher-order effects.
The parameters determining the performance of ARIADNE were over time tuned to ZEUS data, starting from those established in early studies~\cite{Ingelman:1996ge}.
The running of $\alpha_s$ was treated by ARIADNE using its default parameters, $\alpha_{s} = 12\pi/(33-2n_f)\ln(p_\perp^2/\Lambda_{QCD}^2)$, with $n_f = 5$ and $\Lambda_{QCD} = 0.22 \;\mathrm{GeV}$.
The simulation was performed without a diffractive contribution, and is referred to as LO+PS.
The Lund string model, implemented in JETSET 7.4.1 \cite{Lund}, was employed for hadronisation.
Hadronisation parameters were set to those determined from the ALEPH $e^+e^- \rightarrow Z$ data~\cite{ALEPH:1996oqp}.
The simulation included QED radiative corrections (single photon emission from initial- or final-state lepton, self-energy corrections to the exchanged boson, vertex corrections of the lepton-boson vertex) using HERACLES 4.5~\cite{Kwiatkowski:1990es}.
An additional set of simulations was generated using the MEPS model of LEPTO 6.5 \cite{lepto} to evaluate systematic uncertainties due to assumptions made in the ARIADNE model when extracting underlying hadron-level properties from the detector response.
The RAPGAP 3.308~\cite{Jung:1993gf} event generator was used to estimate effects from the initial- and final-state QED radiation to the measurement.
Photoproduction events were simulated using PYTHIA 6.4~\cite{Sjostrand:2006za} to estimate the photoproduction background.

To model the detector response, the generated MC events were processed through detector and trigger simulators, using GEANT 3.21 \cite{Geant,ZGANA:1990}.
The resulting MC samples were normalized to the luminosity of the data.

%\clearpage

% ----------------------------------------------------------------------------
%       Event Selection
% ----------------------------------------------------------------------------
\section{Event selection}
\label{sec-sel}

The ZEUS experiment operated using a three-level trigger system \cite{zeus:1993:bluebook, ZEUS_trigger:1992, ZEUS_trigger:2007} to give a preselection of NC DIS events.
This triggering scheme was based on an energy-deposit pattern in the CAL consistent with an isolated electron, along with additional threshold requirements on the energy and longitudinal momentum of the electron.
Further offline selection of NC DIS events was performed using a methodology that was employed in previous ZEUS analyses related to jet production in DIS \cite{ZEUS_jets:2010, Jets_dis:2012, Jets_dis:2018}.

The following selection criteria were applied to select a clean DIS sample:
\begin{itemize}
\item the inelasticity of the interactions was constrained to $y_\mathrm{JB} > 0.04$ using the Jacquet--Blondel method \cite{Amaldi:1979yh}, and $y_\mathrm{el} < 0.7$ using the electron method \cite{Bentvelsen:1992fu}.
The photon virtuality was determined using the double-angle method \cite{Bentvelsen:1992fu} and was required to satisfy $10 \gev^2 < Q^2_\mathrm{DA} < 350 \gev^2$.
These kinematic selections provided access to a true range of the Bjorken-scaling variable, $x_\mathrm{Bj}$~\cite{H1:2015dma}, $0.0002 < x_\mathrm{Bj} < 0.1$;

\item the event vertex position was required to satisfy $|Z_{\text{vtx}}| < 40 \cm$ in order to reduce background contributions from non-$ep$ collisions;

\item $E - p_Z$ is defined as the difference between the total energy and the $Z$ component of final-state momentum and is measured as $E - p_Z = \sum_{i}E_i(1-\cos \theta_i)$, where $E_i$ is the energy of the $i$-th cell of the CAL and $\theta_i$ is its polar angle.
The sum runs over all cells in the detector.
In a fully contained event, the value of $E - p_Z$ is expected to be around twice the energy of the incoming electron, $\sim 55 \gev$~\cite{ZEUS_empz}.
Events were required to have $45 \gev < E - p_Z < 65 \gev$;

%\item A selection cut of $p_T/\sqrt{E_T} < 2.5 \gev^{1/2}$ was applied to reduce contributions from cosmic rays and beam-related background events.

\item the total transverse momentum of the event was required to be consistent with zero by demanding $p_\mathrm{T}/\sqrt{E_\mathrm{T}} < 2.5 \gev^{1/2}$, where $p_\mathrm{T}$ and $E_\mathrm{T}$ are sums of the individual vectorial transverse momenta and scalar transverse energies of all energy deposits in the CAL, respectively;

\item DIS electrons were selected using a neural network algorithm, SINISTRA~\cite{ZEUS_neural}, based on the energy deposit pattern in the CAL.
The requirements were a probability $> 90\;\%$, an energy $E_{e} > 10 \gev$, a polar angle $\theta_{e} > 140^\circ$, and $r_\mathrm{RCAL} > 20 \cm$, where $r_\mathrm{RCAL}$ is the radius of the impact point on the RCAL\footnote{This effectively imposed an upper bound on electron polar angle at approximately $\theta_{e} \lessapprox 175^\circ$.};

\item electrons typically deposit most of their energy in an isolated region in the CAL.
Excluding the energy in the CAL cell containing the position of the electron candidate and its nearest neighbours, the energy deposit in a cone of radius $\sqrt{\Delta \eta^2 + \Delta \phi^2} < 0.8$ around the position of the electron candidate was required to be less than $10\;\%$ of the total energy in the cone in order to select electrons well separated from hadronic activity;

\item electron candidates whose impact point on the RCAL fell within the rectangular region defined by $-14\;\mathrm{cm} < X_\mathrm{RCAL} < 12\;\mathrm{cm}$ and $Y_\mathrm{RCAL} > 90\;\mathrm{cm}$ were excluded from further analysis.
This region was occupied by the cooling pipe for the solenoid.
\end{itemize}

Jets were reconstructed in the laboratory frame with the $k_\mathrm{T}$-clustering algorithm \cite{Jet_kt:1993a} using the $E$-recombination scheme in the longitudinally-invariant inclusive mode \cite{Jet_kt:1993b} with the jet-radius parameter set to $1.0$.
The reconstruction was carried out using the FastJet 3.4.0 package~\cite{Cacciari:2011ma,Cacciari:2005hq}.
Calorimeter clusters and tracks were combined to form Energy Flow Objects (EFOs)~\cite{EFO:2000}.
Event kinematics were reconstructed based on the EFOs.
Four-vector information of all EFOs, except for SINISTRA electron candidates, was used as input for the jet reconstruction.
Reconstructed jets that satisfied the following criteria were selected for further analysis:
transverse momentum of the jets within the range of $2.5 \gev < p_\mathrm{T,jet} < 30 \gev$, and jet pseudorapidity within $-1.5 < \eta_\mathrm{jet} < 1.8$.
If more than one jet passed these criteria in an event, that with the highest $p_\mathrm{T,jet}$ was chosen as the leading jet.

The final sample consisted of approximately $1.2 \times 10^7$ DIS events with at least one jet that passed both the event-selection and jet-selection criteria.
The contribution from remaining photoproduction events was found to be negligible after the DIS selection, being below $1\%$.
Comparisons of reconstructed DIS kinematic quantities between the data and MC simulations after all cuts are illustrated in Fig.~\ref{QA_event} and for lepton and leading-jet quantities in Fig.~\ref{QA_leptonjet}.
Both ARIADNE and LEPTO describe the data well.
Differences between the two MC simulations were used to determine systematic uncertainties.

%Systematic dependence of the measurement on the exact values used as selection thresholds/windows was estimated by varying these values by their respective reconstruction resolution.

%\clearpage
% ----------------------------------------------------------------------------
%       Cross section measurement
% ----------------------------------------------------------------------------
\section{Signal extraction}
\label{sec-Un}

Distributions of the azimuthal correlation angle obtained with the reconstructed electron and leading jet, $\Delta\phi_\mathrm{det}$, are shown in Fig.~\ref{QA_dphi} for various jet-multiplicity ranges.
The flattening of the event distribution as $\Delta\phi \rightarrow \pi$ for multijet ($N_\mathrm{jet} \geq 2$) events is consistent with the absence of the Born-level DIS process.
The presence of additional jets, arising from $\mathcal{O}(\alpha_s^{k>0})$ processes, such as soft or hard gluon radiation, leads to deviations from a purely back-to-back configuration.

To extract the underlying hadron-level signal, a regularised unfolding was performed using the TUnfold package~\cite{TUnfold}.
The migration matrix describes the detector and reconstruction effects on the hadron-level objects.
The ARIADNE program was chosen to generate the input to the unfolding procedure because it provides a better description of the shapes of the $\Delta\phi$ distributions (see Fig.~\ref{QA_dphi}).
Hadron jets were reconstructed in the laboratory frame using all the final-state ARIADNE particles\footnote{Final-state particles are defined as any stable particle whose lifetime is longer than $10\;\mathrm{ps}$.} except for the scattered electron and neutrinos.
The reconstruction was performed using the $k_\mathrm{T}$-clustering algorithm with the $E$-recombination scheme in the longitudinally invariant inclusive mode \cite{Jet_kt:1993b}, as implemented in the FastJet 3.4.0 package \cite{Cacciari:2011ma,Cacciari:2005hq}.
The jet-radius parameter was set to $R = 1.0$.
The kinematics of each event was obtained based on the scattered electron~\cite{Bentvelsen:1992fu} and the correlation angle $\Delta\phi$ was calculated from the azimuthal angles of the true electron (after both initial- and final-state QED radiations) and the hadron jets.
A detailed description of the unfolding scheme used in this analysis can be found in Appendix~\ref{app_unfold}.

Corrections were applied to account for three different effects arising from the migration of reconstructed quantities.
First, events can falsely enter into the fiducial region of the measurement defined by the reconstructed kinematic quantities, resulting in an impurity in the signal.
The impurity was estimated using the MC simulation and subtracted from the signal.
Secondly, migrations can occur between $\Delta\phi$ and $N_\mathrm{jet}$ bins.
A regularised unfolding, as implemented in TUnfold, was used to correct for this effect.
Lastly, events that falsely fell out of the fiducial region defined by the hadron-level kinematics were corrected bin-wise by factors obtained from the simulation.

% To systematics
%The correlation between detector-level and hadron-level quantities given in the MC simulation relies on the assumptions made in the chosen hadronisation and parton-shower models.
%The impact of these assumptions was evaluated by repeating the unfolding procedure using an alternative MC simulation sample based on the MEPS-LEPTO model.
%The difference in the extracted signal obtained from the two simulation models was considered a systematic uncertainty, reflecting the dependence on these assumptions.
%Also, the dependence of the impurity estimate on data-MC agreement in $\Delta\phi_\mathrm{det}$ and $N_\mathrm{jet,det}$ was considered a systematic uncertainty.

% ----------------------------------------------------------------------------
%       Systematics
% ----------------------------------------------------------------------------
\section{Systematic uncertainties}
\label{sec-sys}

The following sources of systematic uncertainty were investigated:

\begin{itemize}
\item the uncertainty associated with the choice of the regularisation parameter $\tau$ used in the unfolding procedure, as suggested by the TUnfold package, was propagated into the cross section.
Its contribution to the total uncertainty was found to be less than $1\%$ throughout the entire range of $\Delta\phi$, and was neglected;

\item the systematic effect of the $2\%$ uncertainty in the energy scale of the scattered electron measured in the calorimeter was estimated by varying the energy scale in the MC.
Its effect was found to be negligible;

\item the jet-energy scale was varied within its uncertainty of $\pm2.5\%$ for jet transverse energies $E_\mathrm{T,jet}$ < 10 GeV and $\pm1.5\%$ for $E_\mathrm{T,jet}$ > 10 GeV~\cite{Jets_dis:2018} in the MC and found to be negligible;

\item the dependence on the specific values chosen for the event selection was estimated by varying the values by the reconstruction resolution;

\item the uncertainty related to the method used for estimating the impurity background was evaluated by performing the measurement using an alternative approach (see Appendix~\ref{app_unfold}).
The systematic uncertainty associated with the choice of impurity estimation method was determined by comparing the results derived from the nominal and alternative methods;

\item the uncertainty associated with assumptions made in the ARIADNE simulation during the signal extraction process and in the cross-section calculation was evaluated by performing the measurement using an alternative MC sample based on the MEPS-LEPTO model.
The difference in the resulting cross sections was taken as the systematic dependence on the simulation model.

\end{itemize}

All significant uncertainties were symmetrised and added in quadrature to obtain the total systematic uncertainty.
The individual systematic uncertainties compared to the statistical uncertainty are shown in Fig.~\ref{syst_0} for the full fiducial region, while $p_\mathrm{T,jet}^\mathrm{lead} \otimes N_\mathrm{jet}$- and $Q^2 \otimes N_\mathrm{jet}$-dependent comparisons are provided in Appendix~\ref{app_syst}.

%\clearpage

% ----------------------------------------------------------------------------
%       Theory prediction
% ----------------------------------------------------------------------------
\section{Theory predictions}
\label{sec-Th}

%parton jet + hadronization correction.

Fixed-order pQCD predictions for $d\sigma(e + p \rightarrow e + \mathrm{jet}^\mathrm{lead} + X) / d\Delta\phi$ were computed by Borsa et al.~\cite{Ignacio} using the projection-to-Born (P2B) method~\cite{P2B:2015,P2B:2018}, as implemented in the POLDIS framework~\cite{deFlorian:2020,deFlorian:2021}.
The P2B method uses a dijet calculation at $\mathcal{O}(\alpha_s^{k-1})$ accuracy and a fully inclusive calculation at $\mathcal{O}(\alpha_s^{k})$ to produce an $\mathcal{O}(\alpha_s^{k})$ single-inclusive-jet ($N_\mathrm{jet} \geq 1$) prediction.
The $\mathcal{O}(\alpha_s^1)$ results for dijet production, adapted for HERA parameters, were used to produce single-inclusive-jet calculations up to $\mathcal{O}(\alpha_s^2)$ accuracy~\cite{deFlorian:2020,deFlorian:2021}.
These calculations were performed in the laboratory frame.
The unpolarised PDF4LHC15 PDF set~\cite{PDF4LHC15} was used as input to the calculation.
Factorisation and renormalisation scales were chosen as $\mu_F^2=\mu_R^2=Q^2$.
A central value of $\alpha_s = 0.118$ evaluated at $\mu_F = \mu_R = M_Z$ was used.
The theory uncertainty was determined from a seven-point scale variation with rescaling factors [$1/2$, $2$].

These calculations were performed with massless parton jets.
The predicted cross sections were corrected for the effects of hadronisation using results based on an ARIADNE MC study.
A detailed description of the correction procedure is given in Appendix~\ref{app_hadcor}.

% appendix D

%\clearpage

% ----------------------------------------------------------------------------
%       Cross section measurement
% ----------------------------------------------------------------------------
\section{Differential cross sections}
\label{sec-DC}

The differential cross section of inclusive jet production in NC DIS, $d\sigma(e+p \rightarrow e + \mathrm{jet}^\mathrm{lead} + X) / d\Delta\phi$, was measured in the laboratory frame as a function of the azimuthal correlation angle between the scattered lepton and the leading jet, % (leading jet) within ... not (leading) jet within..
within the kinematic space defined by a range of the photon virtuality $10 \gev^2 < Q^2 < 350 \gev^2$;
inelasticity $0.04 < y < 0.7$;
electron energy $E_{e} > 10 \gev$;
electron polar angle $140^\circ < \theta_{e} < 180^\circ$;
jet transverse momentum $2.5 \gev < p_\mathrm{T,jet} < 30 \gev$;
and jet pseudorapidity $-1.5 < \eta_\mathrm{jet} < 1.8$ as follows:

\begin{equation}
\frac{d\sigma}{d\Delta\phi}  (e+p \rightarrow e + \mathrm{jet}^\mathrm{lead} + X)
=          \frac{1}{\mathcal{L}}
\cdot   c_\mathrm{QED}
\cdot   c_\mathrm{L}
\cdot   \frac{N_\mathrm{had}}{\delta \Delta\phi}.
\end{equation}

Here, $\mathcal{L}$ represents the integrated luminosity,
$N_{\mathrm{had}}$ is the extracted signal as a distribution of $\Delta\phi$ corrected for migration effects,
and $\delta \Delta\phi$ is the width of each $\Delta\phi$ bin.
The effects of both initial- and final-state QED radiation off the electron were estimated with RAPGAP (see Appendix~\ref{app_QED}), and the corresponding QED correction factors, $c_{\mathrm{QED}} (\Delta\phi)$, were applied to extract the cross sections before such radiation.
A non-leading jet may be falsely tagged as the leading jet if the true leading jet points too far forward or backward, e.g., through the beam pipe, or its transverse momentum exceeds the upper limit for reconstructed jets.
The effects of incorrectly assigned leading jets were estimated with ARIADNE, and the corresponding correction factors, $c_\mathrm{L}(\Delta\phi)$, were applied.

The inclusive ($N_\mathrm{jet} \geq 1$) differential cross section, $d\sigma(e+p \rightarrow e + \mathrm{jet}^\mathrm{lead} + X) / d\Delta\phi$, of lepton--leading-jet pairs integrated over the studied fiducial region is presented as a function of the lepton--leading-jet correlation angle $\Delta\phi$ in Fig.~\ref{final_1_0}.
Perturbative calculations, treating up to $O(\alpha_s^{2})$ or $O(\alpha_s)$ contributions (see Sec.~\ref{sec-Th}), are compared to the measured cross sections within the $\Delta\phi$ range, $7\pi/15 < \Delta\phi < \pi$.
.

Additional measurements were performed for various ranges of $N_\mathrm{jet}$, $p_\mathrm{T,jet}^\mathrm{lead}$, and $Q^2$.
The $p_\mathrm{T,jet}^\mathrm{lead}$ ranges were divided into three intervals: $2.5$--$7 \gev$, $7$--$12 \gev$, and $12$--$30 \gev$.
The $Q^2$ ranges were $10$--$50 \gev^2$, $50$--$100 \gev^2$, and $100$--$350 \gev^2$.
The jet-multiplicity range was varied as $N_\mathrm{jet} \geq 1$, $\geq 2$, and $\geq 3$ for each $p_\mathrm{T,jet}^\mathrm{lead}$ and $Q^{2}$ range.
Figures~\ref{final_1_1} and~\ref{final_1_2} present the differential cross sections for various ranges of $p_\mathrm{T,jet}^\mathrm{lead} \otimes N_\mathrm{jet}$ and $Q^2 \otimes N_\mathrm{jet}$, and the respective theory predictions.

Figure~\ref{final_2_0} shows a comparison between the measured inclusive cross section and the prediction obtained from the ARIADNE LO+PS simulation.
Comparisons between the data and ARIADNE cover the full range of $\Delta\phi$ from $0$ to $\pi$.
In Figs.~\ref{final_2_1} and~\ref{final_2_2}, comparisons between the data and ARIADNE are presented for various ranges of $p_\mathrm{T,jet}^\mathrm{lead} \otimes N_\mathrm{jet}$ and $Q^2 \otimes N_\mathrm{jet}$, respectively.

Numerical values of the measurements, perturbative calculations, and ARIADNE predictions are summarised in Table~\ref{TAB_00} for the inclusive measurement, Tables~\ref{TAB_10}, ~\ref{TAB_11}, and~\ref{TAB_12} for the studied ranges of $p_\mathrm{T,jet}^\mathrm{lead}$ and $N_\mathrm{jet}$, and Tables~\ref{TAB_20}, ~\ref{TAB_21}, and~\ref{TAB_22} for the $Q^2$ and $N_\mathrm{jet}$ ranges.

%\clearpage

% ----------------------------------------------------------------------------
%       Discussions
% ----------------------------------------------------------------------------
\section{Discussion}
\label{sec-res}

% figure~\ref{final_1_0}
Fixed-order calculations at $\mathcal{O}(\alpha_s^2)$ and $\mathcal{O}(\alpha_s)$ accuracy are compared to the inclusive measurement in Fig.~\ref{final_1_0}.
The $\mathcal{O}(\alpha_s^2)$ corrections are NNLO for the last $\Delta\phi$ bin.
An additional gluon is required for the leading jet to diverge from the back-to-back topology with respect to the scattered lepton,
i.e., $\mathcal{O}(\alpha_s)$ is LO and $\mathcal{O}(\alpha_s^2)$ is NLO for $\Delta\phi < \pi$.
Here, the $\mathcal{O}(\alpha_s^2)$ calculation demonstrates a clear improvement compared to $\mathcal{O}(\alpha_s)$, especially in the region $\Delta\phi < 3\pi/4$ where contributions from additional hard jet production significantly alter lepton--leading-jet production away from the back-to-back topology.
On the other hand, no significant improvement is observed in the region $\Delta\phi \rightarrow \pi$, where substantial contributions from soft gluon radiation and intrinsic parton transverse momentum, $k_\mathrm{T}$, are expected.
This is consistent with the findings of D$\emptyset$~\cite{D0:2005}, CMS~\cite{CMS:2011, CMS:2016, CMS:2018}, ATLAS~\cite{ATLAS:2011}, and H1~\cite{H1:2021wkz}.
Figure~\ref{final_1_0} shows that perturbative predictions up to $\mathcal{O}(\alpha_s^2)$ already provide a good description of the ZEUS data.

The fixed-order calculations were performed at the $\mathcal{O}(\alpha_s)$ and $\mathcal{O}(\alpha_s^2)$ accuracy for $N_\mathrm{jet} \geq 1$ and $N_\mathrm{jet} \geq 2$, while only the $\mathcal{O}(\alpha_s^2)$ calculation applies for $N_\mathrm{jet} \geq 3$, where this is effectively the leading order.
The theory predictions are compared to the measurements in various ranges of $p_\mathrm{T,jet}^\mathrm{lead}$ and $Q^2$ in Figs.~\ref{final_1_1} and ~\ref{final_1_2}, respectively.
In all cases, the $\mathcal{O}(\alpha_s^2)$ corrections significantly improved agreement with the data in the region $\Delta \phi < 3\pi/4$, which is sensitive to additional hard jet production.
This improvement extends into the low-$p_\mathrm{T,jet}$ regime, reaching down to $p_\mathrm{T,jet} > 2.5 \gev$.
An enhancement of events displaying a reduced azimuthal correlation angle with increasing jet multiplicity is observed.
This is expected and consistent with previous findings~\cite{CMS:2018}.

The slope of the measured cross section increases as a function of $Q^{2}$, as the higher-order contributions are suppressed for $N_\mathrm{jet} \geq 1$.
For $N_\mathrm{jet} \geq 2$, the slope only increases with $Q^2$ for $\Delta\phi < 3\pi/4$.
The scattered electron in $ep$ DIS is analogous to one of the two jets in dijet production in hadron collisions if the electron $p_\mathrm{T}$ is larger than the second-highest jet $p_\mathrm{T}$.
Measurements from hadron colliders report a similar trend where the slope of the dijet cross section grows with increasing transverse momentum of the highest-$p_\mathrm{T}$ jet~\cite{CMS:2011,CMS:2016,CMS:2018,ATLAS:2011}.
Furthermore, an improvement in agreement between the data and perturbative calculations is found for single-inclusive ($N_\mathrm{jet} \geq 1$) events near $\Delta\phi \rightarrow \pi$ with increasing $Q^2$.
This finding is consistent with the expected suppression of soft gluon radiation and parton $k_\mathrm{T}$ effects in the high-$Q^{2}$ regime.
No significant dependence of the shape of cross section on $p_\mathrm{T,jet}^\mathrm{lead}$ is observed in the present measurement.

In Fig.~\ref{final_2_0}, the LO+PS prediction derived from the ARIADNE model is compared to the inclusive measurement.
Here, there is a notable success of the LO+PS approach in describing higher-order processes characterised by $\Delta\phi < 3\pi/4$ and $\Delta\phi \rightarrow \pi$, even though they are not fully represented in the LO matrix elements.
In Figs.~\ref{final_2_1} and ~\ref{final_2_2}, $p_\mathrm{T,jet}^\mathrm{lead} \otimes N_\mathrm{jet}$- and $Q^2 \otimes N_\mathrm{jet}$-dependent studies are illustrated.
The performance of ARIADNE is comparable to that of perturbative calculations at $\mathcal{O}(\alpha_s^2)$ accuracy across all ranges of $p_\mathrm{T,jet}^\mathrm{lead}$, $Q^2$, and $N_\mathrm{jet}$.
These observations are consistent with previous findings~\cite{CMS:2016,CMS:2018} and further support the validity of LO+PS approaches in describing a wide range of characteristics of the data.
However, in contrast to the data, ARIADNE predicts an enhancement of events displaying a reduced azimuthal correlation angle with increasing $p_\mathrm{T,jet}^\mathrm{lead}$.
In particular, deviations emerge in $\Delta\phi \rightarrow \pi$ for $N_\mathrm{jet} \geq 2$ and low-$p_\mathrm{T,jet}^\mathrm{lead}$ for all ranges in $Q^{2}$.
This observation might provide information on how parton showering could be improved to describe higher-order processes better, e.g., providing a basis for further tuning of hadronisation parameters.

%\clearpage

% ----------------------------------------------------------------------------
%       Summary
% ----------------------------------------------------------------------------
\section{Summary}
\label{sec-con}

Azimuthal correlations between the scattered lepton and the leading jet in NC DIS at HERA have been measured at ZEUS.
The resulting $\Delta\phi$ distribution was unfolded to hadron level, correcting migration effects in the reconstructed kinematics.
The differential cross section, $d\sigma(e+p \rightarrow e + \mathrm{jet}^\mathrm{lead} + X) / d\Delta\phi$, was derived from the unfolded $\Delta\phi$ distribution in the fiducial region defined by $10 \gev^2 < Q^2 < 350 \gev^2$, $0.04 < y < 0.7$, $E_e > 10 \gev$, $140^\circ < \theta_e < 180^\circ$, $2.5 \gev < p_\mathrm{T,jet} < 30 \gev$, and $-1.5 < \eta_\mathrm{jet} < 1.8$ in the laboratory frame.
The measurement was also performed for various ranges of $p_\mathrm{T,jet}^\mathrm{lead}$, $Q^2$, and $N_\mathrm{jet}$.
The experimental uncertainty was dominated by the systematic dependence on the simulation model used in the unfolding procedure.

Perturbative calculations~\cite{Ignacio,deFlorian:2020,deFlorian:2021} up to $O(\alpha_s^2)$ and MC predictions based on the LO+PS approach implemented in the ARIADNE colour-dipole model have been compared to the data.
The higher-order pQCD corrections show a significant improvement in describing regions that are driven by hard jet production.
In addition, the excellent performance of perturbative calculations has been verified for jet transverse momentum down to $p_\mathrm{T,jet}^\mathrm{lead} > 2.5 \gev$.
The LO+PS predictions by the well-tuned ARIADNE model also describe the data well.
The analysis procedures and results presented in this paper can be important in planning future experiments, e.g., at the Electron-Ion Collider (EIC)~\cite{Accardi:2012qut,AbdulKhalek:2021gbh}.

%\clearpage

% ----------------------------------------------------------------------------
%       Mandatory acknowledgements. You may add your buddies to it.
% ----------------------------------------------------------------------------
\section*{Acknowledgements}
\label{sec-ack}
\Zacknowledge

{
\ifzeusbst
  \ifzmcite
     \bibliographystyle{./BiBTeX/bst/l4z_default3}
  \else
     \bibliographystyle{./BiBTeX/bst/l4z_default3_nomcite}
  \fi
\fi
\ifzdrftbst
  \ifzmcite
    \bibliographystyle{./BiBTeX/bst/l4z_draft3}
  \else
    \bibliographystyle{./BiBTeX/bst/l4z_draft3_nomcite}
  \fi
\fi
\ifzbstepj
  \ifzmcite
    \bibliographystyle{./BiBTeX/bst/l4z_epj3}
  \else
    \bibliographystyle{./BiBTeX/bst/l4z_epj3_nomcite}
  \fi
\fi
\ifzbstjhep
  \ifzmcite
    \bibliographystyle{./BiBTeX/bst/l4z_jhep3}
  \else
    \bibliographystyle{./BiBTeX/bst/l4z_jhep3_nomcite}
  \fi
\fi
\ifzbstnp
  \ifzmcite
    \bibliographystyle{./BiBTeX/bst/l4z_np3}
  \else
    \bibliographystyle{./BiBTeX/bst/l4z_np3_nomcite}
  \fi
\fi
\ifzbstpl
  \ifzmcite
    \bibliographystyle{./BiBTeX/bst/l4z_pl3}
  \else
    \bibliographystyle{./BiBTeX/bst/l4z_pl3_nomcite}
  \fi
\fi
{\raggedright
\bibliography{%./syn.bib,%
              ./myref.bib,%
              ./BiBTeX/bib/l4z_zeus.bib,%
              ./BiBTeX/bib/l4z_h1.bib,%
              ./BiBTeX/bib/l4z_articles.bib,%
              ./BiBTeX/bib/l4z_books.bib,%
              ./BiBTeX/bib/l4z_conferences.bib,%
              ./BiBTeX/bib/l4z_misc.bib,%
              ./BiBTeX/bib/l4z_preprints.bib}}

\providecommand{\urlprefix}{}
\providecommand{\etal}{et al.\xspace}
\providecommand{\coll}{Collab.\xspace}
\catcode`\@=11
\def\@bibitem#1{%
\ifmc@bstsupport
  \mc@iftail{#1}%
    {;\newline\ignorespaces}%
    {\ifmc@first\else.\fi\orig@bibitem{#1}}
  \mc@firstfalse
\else
  \mc@iftail{#1}%
    {\ignorespaces}%
    {\orig@bibitem{#1}}%
\fi}%
\catcode`\@=12
\begin{mcbibliography}{10}

\bibitem{H1:2021xxi}
H1 and ZEUS \coll, I. Abt \emph{et al}.,
\newblock \href{http://dx.doi.org/10.1140/epjc/s10052-022-10083-9}{Eur. Phys.
  J. C{} {\bfseries 82},~243~(2022)}\relax
\relax
\bibitem{ZEUS:2023zie}
ZEUS \coll, I. Abt \emph{et al}.,
\newblock \href{http://dx.doi.org/10.1140/epjc/s10052-023-12180-9}{Eur. Phys.
  J. C{} {\bfseries 83},~1082~(2023)}\relax
\relax
\bibitem{Feng:2019}
X. Liu \emph{et al.},
\newblock \href{http://dx.doi.org/10.1103/PhysRevLett.122.192003}{Phys. Rev.
  Lett.{} {\bfseries 122},~192003~(2019)}\relax
\relax
\bibitem{Liu:2020dct}
X. Liu \emph{et al.},
\newblock \href{http://dx.doi.org/10.1103/PhysRevD.102.094022}{Phys. Rev. D{}
  {\bfseries 102},~094022~(2020)}\relax
\relax
\bibitem{ZEUS:2001lmk}
ZEUS \coll, S. Chekanov \emph{et al.},
\newblock \href{http://dx.doi.org/10.1016/S0370-2693(01)00615-3}{Phys. Lett.
  B{} {\bfseries 511},~19~(2001)}\relax
\relax
\bibitem{ZEUS:2005qyx}
ZEUS \coll, S. Chekanov \emph{et al.},
\newblock \href{http://dx.doi.org/10.1016/j.nuclphysb.2005.09.021}{Nucl. Phys.
  B{} {\bfseries 729},~492~(2005)}\relax
\relax
\bibitem{ZEUS:2007njl}
ZEUS \coll, S. Chekanov \emph{et al.},
\newblock \href{http://dx.doi.org/10.1103/PhysRevD.76.072011}{Phys. Rev. D{}
  {\bfseries 76},~072011~(2007)}\relax
\relax
\bibitem{D0:2005}
{D$\emptyset$ \coll, V. Abazov \emph{et al.}},
\newblock \href{http://dx.doi.org/10.1103/PhysRevLett.94.221801}{Phys. Rev.
  Lett.{} {\bfseries 94},~221801~(2005)}\relax
\relax
\bibitem{CMS:2011}
CMS \coll V. Khachatryan \emph{et al.},
\newblock \href{http://dx.doi.org/10.1103/PhysRevLett.106.122003}{Phys. Rev.
  Lett.{} {\bfseries 106},~122003~(2011)}\relax
\relax
\bibitem{CMS:2016}
CMS \coll V. Khachatryan \emph{et al.},
\newblock \href{http://dx.doi.org/10.1140/epjc/s10052-016-4346-8}{Eur. Phys. J.
  C{} {\bfseries 76},~536~(2016)}\relax
\relax
\bibitem{CMS:2018}
CMS \coll, A. Sirunyan \emph{et al.},
\newblock \href{http://dx.doi.org/10.1140/epjc/s10052-018-6033-4}{Eur. Phys. J.
  C{} {\bfseries 78},~566~(2018)}\relax
\relax
\bibitem{ATLAS:2011}
ATLAS \coll, G. Aad \emph{et al.},
\newblock \href{http://dx.doi.org/10.1103/PhysRevLett.106.172002}{Phys. Rev.
  Lett.{} {\bfseries 106},~172002~(2011)}\relax
\relax
\bibitem{ATLAS:2018}
ATLAS \coll M. Aaboud \emph{et al.},
\newblock \href{http://dx.doi.org/10.1103/PhysRevD.98.092004}{Phys. Rev. D{}
  {\bfseries 98},~092004~(2018)}\relax
\relax
\bibitem{H1:2021wkz}
H1 \coll V. Andreev \emph{et al.},
\newblock \href{http://dx.doi.org/10.1103/PhysRevLett.128.132002}{Phys. Rev.
  Lett.{} {\bfseries 128},~132002~(2022)}\relax
\relax
\bibitem{STAR:2023xvk}
STAR \coll, arXiv:2305.10359\relax
\relax
\bibitem{H1:2015dma}
H1 and ZEUS \coll, H. Abramowicz \emph{et al.},
\newblock \href{http://dx.doi.org/10.1007/JHEP09(2015)149}{JHEP{} {\bfseries
  09},~149~(2015)}\relax
\relax
\bibitem{deFlorian:2020}
I. Borsa \emph{et al.},
\newblock \href{http://dx.doi.org/10.1103/PhysRevLett.125.082001}{Phys. Rev.
  Lett.{} {\bfseries 125},~082001~(2020)}\relax
\relax
\bibitem{deFlorian:2021}
I. Borsa \emph{et al.},
\newblock \href{http://dx.doi.org/10.1103/PhysRevD.103.014008}{Phys. Rev. D{}
  {\bfseries 103},~014008~(2021)}\relax
\relax
\bibitem{Ariadne}
L. L{\"o}nnblad,
\newblock \href{http://dx.doi.org/10.1016/0010-4655(92)90068-A}{Comput. Phys.
  Commun.{} {\bfseries 71},~15~(1992)}\relax
\relax
\bibitem{zeus:1993:bluebook}
ZEUS \coll, U.~Holm~(ed.),
\newblock {\slshape The {ZEUS} Detector}.
\newblock Status Report (unpublished), DESY (1993).
\newblock \urlprefix\url{http://www-zeus.desy.de/bluebook/bluebook.html}\relax
\relax
\bibitem{nim:a279:290}
N.~Harnew \etal,
\newblock \href{http://dx.doi.org/10.1016/0168-9002(89)91096-6}{Nucl.\ Instr.\
  and Meth.{} {\bfseries A~279},~290~(1989)}\relax
\relax
\bibitem{npps:b32:181}
B.~Foster \etal,
\newblock \href{http://dx.doi.org/10.1016/0920-5632(93)90023-Y}{Nucl.\ Phys.\
  Proc.\ Suppl.{} {\bfseries B~32},~181~(1993)}\relax
\relax
\bibitem{nim:a338:254}
B.~Foster \etal,
\newblock \href{http://dx.doi.org/10.1016/0168-9002(94)91313-7}{Nucl.\ Instr.\
  and Meth.{} {\bfseries A~338},~254~(1994)}\relax
\relax
\bibitem{nim:a581:656}
A.~Polini \etal,
\newblock \href{http://dx.doi.org/10.1016/j.nima.2007.08.167}{Nucl.\ Instr.\
  and Meth.{} {\bfseries A~581},~656~(2007)}\relax
\relax
\bibitem{nim:a535:191}
S.~Fourletov,
\newblock \href{http://dx.doi.org/10.1016/j.nima.2004.07.212}{Nucl.\ Instr.\
  and Meth.{} {\bfseries A~535},~191~(2004)}\relax
\relax
\bibitem{nim:a309:77}
M.~Derrick \etal,
\newblock \href{http://dx.doi.org/10.1016/0168-9002(91)90094-7}{Nucl.\ Instr.\
  and Meth.{} {\bfseries A~309},~77~(1991)}\relax
\relax
\bibitem{nim:a309:101}
A.~Andresen \etal,
\newblock \href{http://dx.doi.org/10.1016/0168-9002(91)90095-8}{Nucl.\ Instr.\
  and Meth.{} {\bfseries A~309},~101~(1991)}\relax
\relax
\bibitem{nim:a321:356}
A.~Caldwell \etal,
\newblock \href{http://dx.doi.org/10.1016/0168-9002(92)90413-X}{Nucl.\ Instr.\
  and Meth.{} {\bfseries A~321},~356~(1992)}\relax
\relax
\bibitem{nim:a336:23}
A.~Bernstein \etal,
\newblock \href{http://dx.doi.org/10.1016/0168-9002(93)91078-2}{Nucl.\ Instr.\
  and Meth.{} {\bfseries A~336},~23~(1993)}\relax
\relax
\bibitem{desy-92-066}
J.~Andruszk\'ow \etal,
\newblock Preprint \mbox{DESY-92-066}, DESY, 1992\relax
\relax
\bibitem{zfp:c63:391}
ZEUS \coll, M.~Derrick \etal,
\newblock \href{http://dx.doi.org/10.1007/BF01580320}{Z.\ Phys.{} {\bfseries
  C~63},~391~(1994)}\relax
\relax
\bibitem{acpp:b32:2025}
J.~Andruszk\'ow \etal,
\newblock Acta Phys.\ Pol.{} {\bfseries B~32},~2025~(2001)\relax
\relax
\bibitem{nim:a565:572}
M.~Helbich \etal,
\newblock \href{http://dx.doi.org/10.1016/j.nima.2006.06.049}{Nucl.\ Instr.\
  and Meth.{} {\bfseries A~565},~572~(2006)}\relax
\relax
\bibitem{Djangoh}
H. Spiesberger.
\newblock
  \url{http://wwwthep.physik.uni-mainz.de/~hspiesb/djangoh/djangoh.html}\relax
\relax
\bibitem{CTEQ5}
CTEQ \coll, H. Lai \emph{et al.},
\newblock \href{http://dx.doi.org/10.1007/s100529900196}{Eur. Phys. J. C{}
  {\bfseries 12},~375~(2000)}\relax
\relax
\bibitem{Ingelman:1996ge}
N. Brook \emph{et al.},
\newblock {\slshape {Future Physics at HERA}}.
\newblock  (1996)\relax
\relax
\bibitem{Lund}
T. Sj{\"o}strand,
\newblock \href{http://dx.doi.org/10.1016/0010-4655(94)90132-5}{Comput. Phys.
  Commun.{} {\bfseries 82},~74~(1994)}\relax
\relax
\bibitem{ALEPH:1996oqp}
ALEPH \coll, R. Barate \emph{et al.},
\newblock \href{http://dx.doi.org/10.1016/S0370-1573(97)00045-8}{Phys. Rept.{}
  {\bfseries 294},~1~(1998)}\relax
\relax
\bibitem{Kwiatkowski:1990es}
A. Kwiatkowski \emph{et al.},
\newblock \href{http://dx.doi.org/10.1016/0010-4655(92)90136-M}{Comput. Phys.
  Commun.{} {\bfseries 69},~155~(1992)}\relax
\relax
\bibitem{lepto}
G. Ingelman \emph{et al.},
\newblock \href{http://dx.doi.org/10.1016/S0010-4655(96)00157-9}{Comput. Phys.
  Commun.{} {\bfseries 101},~108~(1997)}\relax
\relax
\bibitem{Jung:1993gf}
H. Jung,
\newblock \href{http://dx.doi.org/10.1016/0010-4655(94)00150-Z}{Comput. Phys.
  Commun.{} {\bfseries 86},~147~(1995)}\relax
\relax
\bibitem{Sjostrand:2006za}
T. Sj{\"o}strand \emph{et al.},
\newblock \href{http://dx.doi.org/10.1088/1126-6708/2006/05/026}{JHEP{}
  {\bfseries 05},~026~(2006)}\relax
\relax
\bibitem{Geant}
R. Brun \emph{et al.},
\newblock Technical Report CERN-DD/EE/84-1{}~(1987)\relax
\relax
\bibitem{ZGANA:1990}
G. Hartner, Y. Iga.
\newblock ZEUS Note 90-084\relax
\relax
\bibitem{ZEUS_trigger:1992}
W.H. Smith \emph{et al.},
\newblock in the proceedings of Computing in High-Energy Physics
  (CHEP){}~(September 21-25, Annecy, France (1992)).
\newblock DESY-92-150B\relax
\relax
\bibitem{ZEUS_trigger:2007}
P. Allfrey \emph{et al.},
\newblock \href{http://dx.doi.org/10.1016/j.nima.2007.06.106}{Nucl. Instrum.
  Meth. A{} {\bfseries 580},~1257~(2007)}\relax
\relax
\bibitem{ZEUS_jets:2010}
ZEUS \coll, H. Abramowicz \emph{et al.},
\newblock \href{http://dx.doi.org/10.1016/j.physletb.2010.06.015}{Phys. Lett.
  B{} {\bfseries 691},~127~(2010)}\relax
\relax
\bibitem{Jets_dis:2012}
ZEUS \coll, H. Abramowicz \emph{et al.},
\newblock \href{http://dx.doi.org/10.1016/j.physletb.2012.07.031}{Phys. Lett.{}
  {\bfseries B715},~88~(2012)}\relax
\relax
\bibitem{Jets_dis:2018}
ZEUS \coll, H. Abramowicz \emph{et al.},
\newblock \href{http://dx.doi.org/10.1007/JHEP01(2018)032}{JHEP{} {\bfseries
  01},~032~(2018)}\relax
\relax
\bibitem{Amaldi:1979yh}
U. Amaldi \emph{et al.},
\newblock {\slshape {ECFA Study of an ep Facility for Europe}}, pp.~377--414.
\newblock  (1979)\relax
\relax
\bibitem{Bentvelsen:1992fu}
S. Bentvelsen \emph{et al.},
\newblock {\slshape {Workshop on Physics at HERA}}.
\newblock  (1992)\relax
\relax
\bibitem{ZEUS_empz}
ZEUS \coll, M. Derrick \emph{et al.},
\newblock \href{http://dx.doi.org/10.1016/0370-2693(93)90065-P}{Phys. Lett. B{}
  {\bfseries 303},~183~(1993)}\relax
\relax
\bibitem{ZEUS_neural}
H. Abramowicz \emph{et al.},
\newblock \href{http://dx.doi.org/10.1016/0168-9002(95)00612-5}{Nucl. Instrum.
  Meth. A{} {\bfseries 365},~508~(1995)}\relax
\relax
\bibitem{Jet_kt:1993a}
S. Catani \emph{et al.},
\newblock \href{http://dx.doi.org/10.1016/0550-3213(93)90166-M}{Nucl. Phys. B{}
  {\bfseries 406},~187~(1993)}\relax
\relax
\bibitem{Jet_kt:1993b}
S. Ellis \emph{et al.},
\newblock \href{http://dx.doi.org/10.1103/PhysRevD.48.3160}{Phys. Rev. D{}
  {\bfseries 48},~3160~(1993)}\relax
\relax
\bibitem{Cacciari:2011ma}
M. Cacciari \emph{et al.},
\newblock \href{http://dx.doi.org/10.1140/epjc/s10052-012-1896-2}{Eur. Phys. J.
  C{} {\bfseries 72},~1896~(2012)}\relax
\relax
\bibitem{Cacciari:2005hq}
M. Cacciari \emph{et al.},
\newblock \href{http://dx.doi.org/10.1016/j.physletb.2006.08.037}{Phys. Lett.
  B{} {\bfseries 641},~57~(2006)}\relax
\relax
\bibitem{EFO:2000}
ZEUS \coll, M. Wing,
\newblock Frascati Phys. Ser.{} {\bfseries 21},~617~(2001)\relax
\relax
\bibitem{TUnfold}
S. Schmitt,
\newblock \href{http://dx.doi.org/10.1088/1748-0221/7/10/T10003}{JINST{}
  {\bfseries 7},~T10003~(2012)}\relax
\relax
\bibitem{Ignacio}
I. Borsa \emph{et al.}
\newblock Personal communications, 2021\relax
\relax
\bibitem{P2B:2015}
M. Cacciari \emph{et al.},
\newblock \href{http://dx.doi.org/10.1103/PhysRevLett.115.082002}{Phys. Rev.
  Lett.{} {\bfseries 115},~082002~(2015)}\relax
\relax
\bibitem{P2B:2018}
M. Cacciari \emph{et al.},
\newblock \href{http://dx.doi.org/10.1103/PhysRevLett.120.139901}{Phys. Rev.
  Lett.{} {\bfseries 120},~139901~(2018)}\relax
\relax
\bibitem{PDF4LHC15}
J. Butterworth \emph{et al.},
\newblock \href{http://dx.doi.org/10.1088/0954-3899/43/2/023001}{J. Phys. G{}
  {\bfseries 43},~023001~(2016)}\relax
\relax
\bibitem{Accardi:2012qut}
A. Accardi \emph{et al.},
\newblock \href{http://dx.doi.org/10.1140/epja/i2016-16268-9}{Eur. Phys. J. A{}
  {\bfseries 52},~268~(2016)}\relax
\relax
\bibitem{AbdulKhalek:2021gbh}
R. Abdul Khalek \emph{et al.},
\newblock \href{http://dx.doi.org/10.1016/j.nuclphysa.2022.122447}{Nucl. Phys.
  A{} {\bfseries 1026},~122447~(2022)}\relax
\relax
\end{mcbibliography}
			
\vfill\eject

\begin{table}
\begin{center}
\resizebox{\textwidth}{!}{
\begin{tabular}{ c | c c c c c c c c c c }
\hline
& \multicolumn{9}{c}{Inclusive}\\
\cline{2-11}
  &  $\Delta\phi^\mathrm{low}$  &  $\Delta\phi^\mathrm{up}$  &  $\frac{d\sigma}{d\Delta\phi} (\mathrm{pb})$  &  $\delta_\mathrm{stat} (\mathrm{frac.})$  &  $\delta_\mathrm{syst} (\mathrm{frac.})$  &  ARIADNE $(\mathrm{pb})$  &  $\mathcal{O}(\alpha_{s}) (\mathrm{pb})$      &  $\delta(\mathcal{O}(\alpha_{s})) (\mathrm{frac.})$  &  $\mathcal{O}(\alpha_{s}^{2}) (\mathrm{pb})$  &  $\delta(\mathcal{O}(\alpha_{s}^{2})) (\mathrm{frac.})$\\
\hline
\multirow{15}{*}{$N_\mathrm{jet} \geq 1$}  &  0.000  &   $ 0.209 $   &  203  &  $\pm$ $ 0.027 $   &  $\pm$ $ 0.25 $   &  186  &  &  &  &\\[0.2em]
  &   $ 0.209 $   &   $ 0.419 $   &  253  &  $\pm$ $ 0.041 $   &  $\pm$ $ 0.14 $   &  252  &  &  &  &\\[0.2em]
  &   $ 0.419 $   &   $ 0.628 $   &  291  &  $\pm$ $ 0.040 $   &  $\pm$ $ 0.17 $   &  291  &  &  &  &\\[0.2em]
  &   $ 0.628 $   &   $ 0.838 $   &  362  &  $\pm$ $ 0.035 $   &  $\pm$ $ 0.19 $   &  343  &  &  &  &\\[0.2em]
  &   $ 0.838 $   &   $ 1.05 $   &  440  &  $\pm$ $ 0.035 $   &  $\pm$ $ 0.17 $   &  438  &  &  &  &\\[0.2em]
  &   $ 1.05 $   &   $ 1.26 $   &  579  &  $\pm$ $ 0.029 $   &  $\pm$ $ 0.16 $   &  579  &  &  &  &\\[0.2em]
  &   $ 1.26 $   &   $ 1.47 $   &  854  &  $\pm$ $ 0.024 $   &  $\pm$ $ 0.16 $   &  853  &  &  &  &\\[0.2em]
  &   $ 1.47 $   &   $ 1.68 $   &  1313  &  $\pm$ $ 0.020 $   &  $\pm$ $ 0.15 $   &  1347  &  467  &  \scriptsize\begin{tabular}{@{}c@{}}$+$ $ 0.098 $ \\$-$ $ 0.12 $ \end{tabular}  &  1677  &  \scriptsize\begin{tabular}{@{}c@{}}$+$ $ 0.66 $ \\$-$ $ 0.34 $ \end{tabular}\\
  &   $ 1.68 $   &   $ 1.88 $   &  2139  &  $\pm$ $ 0.015 $   &  $\pm$ $ 0.14 $   &  2260  &  870  &  \scriptsize\begin{tabular}{@{}c@{}}$+$ $ 0.11 $ \\$-$ $ 0.13 $ \end{tabular}  &  2863  &  \scriptsize\begin{tabular}{@{}c@{}}$+$ $ 0.66 $ \\$-$ $ 0.34 $ \end{tabular}\\
  &   $ 1.88 $   &   $ 2.09 $   &  3770  &  $\pm$ $ 0.011 $   &  $\pm$ $ 0.11 $   &  4053  &  2060  &  \scriptsize\begin{tabular}{@{}c@{}}$+$ $ 0.16 $ \\$-$ $ 0.18 $ \end{tabular}  &  4587  &  \scriptsize\begin{tabular}{@{}c@{}}$+$ $ 0.47 $ \\$-$ $ 0.26 $ \end{tabular}\\
  &   $ 2.09 $   &   $ 2.30 $   &  6665  &  $\pm$ $ 0.0080 $   &  $\pm$ $ 0.091 $   &  7262  &  4790  &  \scriptsize\begin{tabular}{@{}c@{}}$+$ $ 0.20 $ \\$-$ $ 0.24 $ \end{tabular}  &  7608  &  \scriptsize\begin{tabular}{@{}c@{}}$+$ $ 0.29 $ \\$-$ $ 0.21 $ \end{tabular}\\
  &   $ 2.30 $   &   $ 2.51 $   &  12347  &  $\pm$ $ 0.0054 $   &  $\pm$ $ 0.074 $   &  12927  &  10300  &  \scriptsize\begin{tabular}{@{}c@{}}$+$ $ 0.23 $ \\$-$ $ 0.28 $ \end{tabular}  &  13576  &  \scriptsize\begin{tabular}{@{}c@{}}$+$ $ 0.19 $ \\$-$ $ 0.20 $ \end{tabular}\\
  &   $ 2.51 $   &   $ 2.72 $   &  23807  &  $\pm$ $ 0.0035 $   &  $\pm$ $ 0.039 $   &  23742  &  22198  &  \scriptsize\begin{tabular}{@{}c@{}}$+$ $ 0.25 $ \\$-$ $ 0.31 $ \end{tabular}  &  25778  &  \scriptsize\begin{tabular}{@{}c@{}}$+$ $ 0.15 $ \\$-$ $ 0.21 $ \end{tabular}\\
  &   $ 2.72 $   &   $ 2.93 $   &  51160  &  $\pm$ $ 0.0020 $   &  $\pm$ $ 0.045 $   &  46343  &  49672  &  \scriptsize\begin{tabular}{@{}c@{}}$+$ $ 0.23 $ \\$-$ $ 0.29 $ \end{tabular}  &  48284  &  \scriptsize\begin{tabular}{@{}c@{}}$+$ $ 0.12 $ \\$-$ $ 0.22 $ \end{tabular}\\
  &   $ 2.93 $   &   $ 3.14 $   &  103741  &  $\pm$ $ 0.0014 $   &  $\pm$ $ 0.035 $   &  98392  &  111074  &  \scriptsize\begin{tabular}{@{}c@{}}$+$ $ 0.17 $ \\$-$ $ 0.26 $ \end{tabular}  &  78854  &  \scriptsize\begin{tabular}{@{}c@{}}$+$ $ 0.15 $ \\$-$ $ 0.25 $ \end{tabular}\\
\hline
\end{tabular}
}
\end{center}
\caption{
Inclusive measurement of the differential cross sections, $d\sigma/d\Delta\phi$, as obtained from the data, ARIADNE MC simulations, and perturbative calculations at $\mathcal{O}(\alpha_{s})$ and $\mathcal{O}(\alpha_{s}^{2})$ accuracy.
The effect of initial- and final-state radiation has been corrected in data, based on a simulation study performed in the RAPGAP framework.
The quantities $\delta_\mathrm{stat}$ and $\delta_\mathrm{syst}$ represent the statistical and systematic uncertainties relative to the central value, respectively.
The uncertainty in the luminosity measurement ($1.9\%$) is not included in these values.
The quantities $\delta(\mathcal{O}(\alpha_{s}^{k}))$ represent the combined uncertainty of the scale dependence in the calculation and the model dependence in the hadronisation correction in the $\mathcal{O}(\alpha_{s}^{k})$ calculations.
}
\label{TAB_00}
\end{table}

\begin{table}
\begin{center}
\resizebox{\textwidth}{!}{
\begin{tabular}{ c | c c c c c c c c c c }
\hline
& \multicolumn{9}{c}{$2.5 \;\mathrm{GeV} < p_\mathrm{T,jet}^\mathrm{lead} < 7 \;\mathrm{GeV}$}\\
\cline{2-11}
  &  $\Delta\phi^\mathrm{low}$  &  $\Delta\phi^\mathrm{up}$  &  $\frac{d\sigma}{d\Delta\phi} (\mathrm{pb})$  &  $\delta_\mathrm{stat} (\mathrm{frac.})$  &  $\delta_\mathrm{syst} (\mathrm{frac.})$  &  ARIADNE $(\mathrm{pb})$  &  $\mathcal{O}(\alpha_{s}) (\mathrm{pb})$      &  $\delta(\mathcal{O}(\alpha_{s})) (\mathrm{frac.})$  &  $\mathcal{O}(\alpha_{s}^{2}) (\mathrm{pb})$  &  $\delta(\mathcal{O}(\alpha_{s}^{2})) (\mathrm{frac.})$\\
\hline
\multirow{15}{*}{$N_\mathrm{jet} \geq 1$}  &  0.000  &   $ 0.209 $   &  135  &  $\pm$ $ 0.051 $   &  $\pm$ $ 0.27 $   &  112  &  &  &  &\\[0.2em]
  &   $ 0.209 $   &   $ 0.419 $   &  184  &  $\pm$ $ 0.076 $   &  $\pm$ $ 0.18 $   &  146  &  &  &  &\\[0.2em]
  &   $ 0.419 $   &   $ 0.628 $   &  204  &  $\pm$ $ 0.065 $   &  $\pm$ $ 0.21 $   &  172  &  &  &  &\\[0.2em]
  &   $ 0.628 $   &   $ 0.838 $   &  268  &  $\pm$ $ 0.066 $   &  $\pm$ $ 0.23 $   &  211  &  &  &  &\\[0.2em]
  &   $ 0.838 $   &   $ 1.05 $   &  334  &  $\pm$ $ 0.060 $   &  $\pm$ $ 0.22 $   &  280  &  &  &  &\\[0.2em]
  &   $ 1.05 $   &   $ 1.26 $   &  437  &  $\pm$ $ 0.053 $   &  $\pm$ $ 0.19 $   &  390  &  &  &  &\\[0.2em]
  &   $ 1.26 $   &   $ 1.47 $   &  665  &  $\pm$ $ 0.041 $   &  $\pm$ $ 0.20 $   &  608  &  &  &  &\\[0.2em]
  &   $ 1.47 $   &   $ 1.68 $   &  1030  &  $\pm$ $ 0.033 $   &  $\pm$ $ 0.15 $   &  1003  &  385  &  \scriptsize\begin{tabular}{@{}c@{}}$+$ $ 0.11 $ \\$-$ $ 0.14 $ \end{tabular}  &  1231  &  \scriptsize\begin{tabular}{@{}c@{}}$+$ $ 0.60 $ \\$-$ $ 0.33 $ \end{tabular}\\
  &   $ 1.68 $   &   $ 1.88 $   &  1707  &  $\pm$ $ 0.025 $   &  $\pm$ $ 0.15 $   &  1744  &  698  &  \scriptsize\begin{tabular}{@{}c@{}}$+$ $ 0.12 $ \\$-$ $ 0.15 $ \end{tabular}  &  2250  &  \scriptsize\begin{tabular}{@{}c@{}}$+$ $ 0.67 $ \\$-$ $ 0.35 $ \end{tabular}\\
  &   $ 1.88 $   &   $ 2.09 $   &  2980  &  $\pm$ $ 0.016 $   &  $\pm$ $ 0.10 $   &  3212  &  1653  &  \scriptsize\begin{tabular}{@{}c@{}}$+$ $ 0.16 $ \\$-$ $ 0.21 $ \end{tabular}  &  3744  &  \scriptsize\begin{tabular}{@{}c@{}}$+$ $ 0.51 $ \\$-$ $ 0.30 $ \end{tabular}\\
  &   $ 2.09 $   &   $ 2.30 $   &  5400  &  $\pm$ $ 0.013 $   &  $\pm$ $ 0.074 $   &  5936  &  4029  &  \scriptsize\begin{tabular}{@{}c@{}}$+$ $ 0.22 $ \\$-$ $ 0.28 $ \end{tabular}  &  6333  &  \scriptsize\begin{tabular}{@{}c@{}}$+$ $ 0.32 $ \\$-$ $ 0.25 $ \end{tabular}\\
  &   $ 2.30 $   &   $ 2.51 $   &  10246  &  $\pm$ $ 0.0075 $   &  $\pm$ $ 0.063 $   &  10831  &  9071  &  \scriptsize\begin{tabular}{@{}c@{}}$+$ $ 0.26 $ \\$-$ $ 0.33 $ \end{tabular}  &  11517  &  \scriptsize\begin{tabular}{@{}c@{}}$+$ $ 0.20 $ \\$-$ $ 0.24 $ \end{tabular}\\
  &   $ 2.51 $   &   $ 2.72 $   &  20468  &  $\pm$ $ 0.0046 $   &  $\pm$ $ 0.034 $   &  20304  &  20013  &  \scriptsize\begin{tabular}{@{}c@{}}$+$ $ 0.28 $ \\$-$ $ 0.35 $ \end{tabular}  &  22078  &  \scriptsize\begin{tabular}{@{}c@{}}$+$ $ 0.16 $ \\$-$ $ 0.24 $ \end{tabular}\\
  &   $ 2.72 $   &   $ 2.93 $   &  43489  &  $\pm$ $ 0.0028 $   &  $\pm$ $ 0.054 $   &  40022  &  44709  &  \scriptsize\begin{tabular}{@{}c@{}}$+$ $ 0.26 $ \\$-$ $ 0.34 $ \end{tabular}  &  40568  &  \scriptsize\begin{tabular}{@{}c@{}}$+$ $ 0.16 $ \\$-$ $ 0.25 $ \end{tabular}\\
  &   $ 2.93 $   &   $ 3.14 $   &  80832  &  $\pm$ $ 0.0021 $   &  $\pm$ $ 0.020 $   &  79998  &  91331  &  \scriptsize\begin{tabular}{@{}c@{}}$+$ $ 0.20 $ \\$-$ $ 0.32 $ \end{tabular}  &  60442  &  \scriptsize\begin{tabular}{@{}c@{}}$+$ $ 0.20 $ \\$-$ $ 0.29 $ \end{tabular}\\
\hline
\multirow{15}{*}{$N_\mathrm{jet} \geq 2$}  &  0.000  &   $ 0.209 $   &  200  &  $\pm$ $ 0.084 $   &  $\pm$ $ 0.32 $   &  89.9  &  &  &  &\\[0.2em]
  &   $ 0.209 $   &   $ 0.419 $   &  219  &  $\pm$ $ 0.12 $   &  $\pm$ $ 0.25 $   &  99.8  &  &  &  &\\[0.2em]
  &   $ 0.419 $   &   $ 0.628 $   &  213  &  $\pm$ $ 0.10 $   &  $\pm$ $ 0.28 $   &  113  &  &  &  &\\[0.2em]
  &   $ 0.628 $   &   $ 0.838 $   &  292  &  $\pm$ $ 0.10 $   &  $\pm$ $ 0.26 $   &  138  &  &  &  &\\[0.2em]
  &   $ 0.838 $   &   $ 1.05 $   &  341  &  $\pm$ $ 0.095 $   &  $\pm$ $ 0.27 $   &  179  &  &  &  &\\[0.2em]
  &   $ 1.05 $   &   $ 1.26 $   &  387  &  $\pm$ $ 0.088 $   &  $\pm$ $ 0.21 $   &  245  &  &  &  &\\[0.2em]
  &   $ 1.26 $   &   $ 1.47 $   &  533  &  $\pm$ $ 0.068 $   &  $\pm$ $ 0.18 $   &  362  &  &  &  &\\[0.2em]
  &   $ 1.47 $   &   $ 1.68 $   &  756  &  $\pm$ $ 0.058 $   &  $\pm$ $ 0.13 $   &  601  &  319  &  \scriptsize\begin{tabular}{@{}c@{}}$+$ $ 0.12 $ \\$-$ $ 0.15 $ \end{tabular}  &  848  &  \scriptsize\begin{tabular}{@{}c@{}}$+$ $ 0.54 $ \\$-$ $ 0.29 $ \end{tabular}\\
  &   $ 1.68 $   &   $ 1.88 $   &  1036  &  $\pm$ $ 0.046 $   &  $\pm$ $ 0.093 $   &  957  &  524  &  \scriptsize\begin{tabular}{@{}c@{}}$+$ $ 0.15 $ \\$-$ $ 0.19 $ \end{tabular}  &  1476  &  \scriptsize\begin{tabular}{@{}c@{}}$+$ $ 0.69 $ \\$-$ $ 0.37 $ \end{tabular}\\
  &   $ 1.88 $   &   $ 2.09 $   &  1407  &  $\pm$ $ 0.033 $   &  $\pm$ $ 0.11 $   &  1601  &  1147  &  \scriptsize\begin{tabular}{@{}c@{}}$+$ $ 0.24 $ \\$-$ $ 0.30 $ \end{tabular}  &  2195  &  \scriptsize\begin{tabular}{@{}c@{}}$+$ $ 0.59 $ \\$-$ $ 0.37 $ \end{tabular}\\
  &   $ 2.09 $   &   $ 2.30 $   &  1959  &  $\pm$ $ 0.032 $   &  $\pm$ $ 0.17 $   &  2422  &  2478  &  \scriptsize\begin{tabular}{@{}c@{}}$+$ $ 0.37 $ \\$-$ $ 0.46 $ \end{tabular}  &  2990  &  \scriptsize\begin{tabular}{@{}c@{}}$+$ $ 0.38 $ \\$-$ $ 0.32 $ \end{tabular}\\
  &   $ 2.30 $   &   $ 2.51 $   &  2487  &  $\pm$ $ 0.024 $   &  $\pm$ $ 0.19 $   &  3326  &  4011  &  \scriptsize\begin{tabular}{@{}c@{}}$+$ $ 0.45 $ \\$-$ $ 0.55 $ \end{tabular}  &  4241  &  \scriptsize\begin{tabular}{@{}c@{}}$+$ $ 0.34 $ \\$-$ $ 0.34 $ \end{tabular}\\
  &   $ 2.51 $   &   $ 2.72 $   &  2922  &  $\pm$ $ 0.022 $   &  $\pm$ $ 0.14 $   &  3923  &  4420  &  \scriptsize\begin{tabular}{@{}c@{}}$+$ $ 0.40 $ \\$-$ $ 0.48 $ \end{tabular}  &  5553  &  \scriptsize\begin{tabular}{@{}c@{}}$+$ $ 0.39 $ \\$-$ $ 0.38 $ \end{tabular}\\
  &   $ 2.72 $   &   $ 2.93 $   &  1746  &  $\pm$ $ 0.047 $   &  $\pm$ $ 0.42 $   &  3900  &  3165  &  \scriptsize\begin{tabular}{@{}c@{}}$+$ $ 0.48 $ \\$-$ $ 0.55 $ \end{tabular}  &  5640  &  \scriptsize\begin{tabular}{@{}c@{}}$+$ $ 0.99 $ \\$-$ $ 0.66 $ \end{tabular}\\
  &   $ 2.93 $   &   $ 3.14 $   &  766  &  $\pm$ $ 0.081 $   &  $\pm$ $ 0.94 $   &  3549  &  2140  &  \scriptsize\begin{tabular}{@{}c@{}}$+$ $ 0.75 $ \\$-$ $ 0.86 $ \end{tabular}  &  5052  &  \scriptsize\begin{tabular}{@{}c@{}}$+$ $ 2.5 $ \\$-$ $ 1.4 $ \end{tabular}\\
\hline
\multirow{5}{*}{$N_\mathrm{jet} \geq 3$}  &  0.000  &   $ 0.838 $   &  58.3  &  $\pm$ $ 0.15 $   &  $\pm$ $ 0.54 $   &  33.0  &  &  &  &\\[0.2em]
  &   $ 0.838 $   &   $ 1.47 $   &  69.3  &  $\pm$ $ 0.19 $   &  $\pm$ $ 0.34 $   &  60.7  &  &  &  &\\[0.2em]
  &   $ 1.47 $   &   $ 2.09 $   &  99.7  &  $\pm$ $ 0.14 $   &  $\pm$ $ 0.31 $   &  151  &  &  &  178  &  \scriptsize\begin{tabular}{@{}c@{}}$+$ $ 1.2 $ \\$-$ $ 0.55 $ \end{tabular}\\
  &   $ 2.09 $   &   $ 2.72 $   &  50.5  &  $\pm$ $ 0.30 $   &  $\pm$ $ 0.42 $   &  256  &  &  &  297  &  \scriptsize\begin{tabular}{@{}c@{}}$+$ $ 3.7 $ \\$-$ $ 1.8 $ \end{tabular}\\
  &   $ 2.72 $   &   $ 3.14 $   &  412  &  $\pm$ $ 0.074 $   &  $\pm$ $ 0.58 $   &  302  &  &  &  474  &  \scriptsize\begin{tabular}{@{}c@{}}$+$ $ 0.72 $ \\$-$ $ 0.35 $ \end{tabular}\\
\hline
\end{tabular}
}
\end{center}
\caption{
Differential cross sections, $d\sigma/d\Delta\phi$, in the $p_\mathrm{T,jet}^\mathrm{lead}$ region of $2.5\;\mathrm{GeV} < p_\mathrm{T,jet}^\mathrm{lead} < 7\;\mathrm{GeV}$ for $N_\mathrm{jet} \geq$ 1, 2, and 3, as obtained from the data, ARIADNE MC simulations, and perturbative calculations at $\mathcal{O}(\alpha_{s})$ and $\mathcal{O}(\alpha_{s}^{2})$ accuracy.
Other details are as in the caption to Table~\ref{TAB_00}.
}
\label{TAB_10}
\end{table}

\begin{table}
\begin{center}
\resizebox{\textwidth}{!}{
\begin{tabular}{ c | c c c c c c c c c c }
\hline
& \multicolumn{9}{c}{$7 \;\mathrm{GeV} < p_\mathrm{T,jet}^\mathrm{lead} < 12 \;\mathrm{GeV}$}\\
\cline{2-11}
  &  $\Delta\phi^\mathrm{low}$  &  $\Delta\phi^\mathrm{up}$  &  $\frac{d\sigma}{d\Delta\phi} (\mathrm{pb})$  &  $\delta_\mathrm{stat} (\mathrm{frac.})$  &  $\delta_\mathrm{syst} (\mathrm{frac.})$  &  ARIADNE $(\mathrm{pb})$  &  $\mathcal{O}(\alpha_{s}) (\mathrm{pb})$      &  $\delta(\mathcal{O}(\alpha_{s})) (\mathrm{frac.})$  &  $\mathcal{O}(\alpha_{s}^{2}) (\mathrm{pb})$  &  $\delta(\mathcal{O}(\alpha_{s}^{2})) (\mathrm{frac.})$\\
\hline
\multirow{15}{*}{$N_\mathrm{jet} \geq 1$}  &  0.000  &   $ 0.209 $   &  41.8  &  $\pm$ $ 0.050 $   &  $\pm$ $ 0.22 $   &  33.0  &  &  &  &\\[0.2em]
  &   $ 0.209 $   &   $ 0.419 $   &  64.5  &  $\pm$ $ 0.084 $   &  $\pm$ $ 0.12 $   &  61.5  &  &  &  &\\[0.2em]
  &   $ 0.419 $   &   $ 0.628 $   &  79.4  &  $\pm$ $ 0.071 $   &  $\pm$ $ 0.13 $   &  70.8  &  &  &  &\\[0.2em]
  &   $ 0.628 $   &   $ 0.838 $   &  86.3  &  $\pm$ $ 0.073 $   &  $\pm$ $ 0.086 $   &  77.9  &  &  &  &\\[0.2em]
  &   $ 0.838 $   &   $ 1.05 $   &  114  &  $\pm$ $ 0.068 $   &  $\pm$ $ 0.078 $   &  104  &  &  &  &\\[0.2em]
  &   $ 1.05 $   &   $ 1.26 $   &  138  &  $\pm$ $ 0.054 $   &  $\pm$ $ 0.100 $   &  126  &  &  &  &\\[0.2em]
  &   $ 1.26 $   &   $ 1.47 $   &  173  &  $\pm$ $ 0.049 $   &  $\pm$ $ 0.089 $   &  161  &  &  &  &\\[0.2em]
  &   $ 1.47 $   &   $ 1.68 $   &  253  &  $\pm$ $ 0.042 $   &  $\pm$ $ 0.083 $   &  240  &  39.0  &  \scriptsize\begin{tabular}{@{}c@{}}$+$ $ 0.042 $ \\$-$ $ 0.043 $ \end{tabular}  &  286  &  \scriptsize\begin{tabular}{@{}c@{}}$+$ $ 0.67 $ \\$-$ $ 0.30 $ \end{tabular}\\
  &   $ 1.68 $   &   $ 1.88 $   &  387  &  $\pm$ $ 0.032 $   &  $\pm$ $ 0.067 $   &  387  &  125  &  \scriptsize\begin{tabular}{@{}c@{}}$+$ $ 0.094 $ \\$-$ $ 0.070 $ \end{tabular}  &  417  &  \scriptsize\begin{tabular}{@{}c@{}}$+$ $ 0.48 $ \\$-$ $ 0.24 $ \end{tabular}\\
  &   $ 1.88 $   &   $ 2.09 $   &  654  &  $\pm$ $ 0.024 $   &  $\pm$ $ 0.067 $   &  654  &  349  &  \scriptsize\begin{tabular}{@{}c@{}}$+$ $ 0.15 $ \\$-$ $ 0.10 $ \end{tabular}  &  619  &  \scriptsize\begin{tabular}{@{}c@{}}$+$ $ 0.24 $ \\$-$ $ 0.15 $ \end{tabular}\\
  &   $ 2.09 $   &   $ 2.30 $   &  1040  &  $\pm$ $ 0.019 $   &  $\pm$ $ 0.065 $   &  1063  &  650  &  \scriptsize\begin{tabular}{@{}c@{}}$+$ $ 0.16 $ \\$-$ $ 0.11 $ \end{tabular}  &  1049  &  \scriptsize\begin{tabular}{@{}c@{}}$+$ $ 0.23 $ \\$-$ $ 0.14 $ \end{tabular}\\
  &   $ 2.30 $   &   $ 2.51 $   &  1707  &  $\pm$ $ 0.013 $   &  $\pm$ $ 0.051 $   &  1745  &  1102  &  \scriptsize\begin{tabular}{@{}c@{}}$+$ $ 0.16 $ \\$-$ $ 0.11 $ \end{tabular}  &  1771  &  \scriptsize\begin{tabular}{@{}c@{}}$+$ $ 0.24 $ \\$-$ $ 0.14 $ \end{tabular}\\
  &   $ 2.51 $   &   $ 2.72 $   &  2945  &  $\pm$ $ 0.010 $   &  $\pm$ $ 0.068 $   &  2918  &  2033  &  \scriptsize\begin{tabular}{@{}c@{}}$+$ $ 0.16 $ \\$-$ $ 0.12 $ \end{tabular}  &  3149  &  \scriptsize\begin{tabular}{@{}c@{}}$+$ $ 0.22 $ \\$-$ $ 0.13 $ \end{tabular}\\
  &   $ 2.72 $   &   $ 2.93 $   &  6252  &  $\pm$ $ 0.0061 $   &  $\pm$ $ 0.14 $   &  5481  &  4663  &  \scriptsize\begin{tabular}{@{}c@{}}$+$ $ 0.14 $ \\$-$ $ 0.11 $ \end{tabular}  &  6201  &  \scriptsize\begin{tabular}{@{}c@{}}$+$ $ 0.14 $ \\$-$ $ 0.13 $ \end{tabular}\\
  &   $ 2.93 $   &   $ 3.14 $   &  17979  &  $\pm$ $ 0.0033 $   &  $\pm$ $ 0.096 $   &  15793  &  17133  &  \scriptsize\begin{tabular}{@{}c@{}}$+$ $ 0.099 $ \\$-$ $ 0.083 $ \end{tabular}  &  16157  &  \scriptsize\begin{tabular}{@{}c@{}}$+$ $ 0.016 $ \\$-$ $ 0.17 $ \end{tabular}\\
\hline
\multirow{15}{*}{$N_\mathrm{jet} \geq 2$}  &  0.000  &   $ 0.209 $   &  58.9  &  $\pm$ $ 0.076 $   &  $\pm$ $ 0.23 $   &  42.4  &  &  &  &\\[0.2em]
  &   $ 0.209 $   &   $ 0.419 $   &  65.9  &  $\pm$ $ 0.12 $   &  $\pm$ $ 0.23 $   &  58.9  &  &  &  &\\[0.2em]
  &   $ 0.419 $   &   $ 0.628 $   &  84.8  &  $\pm$ $ 0.093 $   &  $\pm$ $ 0.23 $   &  67.2  &  &  &  &\\[0.2em]
  &   $ 0.628 $   &   $ 0.838 $   &  87.4  &  $\pm$ $ 0.10 $   &  $\pm$ $ 0.15 $   &  72.7  &  &  &  &\\[0.2em]
  &   $ 0.838 $   &   $ 1.05 $   &  115  &  $\pm$ $ 0.092 $   &  $\pm$ $ 0.15 $   &  97.5  &  &  &  &\\[0.2em]
  &   $ 1.05 $   &   $ 1.26 $   &  131  &  $\pm$ $ 0.074 $   &  $\pm$ $ 0.14 $   &  112  &  &  &  &\\[0.2em]
  &   $ 1.26 $   &   $ 1.47 $   &  164  &  $\pm$ $ 0.067 $   &  $\pm$ $ 0.13 $   &  147  &  &  &  &\\[0.2em]
  &   $ 1.47 $   &   $ 1.68 $   &  222  &  $\pm$ $ 0.060 $   &  $\pm$ $ 0.073 $   &  210  &  39.8  &  \scriptsize\begin{tabular}{@{}c@{}}$+$ $ 0.049 $ \\$-$ $ 0.051 $ \end{tabular}  &  261  &  \scriptsize\begin{tabular}{@{}c@{}}$+$ $ 0.67 $ \\$-$ $ 0.30 $ \end{tabular}\\
  &   $ 1.68 $   &   $ 1.88 $   &  322  &  $\pm$ $ 0.043 $   &  $\pm$ $ 0.063 $   &  325  &  110  &  \scriptsize\begin{tabular}{@{}c@{}}$+$ $ 0.097 $ \\$-$ $ 0.085 $ \end{tabular}  &  386  &  \scriptsize\begin{tabular}{@{}c@{}}$+$ $ 0.53 $ \\$-$ $ 0.27 $ \end{tabular}\\
  &   $ 1.88 $   &   $ 2.09 $   &  494  &  $\pm$ $ 0.033 $   &  $\pm$ $ 0.069 $   &  524  &  291  &  \scriptsize\begin{tabular}{@{}c@{}}$+$ $ 0.16 $ \\$-$ $ 0.14 $ \end{tabular}  &  542  &  \scriptsize\begin{tabular}{@{}c@{}}$+$ $ 0.28 $ \\$-$ $ 0.18 $ \end{tabular}\\
  &   $ 2.09 $   &   $ 2.30 $   &  744  &  $\pm$ $ 0.027 $   &  $\pm$ $ 0.081 $   &  808  &  541  &  \scriptsize\begin{tabular}{@{}c@{}}$+$ $ 0.19 $ \\$-$ $ 0.16 $ \end{tabular}  &  854  &  \scriptsize\begin{tabular}{@{}c@{}}$+$ $ 0.23 $ \\$-$ $ 0.16 $ \end{tabular}\\
  &   $ 2.30 $   &   $ 2.51 $   &  1025  &  $\pm$ $ 0.020 $   &  $\pm$ $ 0.13 $   &  1195  &  918  &  \scriptsize\begin{tabular}{@{}c@{}}$+$ $ 0.22 $ \\$-$ $ 0.18 $ \end{tabular}  &  1345  &  \scriptsize\begin{tabular}{@{}c@{}}$+$ $ 0.25 $ \\$-$ $ 0.15 $ \end{tabular}\\
  &   $ 2.51 $   &   $ 2.72 $   &  1423  &  $\pm$ $ 0.018 $   &  $\pm$ $ 0.15 $   &  1664  &  1629  &  \scriptsize\begin{tabular}{@{}c@{}}$+$ $ 0.27 $ \\$-$ $ 0.23 $ \end{tabular}  &  2123  &  \scriptsize\begin{tabular}{@{}c@{}}$+$ $ 0.21 $ \\$-$ $ 0.16 $ \end{tabular}\\
  &   $ 2.72 $   &   $ 2.93 $   &  1498  &  $\pm$ $ 0.019 $   &  $\pm$ $ 0.28 $   &  1973  &  2358  &  \scriptsize\begin{tabular}{@{}c@{}}$+$ $ 0.36 $ \\$-$ $ 0.32 $ \end{tabular}  &  2825  &  \scriptsize\begin{tabular}{@{}c@{}}$+$ $ 0.18 $ \\$-$ $ 0.20 $ \end{tabular}\\
  &   $ 2.93 $   &   $ 3.14 $   &  1487  &  $\pm$ $ 0.018 $   &  $\pm$ $ 0.35 $   &  1904  &  1900  &  \scriptsize\begin{tabular}{@{}c@{}}$+$ $ 0.32 $ \\$-$ $ 0.28 $ \end{tabular}  &  2822  &  \scriptsize\begin{tabular}{@{}c@{}}$+$ $ 0.33 $ \\$-$ $ 0.23 $ \end{tabular}\\
\hline
\multirow{5}{*}{$N_\mathrm{jet} \geq 3$}  &  0.000  &   $ 0.838 $   &  24.6  &  $\pm$ $ 0.11 $   &  $\pm$ $ 0.19 $   &  19.6  &  &  &  &\\[0.2em]
  &   $ 0.838 $   &   $ 1.47 $   &  40.7  &  $\pm$ $ 0.11 $   &  $\pm$ $ 0.14 $   &  37.0  &  &  &  &\\[0.2em]
  &   $ 1.47 $   &   $ 2.09 $   &  69.2  &  $\pm$ $ 0.086 $   &  $\pm$ $ 0.19 $   &  79.9  &  &  &  92.4  &  \scriptsize\begin{tabular}{@{}c@{}}$+$ $ 0.83 $ \\$-$ $ 0.40 $ \end{tabular}\\
  &   $ 2.09 $   &   $ 2.72 $   &  88.4  &  $\pm$ $ 0.094 $   &  $\pm$ $ 0.63 $   &  165  &  &  &  196  &  \scriptsize\begin{tabular}{@{}c@{}}$+$ $ 1.3 $ \\$-$ $ 0.65 $ \end{tabular}\\
  &   $ 2.72 $   &   $ 3.14 $   &  129  &  $\pm$ $ 0.090 $   &  $\pm$ $ 0.36 $   &  208  &  &  &  287  &  \scriptsize\begin{tabular}{@{}c@{}}$+$ $ 1.3 $ \\$-$ $ 0.63 $ \end{tabular}\\
\hline
\end{tabular}
}
\end{center}
\caption{
Differential cross sections, $d\sigma/d\Delta\phi$, in the $p_\mathrm{T,jet}^\mathrm{lead}$ region of $7\;\mathrm{GeV} < p_\mathrm{T,jet}^\mathrm{lead} < 12\;\mathrm{GeV}$ for $N_\mathrm{jet} \geq$ 1, 2, and 3, as obtained from the data, ARIADNE MC simulations, and perturbative calculations at $\mathcal{O}(\alpha_{s})$ and $\mathcal{O}(\alpha_{s}^{2})$ accuracy.
Other details are as in the caption to Table~\ref{TAB_00}.
}
\label{TAB_11}
\end{table}

\begin{table}
\begin{center}
\resizebox{\textwidth}{!}{
\begin{tabular}{ c | c c c c c c c c c c }
\hline
& \multicolumn{9}{c}{$12 \;\mathrm{GeV} < p_\mathrm{T,jet}^\mathrm{lead} < 30 \;\mathrm{GeV}$}\\
\cline{2-11}
  &  $\Delta\phi^\mathrm{low}$  &  $\Delta\phi^\mathrm{up}$  &  $\frac{d\sigma}{d\Delta\phi} (\mathrm{pb})$  &  $\delta_\mathrm{stat} (\mathrm{frac.})$  &  $\delta_\mathrm{syst} (\mathrm{frac.})$  &  ARIADNE $(\mathrm{pb})$  &  $\mathcal{O}(\alpha_{s}) (\mathrm{pb})$      &  $\delta(\mathcal{O}(\alpha_{s})) (\mathrm{frac.})$  &  $\mathcal{O}(\alpha_{s}^{2}) (\mathrm{pb})$  &  $\delta(\mathcal{O}(\alpha_{s}^{2})) (\mathrm{frac.})$\\
\hline
\multirow{15}{*}{$N_\mathrm{jet} \geq 1$}  &  0.000  &   $ 0.209 $   &  12.5  &  $\pm$ $ 0.066 $   &  $\pm$ $ 0.36 $   &  23.5  &  &  &  &\\[0.2em]
  &   $ 0.209 $   &   $ 0.419 $   &  17.8  &  $\pm$ $ 0.10 $   &  $\pm$ $ 0.59 $   &  41.2  &  &  &  &\\[0.2em]
  &   $ 0.419 $   &   $ 0.628 $   &  24.8  &  $\pm$ $ 0.087 $   &  $\pm$ $ 0.48 $   &  50.6  &  &  &  &\\[0.2em]
  &   $ 0.628 $   &   $ 0.838 $   &  26.3  &  $\pm$ $ 0.078 $   &  $\pm$ $ 0.40 $   &  50.1  &  &  &  &\\[0.2em]
  &   $ 0.838 $   &   $ 1.05 $   &  27.8  &  $\pm$ $ 0.072 $   &  $\pm$ $ 0.35 $   &  51.6  &  &  &  &\\[0.2em]
  &   $ 1.05 $   &   $ 1.26 $   &  38.7  &  $\pm$ $ 0.066 $   &  $\pm$ $ 0.33 $   &  65.9  &  &  &  &\\[0.2em]
  &   $ 1.26 $   &   $ 1.47 $   &  59.7  &  $\pm$ $ 0.060 $   &  $\pm$ $ 0.23 $   &  92.5  &  &  &  &\\[0.2em]
  &   $ 1.47 $   &   $ 1.68 $   &  68.0  &  $\pm$ $ 0.056 $   &  $\pm$ $ 0.24 $   &  98.8  &  11.4  &  \scriptsize\begin{tabular}{@{}c@{}}$+$ $ 0.045 $ \\$-$ $ 0.033 $ \end{tabular}  &  101  &  \scriptsize\begin{tabular}{@{}c@{}}$+$ $ 0.86 $ \\$-$ $ 0.42 $ \end{tabular}\\
  &   $ 1.68 $   &   $ 1.88 $   &  90.7  &  $\pm$ $ 0.039 $   &  $\pm$ $ 0.19 $   &  125  &  54.1  &  \scriptsize\begin{tabular}{@{}c@{}}$+$ $ 0.16 $ \\$-$ $ 0.098 $ \end{tabular}  &  90.0  &  \scriptsize\begin{tabular}{@{}c@{}}$+$ $ 0.27 $ \\$-$ $ 0.15 $ \end{tabular}\\
  &   $ 1.88 $   &   $ 2.09 $   &  146  &  $\pm$ $ 0.033 $   &  $\pm$ $ 0.18 $   &  189  &  104  &  \scriptsize\begin{tabular}{@{}c@{}}$+$ $ 0.18 $ \\$-$ $ 0.11 $ \end{tabular}  &  144  &  \scriptsize\begin{tabular}{@{}c@{}}$+$ $ 0.19 $ \\$-$ $ 0.12 $ \end{tabular}\\
  &   $ 2.09 $   &   $ 2.30 $   &  209  &  $\pm$ $ 0.027 $   &  $\pm$ $ 0.14 $   &  256  &  150  &  \scriptsize\begin{tabular}{@{}c@{}}$+$ $ 0.17 $ \\$-$ $ 0.11 $ \end{tabular}  &  223  &  \scriptsize\begin{tabular}{@{}c@{}}$+$ $ 0.24 $ \\$-$ $ 0.13 $ \end{tabular}\\
  &   $ 2.30 $   &   $ 2.51 $   &  304  &  $\pm$ $ 0.022 $   &  $\pm$ $ 0.093 $   &  346  &  211  &  \scriptsize\begin{tabular}{@{}c@{}}$+$ $ 0.16 $ \\$-$ $ 0.10 $ \end{tabular}  &  324  &  \scriptsize\begin{tabular}{@{}c@{}}$+$ $ 0.24 $ \\$-$ $ 0.13 $ \end{tabular}\\
  &   $ 2.51 $   &   $ 2.72 $   &  434  &  $\pm$ $ 0.019 $   &  $\pm$ $ 0.078 $   &  476  &  326  &  \scriptsize\begin{tabular}{@{}c@{}}$+$ $ 0.16 $ \\$-$ $ 0.11 $ \end{tabular}  &  500  &  \scriptsize\begin{tabular}{@{}c@{}}$+$ $ 0.23 $ \\$-$ $ 0.14 $ \end{tabular}\\
  &   $ 2.72 $   &   $ 2.93 $   &  796  &  $\pm$ $ 0.013 $   &  $\pm$ $ 0.065 $   &  782  &  635  &  \scriptsize\begin{tabular}{@{}c@{}}$+$ $ 0.14 $ \\$-$ $ 0.10 $ \end{tabular}  &  889  &  \scriptsize\begin{tabular}{@{}c@{}}$+$ $ 0.17 $ \\$-$ $ 0.11 $ \end{tabular}\\
  &   $ 2.93 $   &   $ 3.14 $   &  2832  &  $\pm$ $ 0.0067 $   &  $\pm$ $ 0.087 $   &  2550  &  3103  &  \scriptsize\begin{tabular}{@{}c@{}}$+$ $ 0.051 $ \\$-$ $ 0.046 $ \end{tabular}  &  3183  &  \scriptsize\begin{tabular}{@{}c@{}}$+$ $ 0.042 $ \\$-$ $ 0.0052 $ \end{tabular}\\
\hline
\multirow{12}{*}{$N_\mathrm{jet} \geq 2$}  &  0.000  &   $ 0.628 $   &  16.2  &  $\pm$ $ 0.072 $   &  $\pm$ $ 0.52 $   &  37.5  &  &  &  &\\[0.2em]
  &   $ 0.628 $   &   $ 1.05 $   &  22.1  &  $\pm$ $ 0.077 $   &  $\pm$ $ 0.45 $   &  45.8  &  &  &  &\\[0.2em]
  &   $ 1.05 $   &   $ 1.26 $   &  30.4  &  $\pm$ $ 0.098 $   &  $\pm$ $ 0.44 $   &  59.5  &  &  &  &\\[0.2em]
  &   $ 1.26 $   &   $ 1.47 $   &  51.1  &  $\pm$ $ 0.083 $   &  $\pm$ $ 0.24 $   &  82.6  &  &  &  &\\[0.2em]
  &   $ 1.47 $   &   $ 1.68 $   &  58.1  &  $\pm$ $ 0.077 $   &  $\pm$ $ 0.25 $   &  87.6  &  9.65  &  \scriptsize\begin{tabular}{@{}c@{}}$+$ $ 0.045 $ \\$-$ $ 0.031 $ \end{tabular}  &  92.1  &  \scriptsize\begin{tabular}{@{}c@{}}$+$ $ 0.91 $ \\$-$ $ 0.43 $ \end{tabular}\\
  &   $ 1.68 $   &   $ 1.88 $   &  70.5  &  $\pm$ $ 0.053 $   &  $\pm$ $ 0.24 $   &  105  &  42.6  &  \scriptsize\begin{tabular}{@{}c@{}}$+$ $ 0.16 $ \\$-$ $ 0.099 $ \end{tabular}  &  86.0  &  \scriptsize\begin{tabular}{@{}c@{}}$+$ $ 0.34 $ \\$-$ $ 0.20 $ \end{tabular}\\
  &   $ 1.88 $   &   $ 2.09 $   &  115  &  $\pm$ $ 0.044 $   &  $\pm$ $ 0.20 $   &  157  &  81.2  &  \scriptsize\begin{tabular}{@{}c@{}}$+$ $ 0.18 $ \\$-$ $ 0.11 $ \end{tabular}  &  126  &  \scriptsize\begin{tabular}{@{}c@{}}$+$ $ 0.20 $ \\$-$ $ 0.14 $ \end{tabular}\\
  &   $ 2.09 $   &   $ 2.30 $   &  154  &  $\pm$ $ 0.037 $   &  $\pm$ $ 0.19 $   &  204  &  116  &  \scriptsize\begin{tabular}{@{}c@{}}$+$ $ 0.18 $ \\$-$ $ 0.12 $ \end{tabular}  &  187  &  \scriptsize\begin{tabular}{@{}c@{}}$+$ $ 0.25 $ \\$-$ $ 0.16 $ \end{tabular}\\
  &   $ 2.30 $   &   $ 2.51 $   &  214  &  $\pm$ $ 0.031 $   &  $\pm$ $ 0.13 $   &  262  &  160  &  \scriptsize\begin{tabular}{@{}c@{}}$+$ $ 0.17 $ \\$-$ $ 0.11 $ \end{tabular}  &  256  &  \scriptsize\begin{tabular}{@{}c@{}}$+$ $ 0.24 $ \\$-$ $ 0.15 $ \end{tabular}\\
  &   $ 2.51 $   &   $ 2.72 $   &  266  &  $\pm$ $ 0.029 $   &  $\pm$ $ 0.16 $   &  333  &  242  &  \scriptsize\begin{tabular}{@{}c@{}}$+$ $ 0.20 $ \\$-$ $ 0.13 $ \end{tabular}  &  369  &  \scriptsize\begin{tabular}{@{}c@{}}$+$ $ 0.25 $ \\$-$ $ 0.16 $ \end{tabular}\\
  &   $ 2.72 $   &   $ 2.93 $   &  333  &  $\pm$ $ 0.025 $   &  $\pm$ $ 0.24 $   &  446  &  413  &  \scriptsize\begin{tabular}{@{}c@{}}$+$ $ 0.26 $ \\$-$ $ 0.18 $ \end{tabular}  &  558  &  \scriptsize\begin{tabular}{@{}c@{}}$+$ $ 0.23 $ \\$-$ $ 0.15 $ \end{tabular}\\
  &   $ 2.93 $   &   $ 3.14 $   &  330  &  $\pm$ $ 0.022 $   &  $\pm$ $ 0.17 $   &  416  &  487  &  \scriptsize\begin{tabular}{@{}c@{}}$+$ $ 0.31 $ \\$-$ $ 0.21 $ \end{tabular}  &  649  &  \scriptsize\begin{tabular}{@{}c@{}}$+$ $ 0.23 $ \\$-$ $ 0.18 $ \end{tabular}\\
\hline
\multirow{5}{*}{$N_\mathrm{jet} \geq 3$}  &  0.000  &   $ 0.838 $   &  7.26  &  $\pm$ $ 0.11 $   &  $\pm$ $ 0.27 $   &  11.7  &  &  &  &\\[0.2em]
  &   $ 0.838 $   &   $ 1.47 $   &  10.8  &  $\pm$ $ 0.11 $   &  $\pm$ $ 0.20 $   &  16.0  &  &  &  &\\[0.2em]
  &   $ 1.47 $   &   $ 2.09 $   &  19.4  &  $\pm$ $ 0.096 $   &  $\pm$ $ 0.21 $   &  28.6  &  &  &  28.1  &  \scriptsize\begin{tabular}{@{}c@{}}$+$ $ 0.87 $ \\$-$ $ 0.42 $ \end{tabular}\\
  &   $ 2.09 $   &   $ 2.72 $   &  28.3  &  $\pm$ $ 0.092 $   &  $\pm$ $ 0.35 $   &  48.5  &  &  &  55.7  &  \scriptsize\begin{tabular}{@{}c@{}}$+$ $ 1.1 $ \\$-$ $ 0.56 $ \end{tabular}\\
  &   $ 2.72 $   &   $ 3.14 $   &  36.2  &  $\pm$ $ 0.096 $   &  $\pm$ $ 0.26 $   &  59.4  &  &  &  84.9  &  \scriptsize\begin{tabular}{@{}c@{}}$+$ $ 1.2 $ \\$-$ $ 0.65 $ \end{tabular}\\
\hline
\end{tabular}
}
\end{center}
\caption{
Differential cross sections, $d\sigma/d\Delta\phi$, in the $p_\mathrm{T,jet}^\mathrm{lead}$ region of $12\;\mathrm{GeV} < p_\mathrm{T,jet}^\mathrm{lead} < 30\;\mathrm{GeV}$ for $N_\mathrm{jet} \geq$ 1, 2, and 3, as obtained from the data, ARIADNE MC simulations, and perturbative calculations at $\mathcal{O}(\alpha_{s})$ and $\mathcal{O}(\alpha_{s}^{2})$ accuracy.
Other details are as in the caption to Table~\ref{TAB_00}.
}
\label{TAB_12}
\end{table}

\begin{table}
\begin{center}
\resizebox{\textwidth}{!}{
\begin{tabular}{ c | c c c c c c c c c c }
\hline
& \multicolumn{9}{c}{$10 \;\mathrm{GeV}^{2} < Q^{2} < 50 \;\mathrm{GeV}^{2}$}\\
\cline{2-11}
  &  $\Delta\phi^\mathrm{low}$  &  $\Delta\phi^\mathrm{up}$  &  $\frac{d\sigma}{d\Delta\phi} (\mathrm{pb})$  &  $\delta_\mathrm{stat} (\mathrm{frac.})$  &  $\delta_\mathrm{syst} (\mathrm{frac.})$  &  ARIADNE $(\mathrm{pb})$  &  $\mathcal{O}(\alpha_{s}) (\mathrm{pb})$      &  $\delta(\mathcal{O}(\alpha_{s})) (\mathrm{frac.})$  &  $\mathcal{O}(\alpha_{s}^{2}) (\mathrm{pb})$  &  $\delta(\mathcal{O}(\alpha_{s}^{2})) (\mathrm{frac.})$\\
\hline
\multirow{15}{*}{$N_\mathrm{jet} \geq 1$}  &  0.000  &   $ 0.209 $   &  239  &  $\pm$ $ 0.032 $   &  $\pm$ $ 0.22 $   &  224  &  &  &  &\\[0.2em]
  &   $ 0.209 $   &   $ 0.419 $   &  256  &  $\pm$ $ 0.042 $   &  $\pm$ $ 0.13 $   &  245  &  &  &  &\\[0.2em]
  &   $ 0.419 $   &   $ 0.628 $   &  287  &  $\pm$ $ 0.042 $   &  $\pm$ $ 0.18 $   &  281  &  &  &  &\\[0.2em]
  &   $ 0.628 $   &   $ 0.838 $   &  349  &  $\pm$ $ 0.040 $   &  $\pm$ $ 0.19 $   &  329  &  &  &  &\\[0.2em]
  &   $ 0.838 $   &   $ 1.05 $   &  435  &  $\pm$ $ 0.037 $   &  $\pm$ $ 0.17 $   &  421  &  &  &  &\\[0.2em]
  &   $ 1.05 $   &   $ 1.26 $   &  560  &  $\pm$ $ 0.032 $   &  $\pm$ $ 0.15 $   &  553  &  &  &  &\\[0.2em]
  &   $ 1.26 $   &   $ 1.47 $   &  817  &  $\pm$ $ 0.027 $   &  $\pm$ $ 0.17 $   &  814  &  &  &  &\\[0.2em]
  &   $ 1.47 $   &   $ 1.68 $   &  1247  &  $\pm$ $ 0.022 $   &  $\pm$ $ 0.15 $   &  1273  &  426  &  \scriptsize\begin{tabular}{@{}c@{}}$+$ $ 0.10 $ \\$-$ $ 0.12 $ \end{tabular}  &  1564  &  \scriptsize\begin{tabular}{@{}c@{}}$+$ $ 0.66 $ \\$-$ $ 0.35 $ \end{tabular}\\
  &   $ 1.68 $   &   $ 1.88 $   &  2010  &  $\pm$ $ 0.017 $   &  $\pm$ $ 0.14 $   &  2105  &  791  &  \scriptsize\begin{tabular}{@{}c@{}}$+$ $ 0.12 $ \\$-$ $ 0.13 $ \end{tabular}  &  2641  &  \scriptsize\begin{tabular}{@{}c@{}}$+$ $ 0.67 $ \\$-$ $ 0.34 $ \end{tabular}\\
  &   $ 1.88 $   &   $ 2.09 $   &  3448  &  $\pm$ $ 0.013 $   &  $\pm$ $ 0.11 $   &  3698  &  1849  &  \scriptsize\begin{tabular}{@{}c@{}}$+$ $ 0.16 $ \\$-$ $ 0.19 $ \end{tabular}  &  4157  &  \scriptsize\begin{tabular}{@{}c@{}}$+$ $ 0.49 $ \\$-$ $ 0.27 $ \end{tabular}\\
  &   $ 2.09 $   &   $ 2.30 $   &  5982  &  $\pm$ $ 0.0093 $   &  $\pm$ $ 0.094 $   &  6465  &  4221  &  \scriptsize\begin{tabular}{@{}c@{}}$+$ $ 0.21 $ \\$-$ $ 0.25 $ \end{tabular}  &  6746  &  \scriptsize\begin{tabular}{@{}c@{}}$+$ $ 0.31 $ \\$-$ $ 0.23 $ \end{tabular}\\
  &   $ 2.30 $   &   $ 2.51 $   &  10842  &  $\pm$ $ 0.0062 $   &  $\pm$ $ 0.084 $   &  11275  &  8932  &  \scriptsize\begin{tabular}{@{}c@{}}$+$ $ 0.24 $ \\$-$ $ 0.30 $ \end{tabular}  &  11815  &  \scriptsize\begin{tabular}{@{}c@{}}$+$ $ 0.21 $ \\$-$ $ 0.22 $ \end{tabular}\\
  &   $ 2.51 $   &   $ 2.72 $   &  20629  &  $\pm$ $ 0.0040 $   &  $\pm$ $ 0.043 $   &  20400  &  18943  &  \scriptsize\begin{tabular}{@{}c@{}}$+$ $ 0.27 $ \\$-$ $ 0.33 $ \end{tabular}  &  21961  &  \scriptsize\begin{tabular}{@{}c@{}}$+$ $ 0.16 $ \\$-$ $ 0.23 $ \end{tabular}\\
  &   $ 2.72 $   &   $ 2.93 $   &  42923  &  $\pm$ $ 0.0023 $   &  $\pm$ $ 0.053 $   &  39260  &  41570  &  \scriptsize\begin{tabular}{@{}c@{}}$+$ $ 0.25 $ \\$-$ $ 0.32 $ \end{tabular}  &  39374  &  \scriptsize\begin{tabular}{@{}c@{}}$+$ $ 0.14 $ \\$-$ $ 0.25 $ \end{tabular}\\
  &   $ 2.93 $   &   $ 3.14 $   &  77642  &  $\pm$ $ 0.0019 $   &  $\pm$ $ 0.035 $   &  75775  &  83056  &  \scriptsize\begin{tabular}{@{}c@{}}$+$ $ 0.19 $ \\$-$ $ 0.29 $ \end{tabular}  &  56349  &  \scriptsize\begin{tabular}{@{}c@{}}$+$ $ 0.18 $ \\$-$ $ 0.30 $ \end{tabular}\\
\hline
\multirow{15}{*}{$N_\mathrm{jet} \geq 2$}  &  0.000  &   $ 0.209 $   &  233  &  $\pm$ $ 0.048 $   &  $\pm$ $ 0.17 $   &  176  &  &  &  &\\[0.2em]
  &   $ 0.209 $   &   $ 0.419 $   &  242  &  $\pm$ $ 0.064 $   &  $\pm$ $ 0.23 $   &  190  &  &  &  &\\[0.2em]
  &   $ 0.419 $   &   $ 0.628 $   &  260  &  $\pm$ $ 0.064 $   &  $\pm$ $ 0.21 $   &  215  &  &  &  &\\[0.2em]
  &   $ 0.628 $   &   $ 0.838 $   &  309  &  $\pm$ $ 0.062 $   &  $\pm$ $ 0.15 $   &  245  &  &  &  &\\[0.2em]
  &   $ 0.838 $   &   $ 1.05 $   &  391  &  $\pm$ $ 0.055 $   &  $\pm$ $ 0.16 $   &  306  &  &  &  &\\[0.2em]
  &   $ 1.05 $   &   $ 1.26 $   &  436  &  $\pm$ $ 0.050 $   &  $\pm$ $ 0.13 $   &  390  &  &  &  &\\[0.2em]
  &   $ 1.26 $   &   $ 1.47 $   &  611  &  $\pm$ $ 0.041 $   &  $\pm$ $ 0.085 $   &  543  &  &  &  &\\[0.2em]
  &   $ 1.47 $   &   $ 1.68 $   &  867  &  $\pm$ $ 0.036 $   &  $\pm$ $ 0.085 $   &  837  &  328  &  \scriptsize\begin{tabular}{@{}c@{}}$+$ $ 0.11 $ \\$-$ $ 0.13 $ \end{tabular}  &  1158  &  \scriptsize\begin{tabular}{@{}c@{}}$+$ $ 0.71 $ \\$-$ $ 0.36 $ \end{tabular}\\
  &   $ 1.68 $   &   $ 1.88 $   &  1199  &  $\pm$ $ 0.028 $   &  $\pm$ $ 0.071 $   &  1259  &  572  &  \scriptsize\begin{tabular}{@{}c@{}}$+$ $ 0.14 $ \\$-$ $ 0.16 $ \end{tabular}  &  1835  &  \scriptsize\begin{tabular}{@{}c@{}}$+$ $ 0.76 $ \\$-$ $ 0.39 $ \end{tabular}\\
  &   $ 1.88 $   &   $ 2.09 $   &  1722  &  $\pm$ $ 0.024 $   &  $\pm$ $ 0.087 $   &  1993  &  1311  &  \scriptsize\begin{tabular}{@{}c@{}}$+$ $ 0.23 $ \\$-$ $ 0.27 $ \end{tabular}  &  2523  &  \scriptsize\begin{tabular}{@{}c@{}}$+$ $ 0.52 $ \\$-$ $ 0.32 $ \end{tabular}\\
  &   $ 2.09 $   &   $ 2.30 $   &  2343  &  $\pm$ $ 0.019 $   &  $\pm$ $ 0.086 $   &  2825  &  2726  &  \scriptsize\begin{tabular}{@{}c@{}}$+$ $ 0.35 $ \\$-$ $ 0.41 $ \end{tabular}  &  3243  &  \scriptsize\begin{tabular}{@{}c@{}}$+$ $ 0.28 $ \\$-$ $ 0.25 $ \end{tabular}\\
  &   $ 2.30 $   &   $ 2.51 $   &  2898  &  $\pm$ $ 0.017 $   &  $\pm$ $ 0.11 $   &  3672  &  4161  &  \scriptsize\begin{tabular}{@{}c@{}}$+$ $ 0.42 $ \\$-$ $ 0.49 $ \end{tabular}  &  4455  &  \scriptsize\begin{tabular}{@{}c@{}}$+$ $ 0.26 $ \\$-$ $ 0.28 $ \end{tabular}\\
  &   $ 2.51 $   &   $ 2.72 $   &  3121  &  $\pm$ $ 0.017 $   &  $\pm$ $ 0.14 $   &  4150  &  4348  &  \scriptsize\begin{tabular}{@{}c@{}}$+$ $ 0.39 $ \\$-$ $ 0.45 $ \end{tabular}  &  5799  &  \scriptsize\begin{tabular}{@{}c@{}}$+$ $ 0.39 $ \\$-$ $ 0.36 $ \end{tabular}\\
  &   $ 2.72 $   &   $ 2.93 $   &  2536  &  $\pm$ $ 0.021 $   &  $\pm$ $ 0.20 $   &  4207  &  3215  &  \scriptsize\begin{tabular}{@{}c@{}}$+$ $ 0.36 $ \\$-$ $ 0.40 $ \end{tabular}  &  6152  &  \scriptsize\begin{tabular}{@{}c@{}}$+$ $ 0.79 $ \\$-$ $ 0.48 $ \end{tabular}\\
  &   $ 2.93 $   &   $ 3.14 $   &  2068  &  $\pm$ $ 0.022 $   &  $\pm$ $ 0.57 $   &  3984  &  2519  &  \scriptsize\begin{tabular}{@{}c@{}}$+$ $ 0.35 $ \\$-$ $ 0.38 $ \end{tabular}  &  5783  &  \scriptsize\begin{tabular}{@{}c@{}}$+$ $ 1.1 $ \\$-$ $ 0.56 $ \end{tabular}\\
\hline
\multirow{5}{*}{$N_\mathrm{jet} \geq 3$}  &  0.000  &   $ 0.838 $   &  67.2  &  $\pm$ $ 0.086 $   &  $\pm$ $ 0.51 $   &  63.1  &  &  &  &\\[0.2em]
  &   $ 0.838 $   &   $ 1.47 $   &  86.1  &  $\pm$ $ 0.099 $   &  $\pm$ $ 0.50 $   &  103  &  &  &  &\\[0.2em]
  &   $ 1.47 $   &   $ 2.09 $   &  142  &  $\pm$ $ 0.088 $   &  $\pm$ $ 0.25 $   &  217  &  &  &  257  &  \scriptsize\begin{tabular}{@{}c@{}}$+$ $ 1.2 $ \\$-$ $ 0.56 $ \end{tabular}\\
  &   $ 2.09 $   &   $ 2.72 $   &  132  &  $\pm$ $ 0.12 $   &  $\pm$ $ 0.81 $   &  334  &  &  &  391  &  \scriptsize\begin{tabular}{@{}c@{}}$+$ $ 2.0 $ \\$-$ $ 0.92 $ \end{tabular}\\
  &   $ 2.72 $   &   $ 3.14 $   &  467  &  $\pm$ $ 0.040 $   &  $\pm$ $ 0.41 $   &  393  &  &  &  593  &  \scriptsize\begin{tabular}{@{}c@{}}$+$ $ 0.84 $ \\$-$ $ 0.39 $ \end{tabular}\\
\hline
\end{tabular}
}
\end{center}
\caption{
Differential cross sections, $d\sigma/d\Delta\phi$, in the $Q^{2}$ region of $10\;\mathrm{GeV}^{2} < Q^{2} < 50\;\mathrm{GeV}^{2}$ for $N_\mathrm{jet} \geq$ 1, 2, and 3, as obtained from the data, ARIADNE MC simulations, and perturbative calculations at $\mathcal{O}(\alpha_{s})$ and $\mathcal{O}(\alpha_{s}^{2})$ accuracy.
Other details are as in the caption to Table~\ref{TAB_00}.
}
\label{TAB_20}
\end{table}

\begin{table}
\begin{center}
\resizebox{\textwidth}{!}{
\begin{tabular}{ c | c c c c c c c c c c }
\hline
& \multicolumn{9}{c}{$50 \;\mathrm{GeV}^{2} < Q^{2} < 100 \;\mathrm{GeV}^{2}$}\\
\cline{2-11}
  &  $\Delta\phi^\mathrm{low}$  &  $\Delta\phi^\mathrm{up}$  &  $\frac{d\sigma}{d\Delta\phi} (\mathrm{pb})$  &  $\delta_\mathrm{stat} (\mathrm{frac.})$  &  $\delta_\mathrm{syst} (\mathrm{frac.})$  &  ARIADNE $(\mathrm{pb})$  &  $\mathcal{O}(\alpha_{s}) (\mathrm{pb})$      &  $\delta(\mathcal{O}(\alpha_{s})) (\mathrm{frac.})$  &  $\mathcal{O}(\alpha_{s}^{2}) (\mathrm{pb})$  &  $\delta(\mathcal{O}(\alpha_{s}^{2})) (\mathrm{frac.})$\\
\hline
\multirow{15}{*}{$N_\mathrm{jet} \geq 1$}  &  0.000  &   $ 0.209 $   &  4.73  &  $\pm$ $ 0.081 $   &  $\pm$ $ 0.35 $   &  4.97  &  &  &  &\\[0.2em]
  &   $ 0.209 $   &   $ 0.419 $   &  9.53  &  $\pm$ $ 0.12 $   &  $\pm$ $ 0.20 $   &  9.33  &  &  &  &\\[0.2em]
  &   $ 0.419 $   &   $ 0.628 $   &  9.84  &  $\pm$ $ 0.11 $   &  $\pm$ $ 0.23 $   &  10.8  &  &  &  &\\[0.2em]
  &   $ 0.628 $   &   $ 0.838 $   &  15.2  &  $\pm$ $ 0.087 $   &  $\pm$ $ 0.28 $   &  13.1  &  &  &  &\\[0.2em]
  &   $ 0.838 $   &   $ 1.05 $   &  16.0  &  $\pm$ $ 0.10 $   &  $\pm$ $ 0.15 $   &  17.1  &  &  &  &\\[0.2em]
  &   $ 1.05 $   &   $ 1.26 $   &  26.0  &  $\pm$ $ 0.081 $   &  $\pm$ $ 0.18 $   &  25.2  &  &  &  &\\[0.2em]
  &   $ 1.26 $   &   $ 1.47 $   &  37.1  &  $\pm$ $ 0.075 $   &  $\pm$ $ 0.11 $   &  37.7  &  &  &  &\\[0.2em]
  &   $ 1.47 $   &   $ 1.68 $   &  70.8  &  $\pm$ $ 0.058 $   &  $\pm$ $ 0.12 $   &  67.9  &  32.0  &  \scriptsize\begin{tabular}{@{}c@{}}$+$ $ 0.088 $ \\$-$ $ 0.082 $ \end{tabular}  &  84.1  &  \scriptsize\begin{tabular}{@{}c@{}}$+$ $ 0.36 $ \\$-$ $ 0.22 $ \end{tabular}\\
  &   $ 1.68 $   &   $ 1.88 $   &  127  &  $\pm$ $ 0.044 $   &  $\pm$ $ 0.098 $   &  136  &  59.8  &  \scriptsize\begin{tabular}{@{}c@{}}$+$ $ 0.094 $ \\$-$ $ 0.073 $ \end{tabular}  &  162  &  \scriptsize\begin{tabular}{@{}c@{}}$+$ $ 0.38 $ \\$-$ $ 0.24 $ \end{tabular}\\
  &   $ 1.88 $   &   $ 2.09 $   &  290  &  $\pm$ $ 0.031 $   &  $\pm$ $ 0.064 $   &  297  &  156  &  \scriptsize\begin{tabular}{@{}c@{}}$+$ $ 0.11 $ \\$-$ $ 0.082 $ \end{tabular}  &  313  &  \scriptsize\begin{tabular}{@{}c@{}}$+$ $ 0.24 $ \\$-$ $ 0.18 $ \end{tabular}\\
  &   $ 2.09 $   &   $ 2.30 $   &  563  &  $\pm$ $ 0.023 $   &  $\pm$ $ 0.047 $   &  636  &  411  &  \scriptsize\begin{tabular}{@{}c@{}}$+$ $ 0.15 $ \\$-$ $ 0.12 $ \end{tabular}  &  623  &  \scriptsize\begin{tabular}{@{}c@{}}$+$ $ 0.15 $ \\$-$ $ 0.14 $ \end{tabular}\\
  &   $ 2.30 $   &   $ 2.51 $   &  1209  &  $\pm$ $ 0.015 $   &  $\pm$ $ 0.029 $   &  1284  &  980  &  \scriptsize\begin{tabular}{@{}c@{}}$+$ $ 0.16 $ \\$-$ $ 0.14 $ \end{tabular}  &  1246  &  \scriptsize\begin{tabular}{@{}c@{}}$+$ $ 0.083 $ \\$-$ $ 0.11 $ \end{tabular}\\
  &   $ 2.51 $   &   $ 2.72 $   &  2526  &  $\pm$ $ 0.0090 $   &  $\pm$ $ 0.045 $   &  2553  &  2261  &  \scriptsize\begin{tabular}{@{}c@{}}$+$ $ 0.18 $ \\$-$ $ 0.16 $ \end{tabular}  &  2572  &  \scriptsize\begin{tabular}{@{}c@{}}$+$ $ 0.060 $ \\$-$ $ 0.087 $ \end{tabular}\\
  &   $ 2.72 $   &   $ 2.93 $   &  6302  &  $\pm$ $ 0.0054 $   &  $\pm$ $ 0.043 $   &  5419  &  5465  &  \scriptsize\begin{tabular}{@{}c@{}}$+$ $ 0.15 $ \\$-$ $ 0.14 $ \end{tabular}  &  5615  &  \scriptsize\begin{tabular}{@{}c@{}}$+$ $ 0.047 $ \\$-$ $ 0.082 $ \end{tabular}\\
  &   $ 2.93 $   &   $ 3.14 $   &  16928  &  $\pm$ $ 0.0031 $   &  $\pm$ $ 0.034 $   &  15663  &  16726  &  \scriptsize\begin{tabular}{@{}c@{}}$+$ $ 0.12 $ \\$-$ $ 0.12 $ \end{tabular}  &  13954  &  \scriptsize\begin{tabular}{@{}c@{}}$+$ $ 0.062 $ \\$-$ $ 0.095 $ \end{tabular}\\
\hline
\multirow{15}{*}{$N_\mathrm{jet} \geq 2$}  &  0.000  &   $ 0.209 $   &  9.07  &  $\pm$ $ 0.12 $   &  $\pm$ $ 0.31 $   &  5.45  &  &  &  &\\[0.2em]
  &   $ 0.209 $   &   $ 0.419 $   &  12.5  &  $\pm$ $ 0.20 $   &  $\pm$ $ 0.28 $   &  9.14  &  &  &  &\\[0.2em]
  &   $ 0.419 $   &   $ 0.628 $   &  11.9  &  $\pm$ $ 0.20 $   &  $\pm$ $ 0.47 $   &  10.2  &  &  &  &\\[0.2em]
  &   $ 0.628 $   &   $ 0.838 $   &  22.7  &  $\pm$ $ 0.12 $   &  $\pm$ $ 0.30 $   &  12.2  &  &  &  &\\[0.2em]
  &   $ 0.838 $   &   $ 1.05 $   &  19.0  &  $\pm$ $ 0.18 $   &  $\pm$ $ 0.48 $   &  16.2  &  &  &  &\\[0.2em]
  &   $ 1.05 $   &   $ 1.26 $   &  34.0  &  $\pm$ $ 0.11 $   &  $\pm$ $ 0.22 $   &  23.0  &  &  &  &\\[0.2em]
  &   $ 1.26 $   &   $ 1.47 $   &  43.6  &  $\pm$ $ 0.12 $   &  $\pm$ $ 0.21 $   &  34.1  &  &  &  &\\[0.2em]
  &   $ 1.47 $   &   $ 1.68 $   &  80.6  &  $\pm$ $ 0.083 $   &  $\pm$ $ 0.14 $   &  60.3  &  24.3  &  \scriptsize\begin{tabular}{@{}c@{}}$+$ $ 0.070 $ \\$-$ $ 0.053 $ \end{tabular}  &  74.7  &  \scriptsize\begin{tabular}{@{}c@{}}$+$ $ 0.31 $ \\$-$ $ 0.18 $ \end{tabular}\\
  &   $ 1.68 $   &   $ 1.88 $   &  129  &  $\pm$ $ 0.062 $   &  $\pm$ $ 0.14 $   &  117  &  47.2  &  \scriptsize\begin{tabular}{@{}c@{}}$+$ $ 0.084 $ \\$-$ $ 0.062 $ \end{tabular}  &  151  &  \scriptsize\begin{tabular}{@{}c@{}}$+$ $ 0.38 $ \\$-$ $ 0.23 $ \end{tabular}\\
  &   $ 1.88 $   &   $ 2.09 $   &  237  &  $\pm$ $ 0.047 $   &  $\pm$ $ 0.088 $   &  239  &  128  &  \scriptsize\begin{tabular}{@{}c@{}}$+$ $ 0.12 $ \\$-$ $ 0.088 $ \end{tabular}  &  290  &  \scriptsize\begin{tabular}{@{}c@{}}$+$ $ 0.30 $ \\$-$ $ 0.21 $ \end{tabular}\\
  &   $ 2.09 $   &   $ 2.30 $   &  407  &  $\pm$ $ 0.034 $   &  $\pm$ $ 0.079 $   &  479  &  345  &  \scriptsize\begin{tabular}{@{}c@{}}$+$ $ 0.18 $ \\$-$ $ 0.15 $ \end{tabular}  &  556  &  \scriptsize\begin{tabular}{@{}c@{}}$+$ $ 0.21 $ \\$-$ $ 0.19 $ \end{tabular}\\
  &   $ 2.30 $   &   $ 2.51 $   &  692  &  $\pm$ $ 0.025 $   &  $\pm$ $ 0.076 $   &  834  &  805  &  \scriptsize\begin{tabular}{@{}c@{}}$+$ $ 0.24 $ \\$-$ $ 0.21 $ \end{tabular}  &  998  &  \scriptsize\begin{tabular}{@{}c@{}}$+$ $ 0.13 $ \\$-$ $ 0.15 $ \end{tabular}\\
  &   $ 2.51 $   &   $ 2.72 $   &  989  &  $\pm$ $ 0.018 $   &  $\pm$ $ 0.067 $   &  1267  &  1582  &  \scriptsize\begin{tabular}{@{}c@{}}$+$ $ 0.33 $ \\$-$ $ 0.31 $ \end{tabular}  &  1644  &  \scriptsize\begin{tabular}{@{}c@{}}$+$ $ 0.13 $ \\$-$ $ 0.15 $ \end{tabular}\\
  &   $ 2.72 $   &   $ 2.93 $   &  855  &  $\pm$ $ 0.028 $   &  $\pm$ $ 0.17 $   &  1413  &  1776  &  \scriptsize\begin{tabular}{@{}c@{}}$+$ $ 0.43 $ \\$-$ $ 0.39 $ \end{tabular}  &  2005  &  \scriptsize\begin{tabular}{@{}c@{}}$+$ $ 0.18 $ \\$-$ $ 0.25 $ \end{tabular}\\
  &   $ 2.93 $   &   $ 3.14 $   &  593  &  $\pm$ $ 0.035 $   &  $\pm$ $ 0.72 $   &  1279  &  1176  &  \scriptsize\begin{tabular}{@{}c@{}}$+$ $ 0.42 $ \\$-$ $ 0.36 $ \end{tabular}  &  1817  &  \scriptsize\begin{tabular}{@{}c@{}}$+$ $ 0.43 $ \\$-$ $ 0.42 $ \end{tabular}\\
\hline
\multirow{5}{*}{$N_\mathrm{jet} \geq 3$}  &  0.000  &   $ 0.838 $   &  4.14  &  $\pm$ $ 0.19 $   &  $\pm$ $ 0.50 $   &  3.33  &  &  &  &\\[0.2em]
  &   $ 0.838 $   &   $ 1.47 $   &  11.7  &  $\pm$ $ 0.17 $   &  $\pm$ $ 0.38 $   &  10.3  &  &  &  &\\[0.2em]
  &   $ 1.47 $   &   $ 2.09 $   &  34.8  &  $\pm$ $ 0.10 $   &  $\pm$ $ 0.19 $   &  36.5  &  &  &  44.2  &  \scriptsize\begin{tabular}{@{}c@{}}$+$ $ 0.59 $ \\$-$ $ 0.34 $ \end{tabular}\\
  &   $ 2.09 $   &   $ 2.72 $   &  27.1  &  $\pm$ $ 0.18 $   &  $\pm$ $ 0.65 $   &  95.5  &  &  &  113  &  \scriptsize\begin{tabular}{@{}c@{}}$+$ $ 1.9 $ \\$-$ $ 1.1 $ \end{tabular}\\
  &   $ 2.72 $   &   $ 3.14 $   &  89.1  &  $\pm$ $ 0.093 $   &  $\pm$ $ 0.16 $   &  122  &  &  &  185  &  \scriptsize\begin{tabular}{@{}c@{}}$+$ $ 0.96 $ \\$-$ $ 0.55 $ \end{tabular}\\
\hline
\end{tabular}
}
\end{center}
\caption{
Differential cross sections, $d\sigma/d\Delta\phi$, in the $Q^{2}$ region of $50\;\mathrm{GeV}^{2} < Q^{2} < 100\;\mathrm{GeV}^{2}$ for $N_\mathrm{jet} \geq$ 1, 2, and 3, as obtained from the data, ARIADNE MC simulations, and perturbative calculations at $\mathcal{O}(\alpha_{s})$ and $\mathcal{O}(\alpha_{s}^{2})$ accuracy.
Other details are as in the caption to Table~\ref{TAB_00}.
}
\label{TAB_21}
\end{table}

\begin{table}
\begin{center}
\resizebox{\textwidth}{!}{
\begin{tabular}{ c | c c c c c c c c c c }
\hline
& \multicolumn{9}{c}{$100 \;\mathrm{GeV}^{2} < Q^{2} < 350 \;\mathrm{GeV}^{2}$}\\
\cline{2-11}
  &  $\Delta\phi^\mathrm{low}$  &  $\Delta\phi^\mathrm{up}$  &  $\frac{d\sigma}{d\Delta\phi} (\mathrm{pb})$  &  $\delta_\mathrm{stat} (\mathrm{frac.})$  &  $\delta_\mathrm{syst} (\mathrm{frac.})$  &  ARIADNE $(\mathrm{pb})$  &  $\mathcal{O}(\alpha_{s}) (\mathrm{pb})$      &  $\delta(\mathcal{O}(\alpha_{s})) (\mathrm{frac.})$  &  $\mathcal{O}(\alpha_{s}^{2}) (\mathrm{pb})$  &  $\delta(\mathcal{O}(\alpha_{s}^{2})) (\mathrm{frac.})$\\
\hline
\multirow{15}{*}{$N_\mathrm{jet} \geq 1$}  &  0.000  &   $ 0.209 $   &  0.486  &  $\pm$ $ 0.045 $   &  $\pm$ $ 0.36 $   &  0.232  &  &  &  &\\[0.2em]
  &   $ 0.209 $   &   $ 0.419 $   &  0.771  &  $\pm$ $ 0.19 $   &  $\pm$ $ 0.81 $   &  1.33  &  &  &  &\\[0.2em]
  &   $ 0.419 $   &   $ 0.628 $   &  1.37  &  $\pm$ $ 0.15 $   &  $\pm$ $ 0.56 $   &  1.79  &  &  &  &\\[0.2em]
  &   $ 0.628 $   &   $ 0.838 $   &  1.92  &  $\pm$ $ 0.15 $   &  $\pm$ $ 0.45 $   &  2.28  &  &  &  &\\[0.2em]
  &   $ 0.838 $   &   $ 1.05 $   &  2.78  &  $\pm$ $ 0.099 $   &  $\pm$ $ 0.32 $   &  2.79  &  &  &  &\\[0.2em]
  &   $ 1.05 $   &   $ 1.26 $   &  3.46  &  $\pm$ $ 0.12 $   &  $\pm$ $ 0.30 $   &  4.07  &  &  &  &\\[0.2em]
  &   $ 1.26 $   &   $ 1.47 $   &  7.75  &  $\pm$ $ 0.088 $   &  $\pm$ $ 0.30 $   &  6.68  &  &  &  &\\[0.2em]
  &   $ 1.47 $   &   $ 1.68 $   &  13.4  &  $\pm$ $ 0.065 $   &  $\pm$ $ 0.30 $   &  12.4  &  10.3  &  \scriptsize\begin{tabular}{@{}c@{}}$+$ $ 0.16 $ \\$-$ $ 0.16 $ \end{tabular}  &  23.6  &  \scriptsize\begin{tabular}{@{}c@{}}$+$ $ 0.38 $ \\$-$ $ 0.33 $ \end{tabular}\\
  &   $ 1.68 $   &   $ 1.88 $   &  25.4  &  $\pm$ $ 0.066 $   &  $\pm$ $ 0.29 $   &  27.9  &  19.0  &  \scriptsize\begin{tabular}{@{}c@{}}$+$ $ 0.14 $ \\$-$ $ 0.12 $ \end{tabular}  &  45.8  &  \scriptsize\begin{tabular}{@{}c@{}}$+$ $ 0.41 $ \\$-$ $ 0.32 $ \end{tabular}\\
  &   $ 1.88 $   &   $ 2.09 $   &  62.3  &  $\pm$ $ 0.045 $   &  $\pm$ $ 0.14 $   &  68.5  &  50.5  &  \scriptsize\begin{tabular}{@{}c@{}}$+$ $ 0.14 $ \\$-$ $ 0.11 $ \end{tabular}  &  95.1  &  \scriptsize\begin{tabular}{@{}c@{}}$+$ $ 0.28 $ \\$-$ $ 0.21 $ \end{tabular}\\
  &   $ 2.09 $   &   $ 2.30 $   &  158  &  $\pm$ $ 0.030 $   &  $\pm$ $ 0.14 $   &  168  &  142  &  \scriptsize\begin{tabular}{@{}c@{}}$+$ $ 0.16 $ \\$-$ $ 0.12 $ \end{tabular}  &  208  &  \scriptsize\begin{tabular}{@{}c@{}}$+$ $ 0.16 $ \\$-$ $ 0.13 $ \end{tabular}\\
  &   $ 2.30 $   &   $ 2.51 $   &  337  &  $\pm$ $ 0.020 $   &  $\pm$ $ 0.034 $   &  377  &  365  &  \scriptsize\begin{tabular}{@{}c@{}}$+$ $ 0.19 $ \\$-$ $ 0.14 $ \end{tabular}  &  455  &  \scriptsize\begin{tabular}{@{}c@{}}$+$ $ 0.10 $ \\$-$ $ 0.10 $ \end{tabular}\\
  &   $ 2.51 $   &   $ 2.72 $   &  758  &  $\pm$ $ 0.014 $   &  $\pm$ $ 0.021 $   &  820  &  898  &  \scriptsize\begin{tabular}{@{}c@{}}$+$ $ 0.21 $ \\$-$ $ 0.16 $ \end{tabular}  &  1029  &  \scriptsize\begin{tabular}{@{}c@{}}$+$ $ 0.057 $ \\$-$ $ 0.095 $ \end{tabular}\\
  &   $ 2.72 $   &   $ 2.93 $   &  2120  &  $\pm$ $ 0.0064 $   &  $\pm$ $ 0.060 $   &  1850  &  2307  &  \scriptsize\begin{tabular}{@{}c@{}}$+$ $ 0.17 $ \\$-$ $ 0.13 $ \end{tabular}  &  2472  &  \scriptsize\begin{tabular}{@{}c@{}}$+$ $ 0.042 $ \\$-$ $ 0.068 $ \end{tabular}\\
  &   $ 2.93 $   &   $ 3.14 $   &  8500  &  $\pm$ $ 0.0034 $   &  $\pm$ $ 0.048 $   &  7567  &  10525  &  \scriptsize\begin{tabular}{@{}c@{}}$+$ $ 0.11 $ \\$-$ $ 0.12 $ \end{tabular}  &  9396  &  \scriptsize\begin{tabular}{@{}c@{}}$+$ $ 0.046 $ \\$-$ $ 0.067 $ \end{tabular}\\
\hline
\multirow{12}{*}{$N_\mathrm{jet} \geq 2$}  &  0.000  &   $ 0.628 $   &  0.621  &  $\pm$ $ 0.32 $   &  $\pm$ $ 1.0 $   &  0.912  &  &  &  &\\[0.2em]
  &   $ 0.628 $   &   $ 1.05 $   &  2.84  &  $\pm$ $ 0.21 $   &  $\pm$ $ 0.61 $   &  2.54  &  &  &  &\\[0.2em]
  &   $ 1.05 $   &   $ 1.26 $   &  6.29  &  $\pm$ $ 0.18 $   &  $\pm$ $ 0.42 $   &  4.05  &  &  &  &\\[0.2em]
  &   $ 1.26 $   &   $ 1.47 $   &  9.61  &  $\pm$ $ 0.18 $   &  $\pm$ $ 0.39 $   &  6.41  &  &  &  &\\[0.2em]
  &   $ 1.47 $   &   $ 1.68 $   &  20.5  &  $\pm$ $ 0.088 $   &  $\pm$ $ 0.31 $   &  11.7  &  6.18  &  \scriptsize\begin{tabular}{@{}c@{}}$+$ $ 0.057 $ \\$-$ $ 0.040 $ \end{tabular}  &  18.9  &  \scriptsize\begin{tabular}{@{}c@{}}$+$ $ 0.24 $ \\$-$ $ 0.18 $ \end{tabular}\\
  &   $ 1.68 $   &   $ 1.88 $   &  24.6  &  $\pm$ $ 0.12 $   &  $\pm$ $ 0.25 $   &  25.6  &  12.6  &  \scriptsize\begin{tabular}{@{}c@{}}$+$ $ 0.096 $ \\$-$ $ 0.067 $ \end{tabular}  &  40.0  &  \scriptsize\begin{tabular}{@{}c@{}}$+$ $ 0.43 $ \\$-$ $ 0.31 $ \end{tabular}\\
  &   $ 1.88 $   &   $ 2.09 $   &  57.2  &  $\pm$ $ 0.069 $   &  $\pm$ $ 0.11 $   &  60.1  &  37.1  &  \scriptsize\begin{tabular}{@{}c@{}}$+$ $ 0.14 $ \\$-$ $ 0.086 $ \end{tabular}  &  88.0  &  \scriptsize\begin{tabular}{@{}c@{}}$+$ $ 0.35 $ \\$-$ $ 0.25 $ \end{tabular}\\
  &   $ 2.09 $   &   $ 2.30 $   &  129  &  $\pm$ $ 0.044 $   &  $\pm$ $ 0.070 $   &  139  &  114  &  \scriptsize\begin{tabular}{@{}c@{}}$+$ $ 0.17 $ \\$-$ $ 0.12 $ \end{tabular}  &  193  &  \scriptsize\begin{tabular}{@{}c@{}}$+$ $ 0.23 $ \\$-$ $ 0.18 $ \end{tabular}\\
  &   $ 2.30 $   &   $ 2.51 $   &  239  &  $\pm$ $ 0.031 $   &  $\pm$ $ 0.055 $   &  290  &  303  &  \scriptsize\begin{tabular}{@{}c@{}}$+$ $ 0.24 $ \\$-$ $ 0.17 $ \end{tabular}  &  405  &  \scriptsize\begin{tabular}{@{}c@{}}$+$ $ 0.17 $ \\$-$ $ 0.14 $ \end{tabular}\\
  &   $ 2.51 $   &   $ 2.72 $   &  449  &  $\pm$ $ 0.023 $   &  $\pm$ $ 0.036 $   &  542  &  742  &  \scriptsize\begin{tabular}{@{}c@{}}$+$ $ 0.31 $ \\$-$ $ 0.23 $ \end{tabular}  &  829  &  \scriptsize\begin{tabular}{@{}c@{}}$+$ $ 0.10 $ \\$-$ $ 0.13 $ \end{tabular}\\
  &   $ 2.72 $   &   $ 2.93 $   &  510  &  $\pm$ $ 0.019 $   &  $\pm$ $ 0.11 $   &  763  &  1294  &  \scriptsize\begin{tabular}{@{}c@{}}$+$ $ 0.47 $ \\$-$ $ 0.34 $ \end{tabular}  &  1342  &  \scriptsize\begin{tabular}{@{}c@{}}$+$ $ 0.15 $ \\$-$ $ 0.19 $ \end{tabular}\\
  &   $ 2.93 $   &   $ 3.14 $   &  382  &  $\pm$ $ 0.032 $   &  $\pm$ $ 0.75 $   &  706  &  991  &  \scriptsize\begin{tabular}{@{}c@{}}$+$ $ 0.48 $ \\$-$ $ 0.34 $ \end{tabular}  &  1312  &  \scriptsize\begin{tabular}{@{}c@{}}$+$ $ 0.38 $ \\$-$ $ 0.31 $ \end{tabular}\\
\hline
\multirow{5}{*}{$N_\mathrm{jet} \geq 3$}  &  0.000  &   $ 0.838 $   &  0.518  &  $\pm$ $ 0.40 $   &  $\pm$ $ 1.7 $   &  0.553  &  &  &  &\\[0.2em]
  &   $ 0.838 $   &   $ 1.47 $   &  2.34  &  $\pm$ $ 0.28 $   &  $\pm$ $ 0.30 $   &  1.96  &  &  &  &\\[0.2em]
  &   $ 1.47 $   &   $ 2.09 $   &  11.4  &  $\pm$ $ 0.12 $   &  $\pm$ $ 0.091 $   &  10.4  &  &  &  14.0  &  \scriptsize\begin{tabular}{@{}c@{}}$+$ $ 0.51 $ \\$-$ $ 0.31 $ \end{tabular}\\
  &   $ 2.09 $   &   $ 2.72 $   &  24.3  &  $\pm$ $ 0.12 $   &  $\pm$ $ 0.77 $   &  45.6  &  &  &  64.9  &  \scriptsize\begin{tabular}{@{}c@{}}$+$ $ 1.1 $ \\$-$ $ 0.67 $ \end{tabular}\\
  &   $ 2.72 $   &   $ 3.14 $   &  24.3  &  $\pm$ $ 0.14 $   &  $\pm$ $ 0.42 $   &  68.2  &  &  &  122  &  \scriptsize\begin{tabular}{@{}c@{}}$+$ $ 2.0 $ \\$-$ $ 1.2 $ \end{tabular}\\
\hline
\end{tabular}
}
\end{center}
\caption{
Differential cross sections, $d\sigma/d\Delta\phi$, in the $Q^{2}$ region of $100\;\mathrm{GeV}^{2} < Q^{2} < 350\;\mathrm{GeV}^{2}$ for $N_\mathrm{jet} \geq$ 1, 2, and 3, as obtained from the data, ARIADNE MC simulations, and perturbative calculations at $\mathcal{O}(\alpha_{s})$ and $\mathcal{O}(\alpha_{s}^{2})$ accuracy.
Other details are as in the caption to Table~\ref{TAB_00}.
}
\label{TAB_22}
\end{table}

%%%%%%%%%%%%%%%%%%%%%%%%%%%%%%%%%%%%%%%%%%%%%
% Control 
%%%%%%%%%%%%%%%%%%%%%%%%%%%%%%%%%%%%%%%%%%%%%
\begin{figure}
\centering
\includegraphics[width=\textwidth]{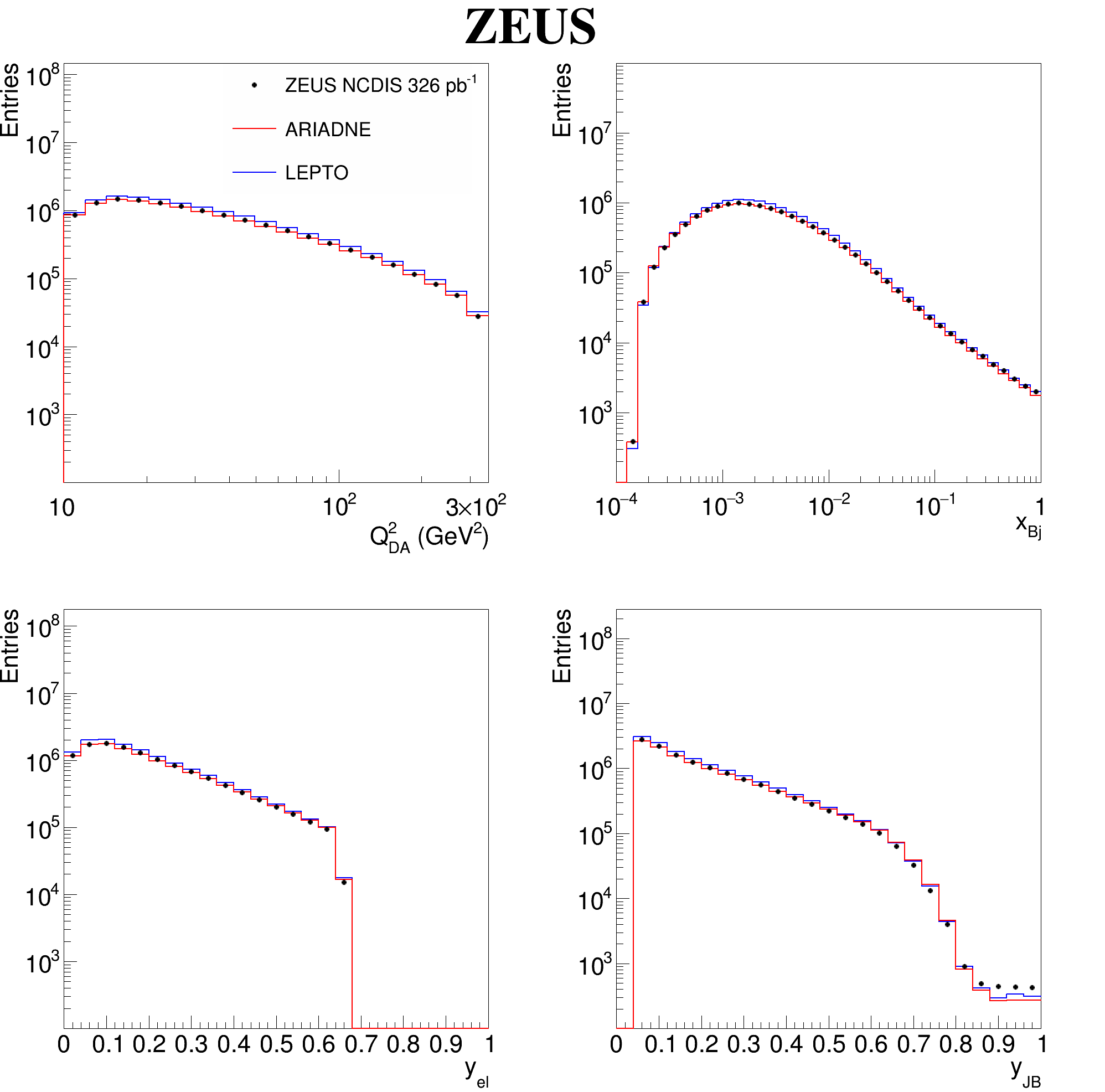}
\caption{
Comparison between data (dots) and ARIADNE and LEPTO MC simulations (histograms) for event level quantities:
photon virtuality reconstructed with the double angle method $Q_{\mathrm{DA}}^2$ (top left), Bjorken-x (top right) and lepton inelasticity reconstructed with the electron method $y_{el}$ (bottom left), and that with the Jacquet--Blondel method $y_\mathrm{JB}$ (bottom right).
The MC simulations are normalised to the luminosity of the data.
}
\label{QA_event}
\end{figure}
\clearpage

\begin{figure}
\centering
\includegraphics[width=\textwidth]{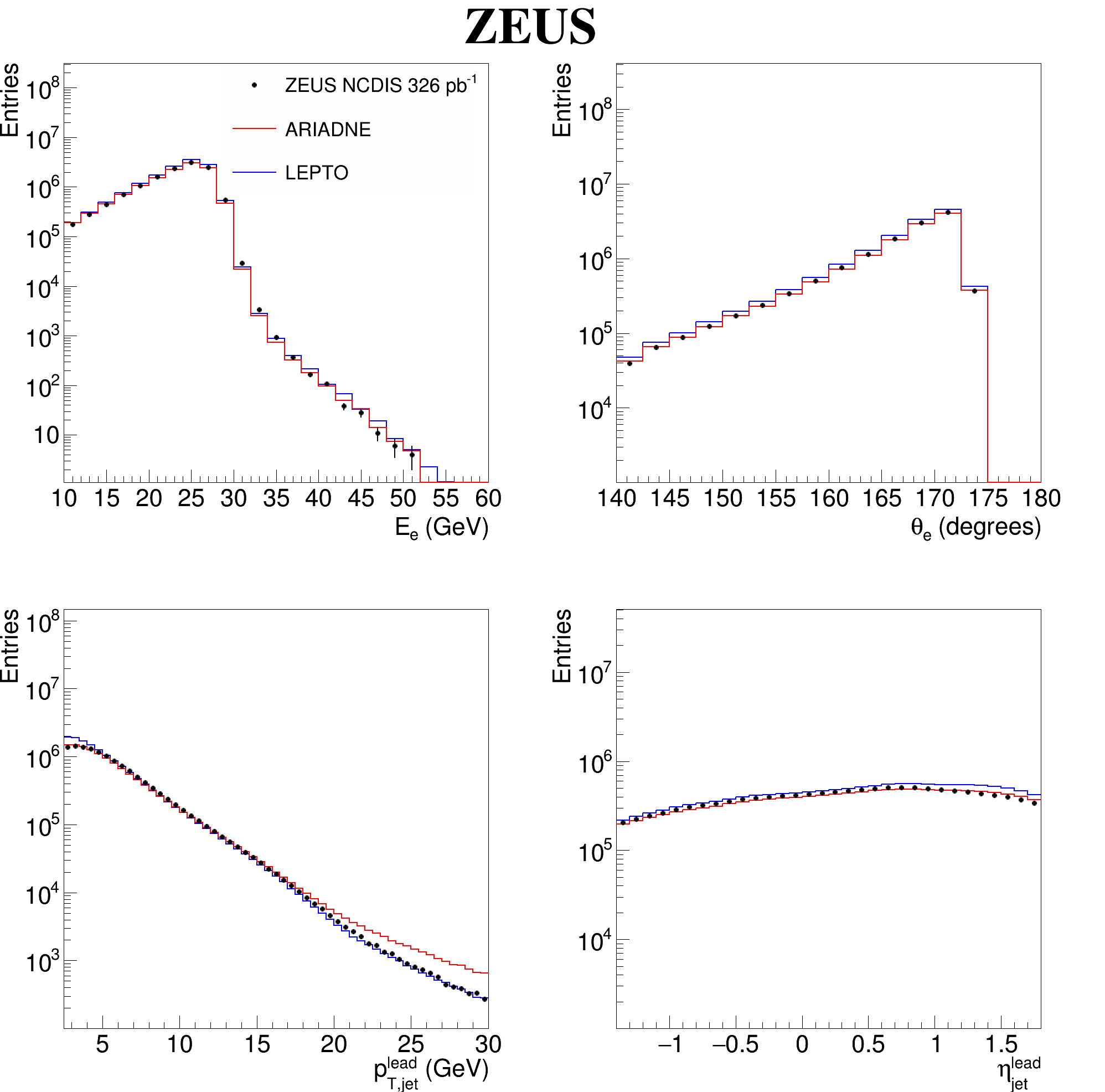}
\caption{
Comparison between data (dots) and ARIADNE and LEPTO MC simulations (histograms) for lepton (top) and jet (bottom) quantities:
lepton energy $E_{e}$  (top left), lepton polar angle $\theta_{e}$ (top right), leading-jet transverse momentum $p_\mathrm{T,jet}^\mathrm{lead}$ (bottom left), and leading-jet pseudorapidity $\eta_\mathrm{jet}^\mathrm{lead}$ (bottom right).
The MC simulations are normalised to the luminosity of the data.
}
\label{QA_leptonjet}
\end{figure}
\clearpage

\begin{figure}
\centering
\includegraphics[width=\textwidth]{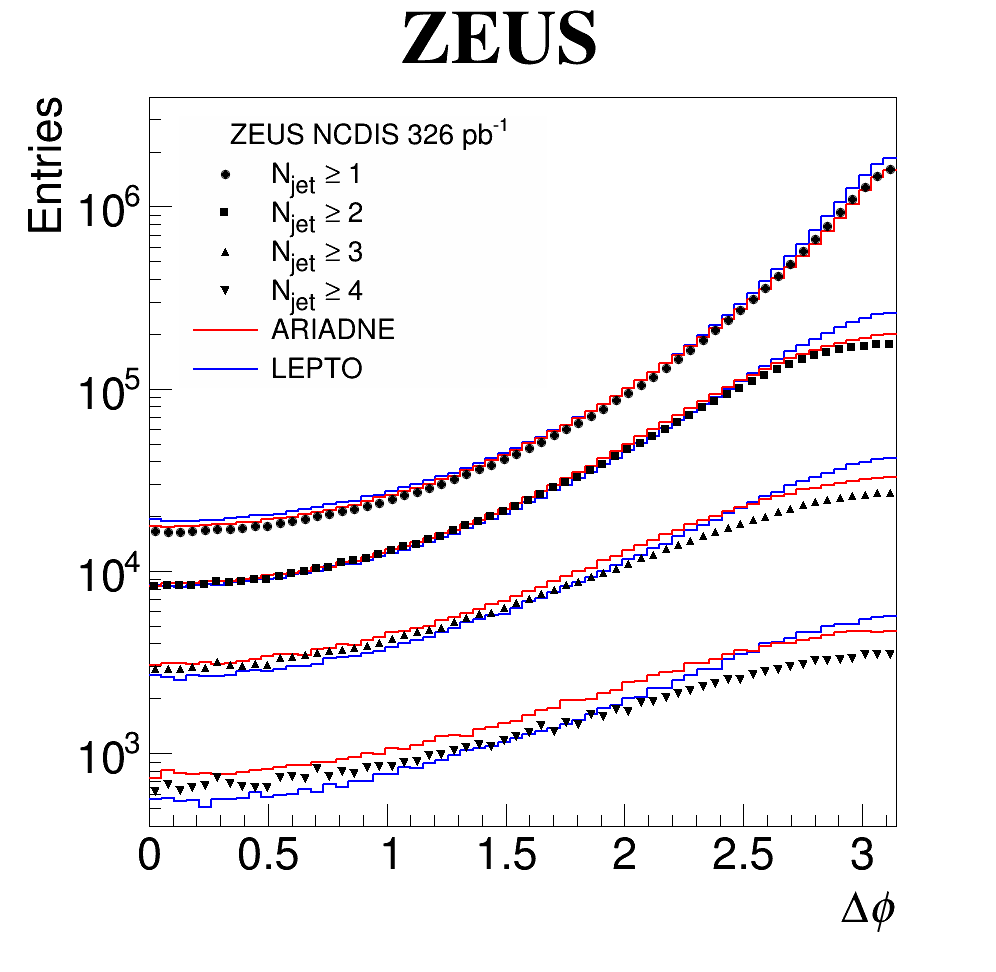}
\caption{
The distributions of the reconstructed lepton--leading-jet pair as functions of the the correlation angle, $\Delta\phi$, without corrections for detector effects.
The points represent the yield from HERA II data with each shape representing different jet multiplicity range, as described in the legend.
The histograms represent the distributions obtained from the ARIADNE and LEPTO MC simulations.
}
\label{QA_dphi}
\end{figure}
\clearpage

%%%%%%%%%%%%%%%%%%%%%%%%%%%%%%%%%%%%%%%%%%%%%
% Systematics
%%%%%%%%%%%%%%%%%%%%%%%%%%%%%%%%%%%%%%%%%%%%%
\begin{figure}[h]
\centering
\includegraphics[width=1\textwidth]{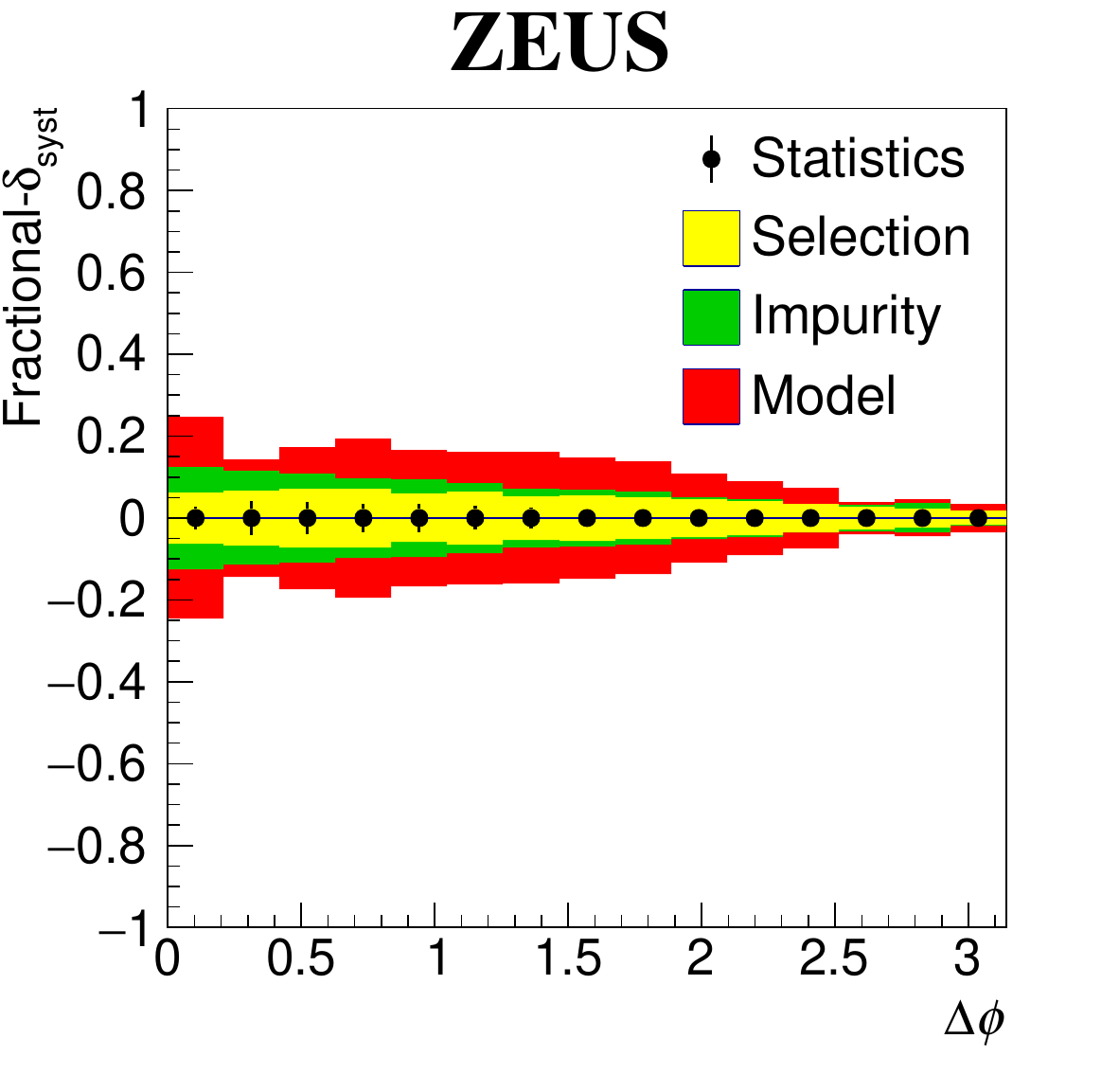}
\caption{ 
Breakdown of the systematic uncertainty in the inclusive measurement compared to the statistical uncertainty.
Each band represents the addition in quadrature to all of the prior contributions, so that "Selection" only represents the systematic uncertainty obtained by varying the selection-cut values, "Impurity" represents the selection uncertainty added in quadrature to the uncertainty associated with the impurity background calculation, and "Model" includes both selection and impurity uncertainties in addition to the systematic dependence in the assumptions made in the ARIADNE model.
}
\label{syst_0}
\end{figure}

%%%%%%%%%%%%%%%%%%%%%%%%%%%%%%%%%%%%%%%%%%%%%
% Differential cross section
%%%%%%%%%%%%%%%%%%%%%%%%%%%%%%%%%%%%%%%%%%%%%
\begin{figure}
\centering
\includegraphics[width=\textwidth]{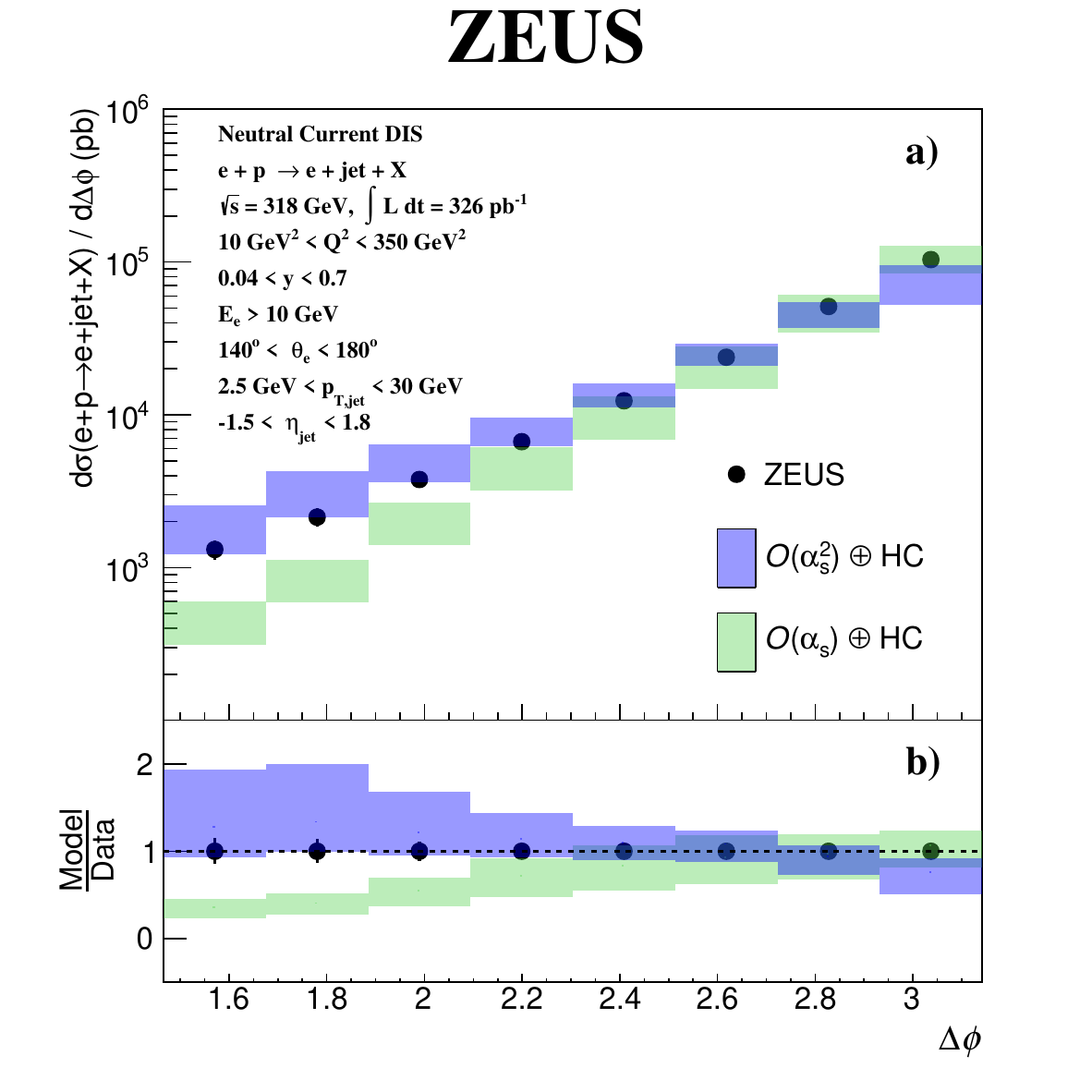}
\caption{ 
a) The differential cross section $d\sigma(e + p \rightarrow e + \mathrm{jet}^\mathrm{lead} + X) / d\Delta\phi$ as a function of the azimuthal correlation angle $\Delta\phi$ in the full fiducial region.
The vertical error bars represent the statistical and systematic uncertainties added in quadrature.
The green and blue bands represent the perturbative QCD calculations at $\mathcal{O}(\alpha)$ and $\mathcal{O}(\alpha^2)$ accuracy, respectively, corrected for hadronisation effects (HC).
b) Ratio of model/data.
}
\label{final_1_0}
\end{figure}
\clearpage

\begin{figure}
\centering
\includegraphics[width=\textwidth]{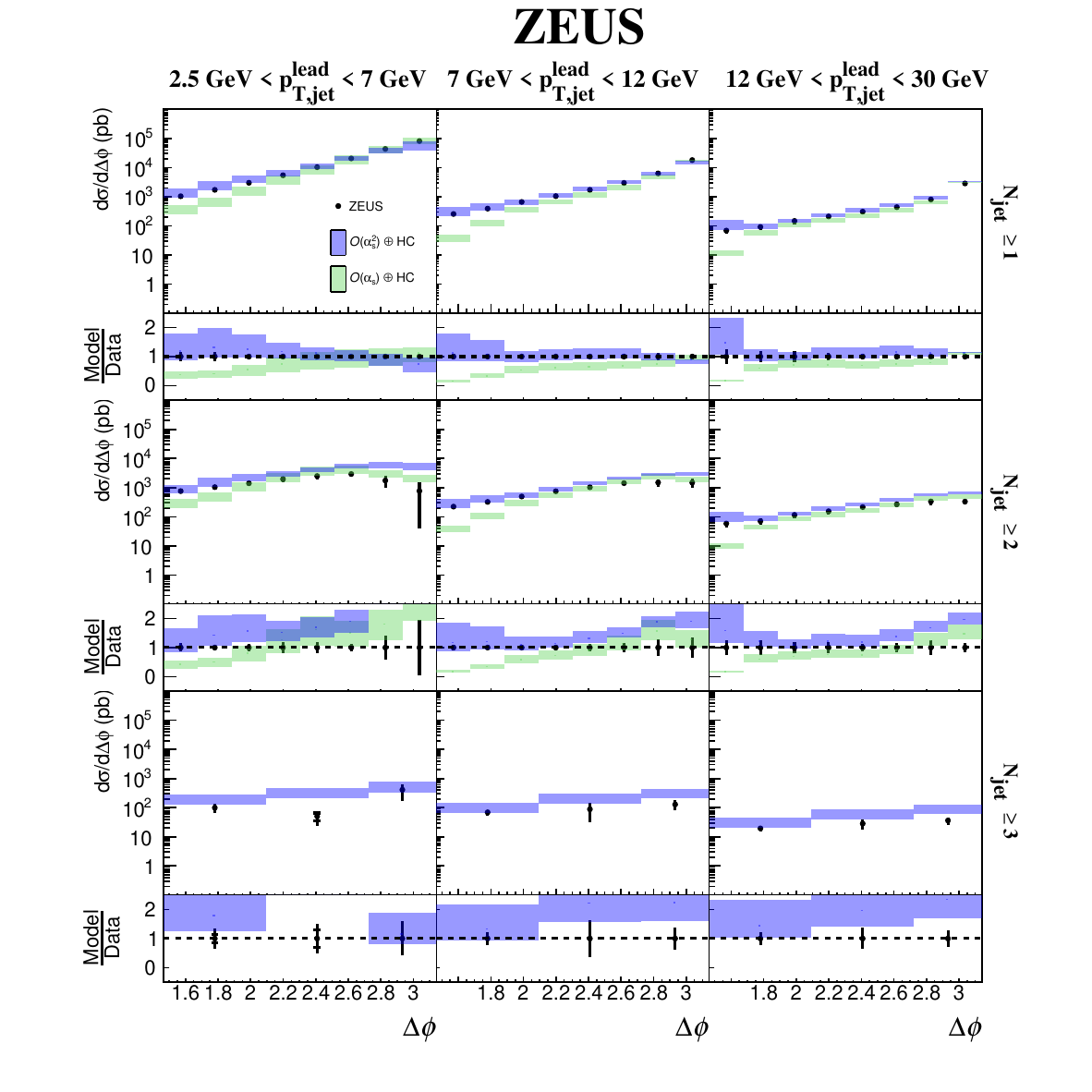}
\caption{ 
The differential cross sections $d\sigma(e + p \rightarrow e + \mathrm{jet}^\mathrm{lead} + X) / d\Delta\phi$ as functions of the azimuthal correlation angle $\Delta\phi$, while varying the $N_\mathrm{jet}$ and $p_\mathrm{T,jet}^\mathrm{lead}$ ranges.
The dots denote the ZEUS measurement.
The vertical error bars represent the statistical and systematic uncertainties added in quadrature. 
When horizontal ticks are visible, they represent the statistical uncertainty.
The green and blue bands represent the perturbative calculations at $\mathcal{O}(\alpha)$ and $\mathcal{O}(\alpha^2)$ accuracy, respectively, corrected for hadronisation effects (HC).
}
\label{final_1_1}
\end{figure}
\clearpage

\begin{figure}
\centering
\includegraphics[width=\textwidth]{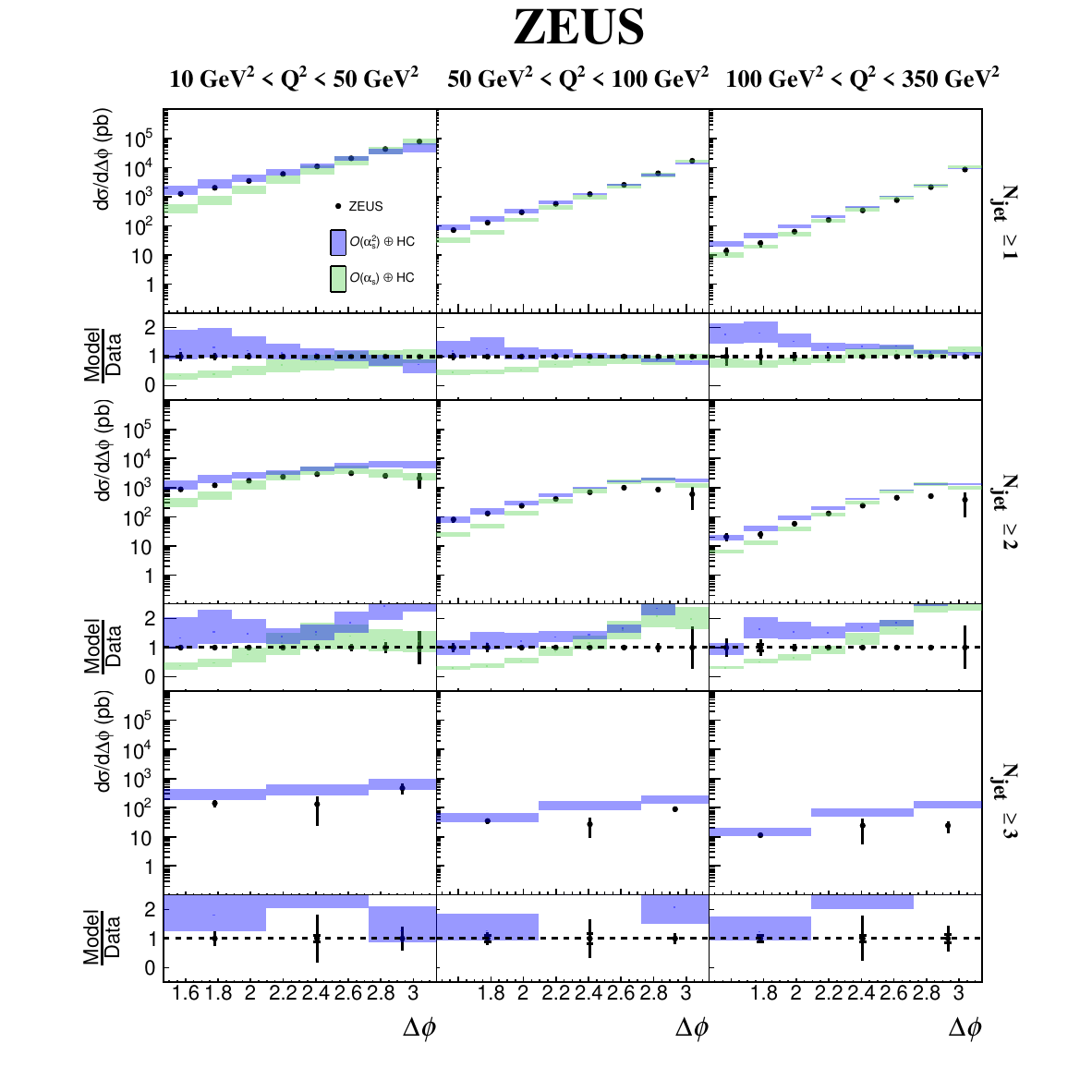}
\caption{ 
The differential cross sections $d\sigma(e + p \rightarrow e + \mathrm{jet}^\mathrm{lead} + X) / d\Delta\phi$ as functions of the azimuthal correlation angle $\Delta\phi$, while varying the $N_\mathrm{jet}$ and $Q^2$ ranges.
The dots denote the ZEUS measurement.
The vertical error bars represent the statistical and systematic uncertainties added in quadrature. 
When horizontal ticks are visible, they represent the statistical uncertainty.
The green and blue bands represent the perturbative calculations at $\mathcal{O}(\alpha)$ and $\mathcal{O}(\alpha^2)$ accuracy, respectively, corrected for hadronisation effects (HC).
}
\label{final_1_2}
\end{figure}
\clearpage

\begin{figure}
\centering
\includegraphics[width=\textwidth]{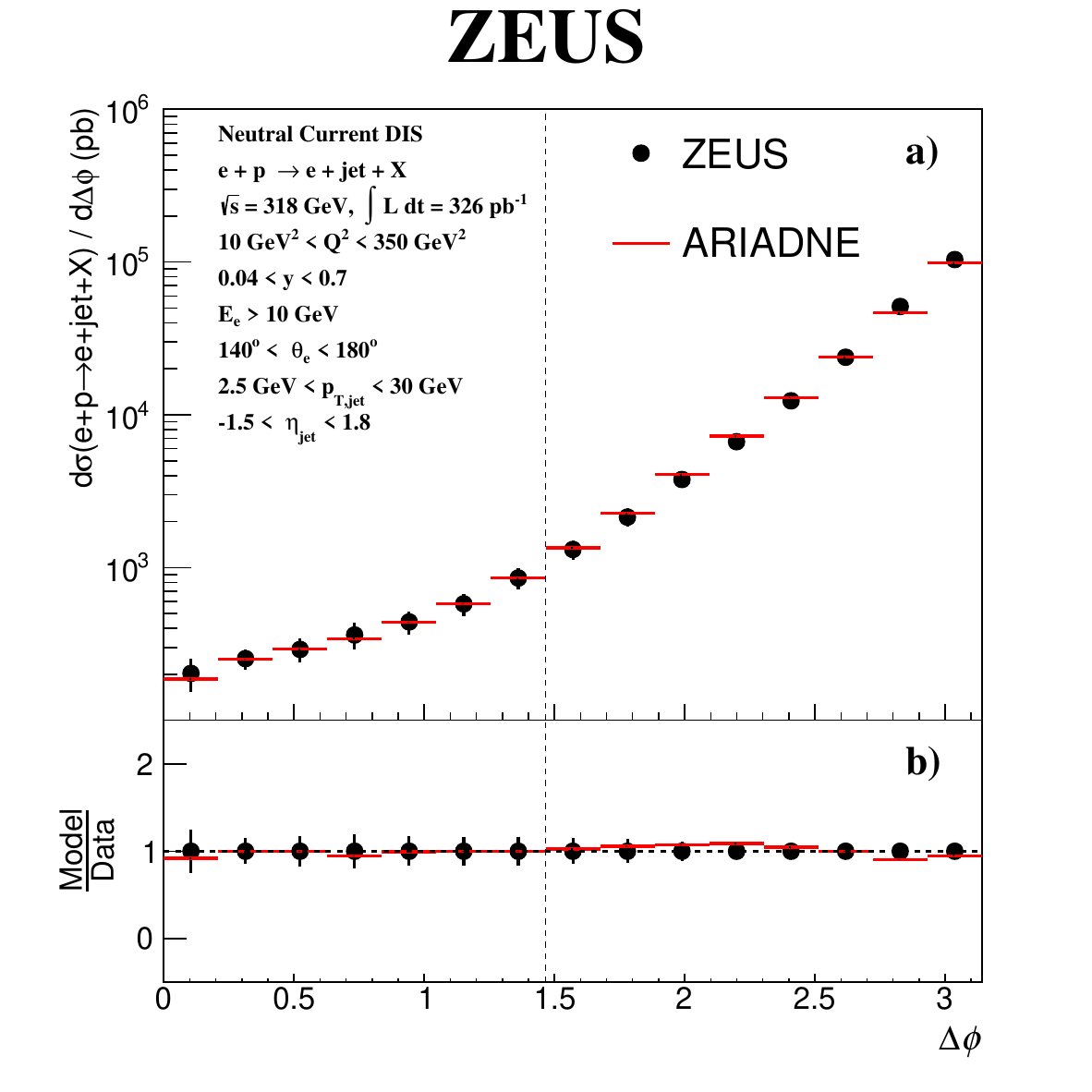}
\caption{ 
a) The differential cross section $d\sigma(e + p \rightarrow e + \mathrm{jet}^\mathrm{lead} + X) / d\Delta\phi$ as a function of the azimuthal correlation angle $\Delta\phi$ in the full fiducial region.
The vertical error bars represent the statistical and systematic uncertainties added in quadrature.
The red line represents the prediction from the ARIADNE MC generator.
b) Ratio of model/data.
The vertical dashed line indicates the $\Delta\phi$ range of the perturbative QCD predictions in Fig.~\ref{final_1_0}.
}
\label{final_2_0}
\end{figure}
\clearpage

\begin{figure}
\centering
\includegraphics[width=\textwidth]{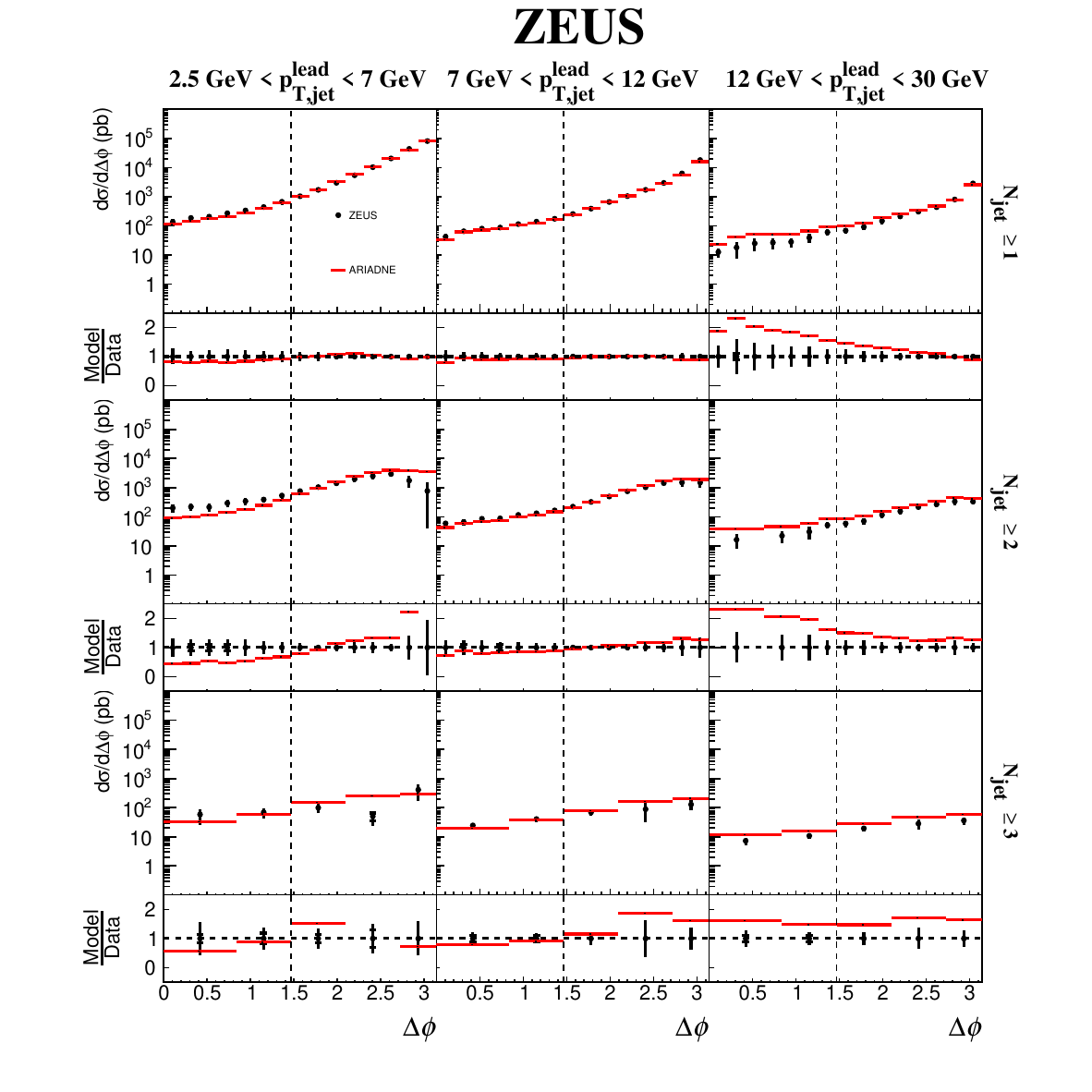}
\caption{ 
The differential cross sections $d\sigma(e + p \rightarrow e + \mathrm{jet}^\mathrm{lead} + X) / d\Delta\phi$ as functions of the azimuthal correlation angle $\Delta\phi$, while varying the $N_\mathrm{jet}$ and $p_\mathrm{T,jet}^\mathrm{lead}$ ranges.
The vertical error bars represent the statistical and systematic uncertainties added in quadrature. 
When horizontal ticks are visible, they represent the statistical uncertainty.
Other details are as in the caption to Fig.~\ref{final_2_0}.
}
\label{final_2_1}
\end{figure}
\clearpage

\begin{figure}
\centering
\includegraphics[width=\textwidth]{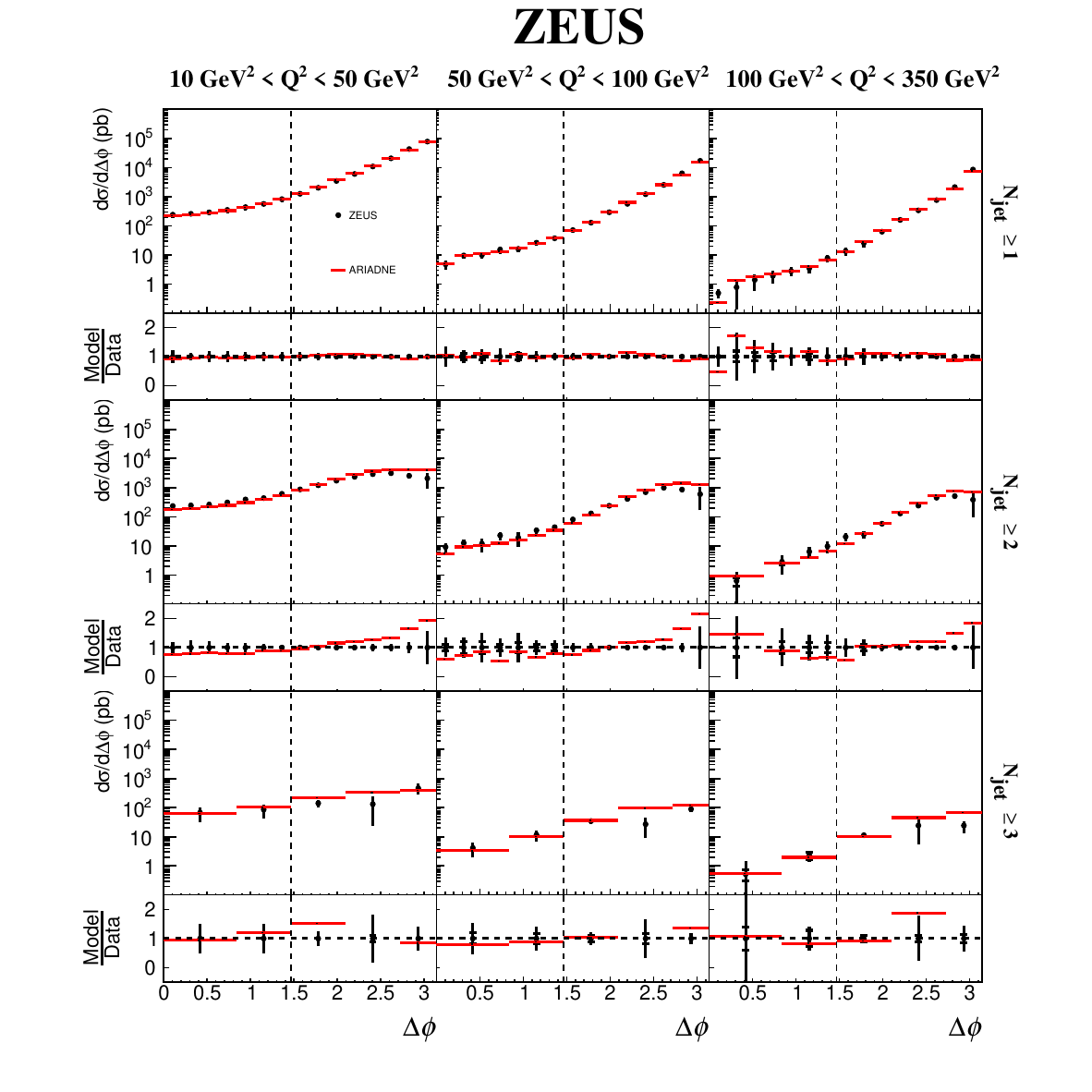}
\caption{ 
The differential cross sections $d\sigma(e + p \rightarrow e + \mathrm{jet}^\mathrm{lead} + X) / d\Delta\phi$ as functions of the azimuthal correlation angle $\Delta\phi$, while varying the $N_\mathrm{jet}$ and $Q^2$ ranges.
The vertical error bars represent the statistical and systematic uncertainties added in quadrature. 
When horizontal ticks are visible, they represent the statistical uncertainty.
Other details are as in the caption to Fig.~\ref{final_2_0}.
}
\label{final_2_2}
\end{figure}
\clearpage

%%%%%%%%%%%%%%%%%%%%%%%%%%%%%%%%%%%%%%%%%%%%%

%%%%%%%%%%%%%%%%%%%%%%%%%%%%%%%%%%%%%%%%%%%%%

%%%%%%%%%%%%%%%%%%%%%%%%%%%%%%%%%%%%%%%%%%%%%

%%%%%%%%%%%%%%%%%%%%%%%%%%%%%%%%%%%%%%%%%%%%%

%%%%%%%%%%%%%%%%%%%%%%%%%%%%%%%%%%%%%%%%%%%%%

%%%%%%%%%%%%%%%%%%%%%%%%%%%%%%%%%%%%%%%%%%%%%

%%%%%%%%%%%%%%%%%%%%%%%%%%%%%%%%%%%%%%%%%%%%%

%%%%%%%%%%%%%%%%%%%%%%%%%%%%%%%%%%%%%%%%%%%%%

%%%%%%%%%%%%%%%%%%%%%%%%%%%%%%%%%%%%%%%%%%%%%

%%%%%%%%%%%%%%%%%%%%%%%%%%%%%%%%%%%%%%%%%%%%%

%%%%%%%%%%%%%%%%%%%%%%%%%%%%%%%%%%%%%%%%%%%%%

%%%%%%%%%%%%%%%%%%%%%%%%%%%%%%%%%%%%%%%%%%%%%

\begin{appendices}

%%%%%%%%%%%%%%%%%%%%%%%%%%%%%%%%%%%%%%%%%%%%%
% Unfolding
%%%%%%%%%%%%%%%%%%%%%%%%%%%%%%%%%%%%%%%%%%%%%
\section{Unfolding}
\label{app_unfold}

The event kinematics obtained from the detector response is subject to effects arising from imperfect resolution and inefficiency of the chosen reconstruction scheme.
A three-level unfolding scheme was employed in order to map the yield of reconstructed lepton--leading-jet pairs to that of true electrons and hadron jets that are free of these effects.

Each DIS event can be categorised into one of four groups:

\begin{itemize}
\item $n_{11}$ --- the event enters into both the fiducial region defined by the kinematics reconstructed with the detector response and the region defined by the true electron and final-state hadrons;

\item $n_{10}$ --- it falsely falls outside the fiducial region defined by the detector-level kinematics;

\item $n_{01}$ --- it falsely falls into the detector-level fiducial region;

\item $n_{00}$ --- it correctly falls outside the detector-level fiducial region.
\end{itemize}

The $\Delta \phi \otimes N_\mathrm{jet}$ distribution of $n_{01}$, also referred to as impurity background, was first subtracted from the measured signal ($n_{01} + n_{11}$) to extract $n_{11}$.
The impurity background was estimated using two different methods:
a) the relative fraction of the impurity background in the MC simulation was taken as the background in data,
$n_{01}^\mathrm{data} = n_{01}^\mathrm{MC}/(n_{01}^\mathrm{MC}+ n_{11}^\mathrm{MC}) \cdot (n_{01}^\mathrm{data} + n_{11}^\mathrm{data})$.
The resulting $\Delta \phi \otimes N_\mathrm{jet}$ distribution was considered the nominal impurity background,
b) the absolute yield derived from the simulation was directly taken as the background in data, $n_{01}^\mathrm{data} = n_{01}^\mathrm{MC}$.
The difference in the background extracted with these methods was taken as the systematic uncertainty in the impurity estimate.

This was followed by a two-dimensional regularised unfolding technique implemented in the TUnfold package~\cite{TUnfold} to account for migration of $\Delta\phi$ and $N_\mathrm{jet}$ within the extracted $n_{11}$ events.
With $\tilde{n}$ defined as a distribution of a chosen quantity reconstructed from the detector response, $n$ as the distribution of the same quantity defined at the hadron level, and $\mathbf{A}$ as the response matrix, a folding equation can be formed as $\tilde{n} = \mathbf{A}n$.
A regularised unfolding is performed by minimising the following expression:

\begin{equation}
\label{eq-un1}
\chi^2 
= (\tilde{n}-\mathbf{A}n)^\mathrm{T}   \mathbf{V}_{\tilde{n}}^{-1}   (\tilde{n}-\mathbf{A}n)
+ \tau^2    (n - n_0)^\mathrm{T}     \mathbf{L}^\mathrm{T} \mathbf{L}     (n - n_0),
\end{equation}

where $\mathbf{V}_{\tilde{n}}$ represents the covariance matrix of the quantity $\tilde{n}$, the parameter $\tau$ is referred to as the regularisation parameter, $n_0$ is the normalised truth-level distribution obtained from $\mathbf{A}$, and the matrix $\mathbf{L}$ contains the regularisation conditions.

The first term in equation~\ref{eq-un1} represents the least-square minimisation of the folding equation.
In this measurement, the response matrix was constructed from the migration matrix, $\mathbf{M}$.
The two-dimensional information of $\Delta\phi$ and $N_\mathrm{jet}$ from each $n_{11}$ event in the MC simulation was mapped onto a one-dimensional axis.
The values in $\mathbf{M}$ were determined by mapping the hadron-level distribution of $\Delta \phi \otimes N_\mathrm{jet}$ to the detector-level distribution.
Figures~\ref{unfold_1} and ~\ref{unfold_2} represent the migration matrices for the inclusive and $p_\mathrm{T,jet}^\mathrm{lead}$/$Q^2$-dependent measurements, respectively.
Matrices of statistical correlation coefficients of the unfolded lepton--leading-jet yield for the inclusive and $p_\mathrm{T,jet}^\mathrm{lead}$/$Q^2$-dependent measurements are shown in Figs.~\ref{correlation_1} and ~\ref{correlation_2}, respectively.
Typically, negative correlations of $\sim-0.5$ are observed in the adjacent $\Delta\phi$ and $N_\mathrm{jet}$ bins due to detector and reconstruction effects.

The least-square approach is prone to large fluctuations in the resulting $n$ distribution.
The regularisation term, shown as the second term in equation~\ref{eq-un1}, dampens this fluctuation based on the regularisation parameter and the regularisation conditions.
In this measurement, the regularisation was performed on the second derivatives; the diagonal elements of $\mathbf{L}$, $L_{i,i}$, were set to $1$, $L_{i,i+1} = -2$, and $L_{i,i+2} = 1$. 
The regularisation parameter, $\tau$, was obtained using the $L$-curve scan method, i.e., the value of $\tau$ is chosen from the point where the curvature is maximal in the graph of $L_y (L_x)$ from a simplified form of equation~\ref{eq-un1}, $\chi^2 = e^{L_x} + \tau^2 e^{L_y}$.
The term $\tau^2 e^{L_y}$ typically contributes $10$ to $20\;\%$ to the value of $\chi^2$.

Finally, the efficiency of the ZEUS detector and reconstruction procedure was estimated with the MC simulation as $\epsilon = \frac{n_{11}^\mathrm{MC}}{n_{11}^\mathrm{MC} + n_{10}^\mathrm{MC}}$.
The unfolding procedure described above can be expressed as follows:

\begin{equation}
\label{eq-yield}
n_{\mathrm{had}} =
\frac{1}{\epsilon} \cdot \mathbf{A'}^{-1} (\tilde{n}_\mathrm{sig} - \tilde{n}_\mathrm{imp}),
\end{equation}

where $\mathbf{A'}^{-1}$ represents the regularised unfolding, $\tilde{n}_\mathrm{sig}$ is the $\Delta \phi \otimes N_\mathrm{jet}$ distribution of $n_{01} + n_{11}$ as reconstructed from the detector response, $\tilde{n}_\mathrm{imp}$ is the estimated impurity background, and $n_{\mathrm{had}}$ is the hadron-level $\Delta \phi \otimes N_\mathrm{jet}$ distribution of $n_{11} + n_{10}$ that is free of detector and reconstruction effects.

\begin{figure}[h]
\centering
\includegraphics[width=\textwidth]{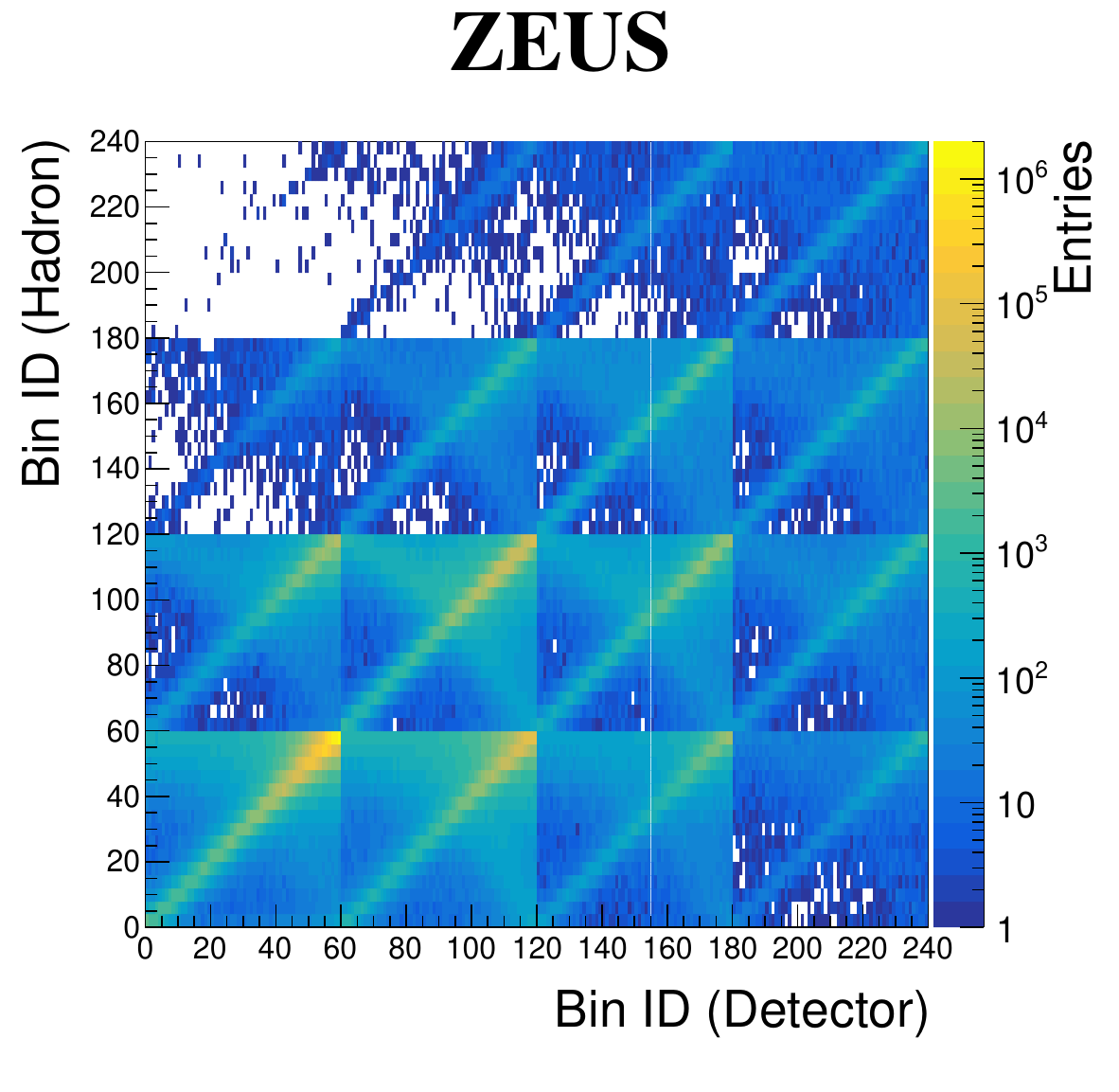}
\caption{
The migration matrix as input for the unfolding of the inclusive measurement.
This two-dimensional distribution describes the migration of $\Delta\phi$ and $N_\mathrm{jet}$ during the detection and reconstruction processes.
The azimuthal correlation angle $\Delta\phi$ of each lepton--leading-jet pair was assigned a bin ID between 0 to 59 segmented uniformly from 0 to $\pi$.
This bin ID was offset by multiples of 60 based on the jet multiplicity so that the pair from a single jet event, as suggested by the MC simulation, was assigned a value between 0 to 59, dijet events were given 60 to 119, trijet 120 to 179, and four or more jets 180 to 239.
The vertical axis represents the distribution at the hadron level, while the horizontal axis represents the one at the detector level.
}
\label{unfold_1}
\end{figure}

\begin{figure}[h]
\centering
\includegraphics[width=\textwidth]{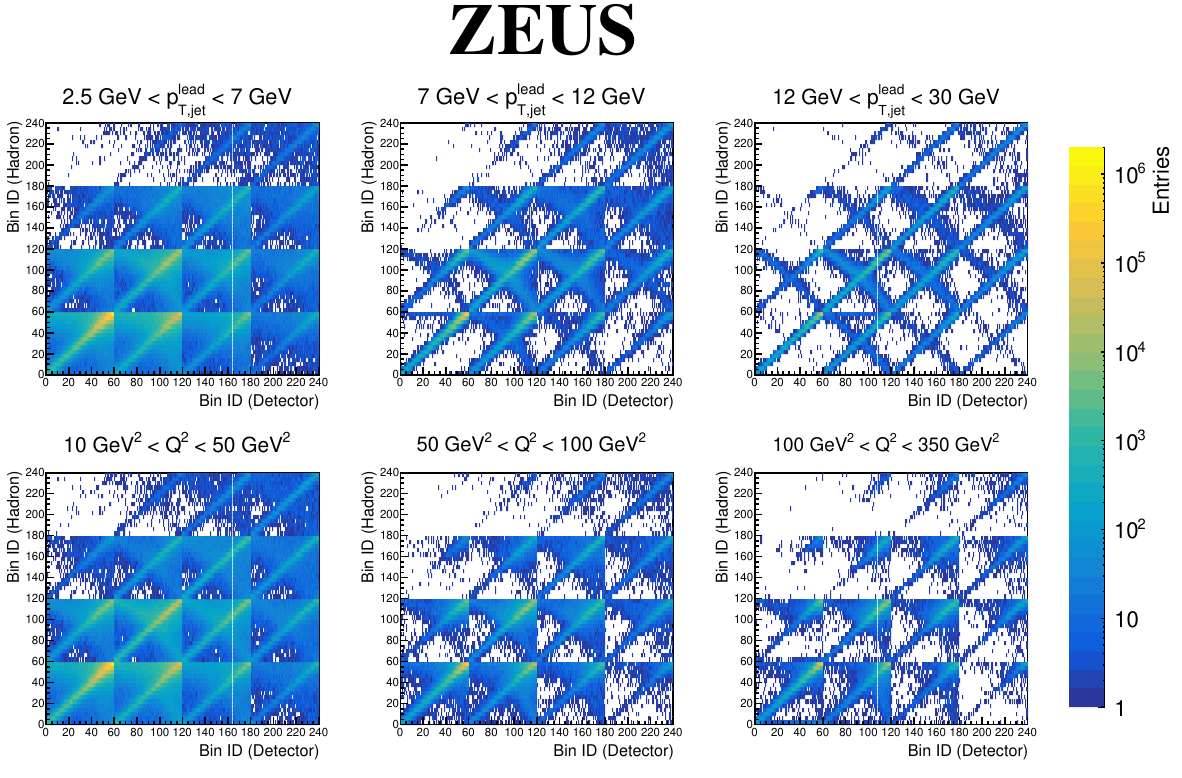}
\caption{
The migration matrices as input for the unfolding of the measurements in various $p_\mathrm{T,jet}$ (top) and $Q^2$ (bottom) ranges.
These two-dimensional distributions describe the migration of $\Delta\phi$ and $N_\mathrm{jet}$ during the detection and reconstruction stage.
Other details are as in the caption to Fig.~\ref{unfold_1}.
}
\label{unfold_2}
\end{figure}

\begin{figure}[h]
\centering
\includegraphics[width=\textwidth]{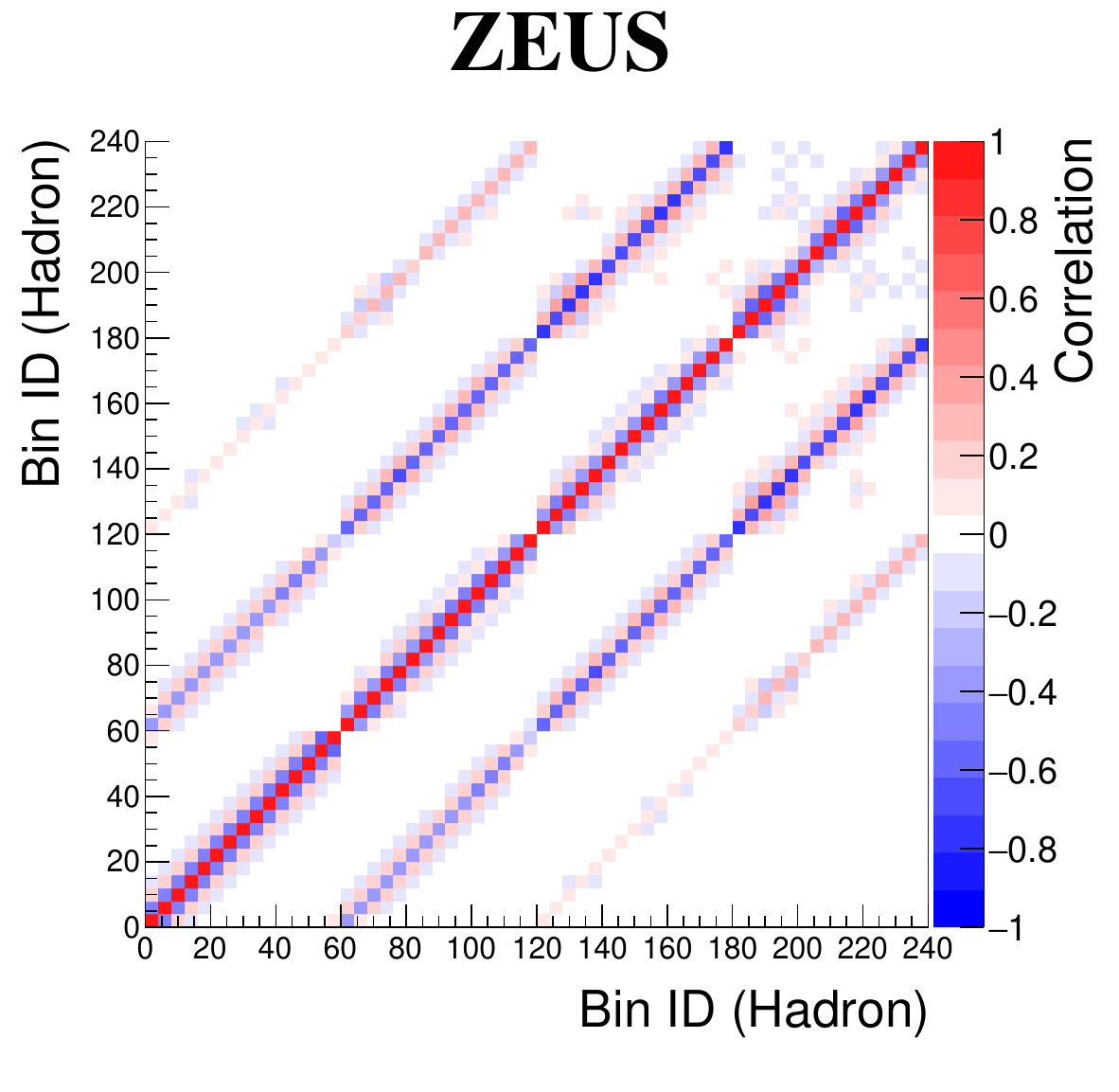}
\caption{
The correlation matrix for the inclusive measurement.
The azimuthal correlation angle $\Delta\phi$ of each lepton--leading-jet pair was assigned a bin ID between 0 to 59 segmented uniformly from 0 to $\pi$.
This bin ID was offset by multiples of 60 based on the jet multiplicity so that the pair from a single jet event, as suggested by the MC simulation, was assigned a value between 0 to 59, dijet events were given 60 to 119, trijet 120 to 179, and four or more jets 180 to 239.
}
\label{correlation_1}
\end{figure}

\begin{figure}[h]
\centering
\includegraphics[width=\textwidth]{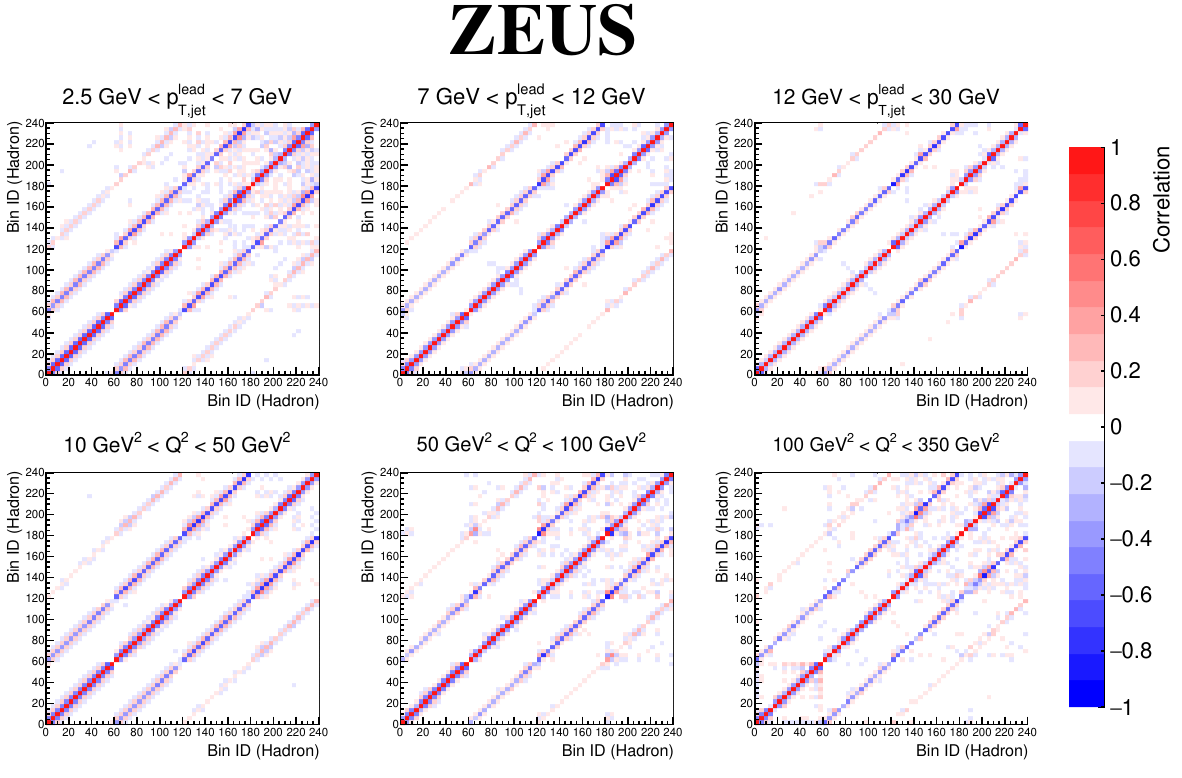}
\caption{
The correlation matrix for $p_\mathrm{T,jet}$- (top) and $Q^2$-dependent (bottom) measurements.
Other details are as in the caption to Fig.~\ref{correlation_1}.
}
\label{correlation_2}
\end{figure}

\clearpage
%%%%%%%%%%%%%%%%%%%%%%%%%%%%%%%%%%%%%%%%%%%%%
% Syst
%%%%%%%%%%%%%%%%%%%%%%%%%%%%%%%%%%%%%%%%%%%%%
\section{Systematic uncertainties}
\label{app_syst}
Comparisons of the estimated systematic uncertanties to the statistical uncertainty are shown in Figs.~\ref{syst_1} and~\ref{syst_2} for the $p_\mathrm{T,jet} \otimes N_\mathrm{jet}$- and $Q^2 \otimes N_\mathrm{jet}$-dependent measurements, respectively.

\begin{figure}[h]
\centering
\includegraphics[width=\textwidth]{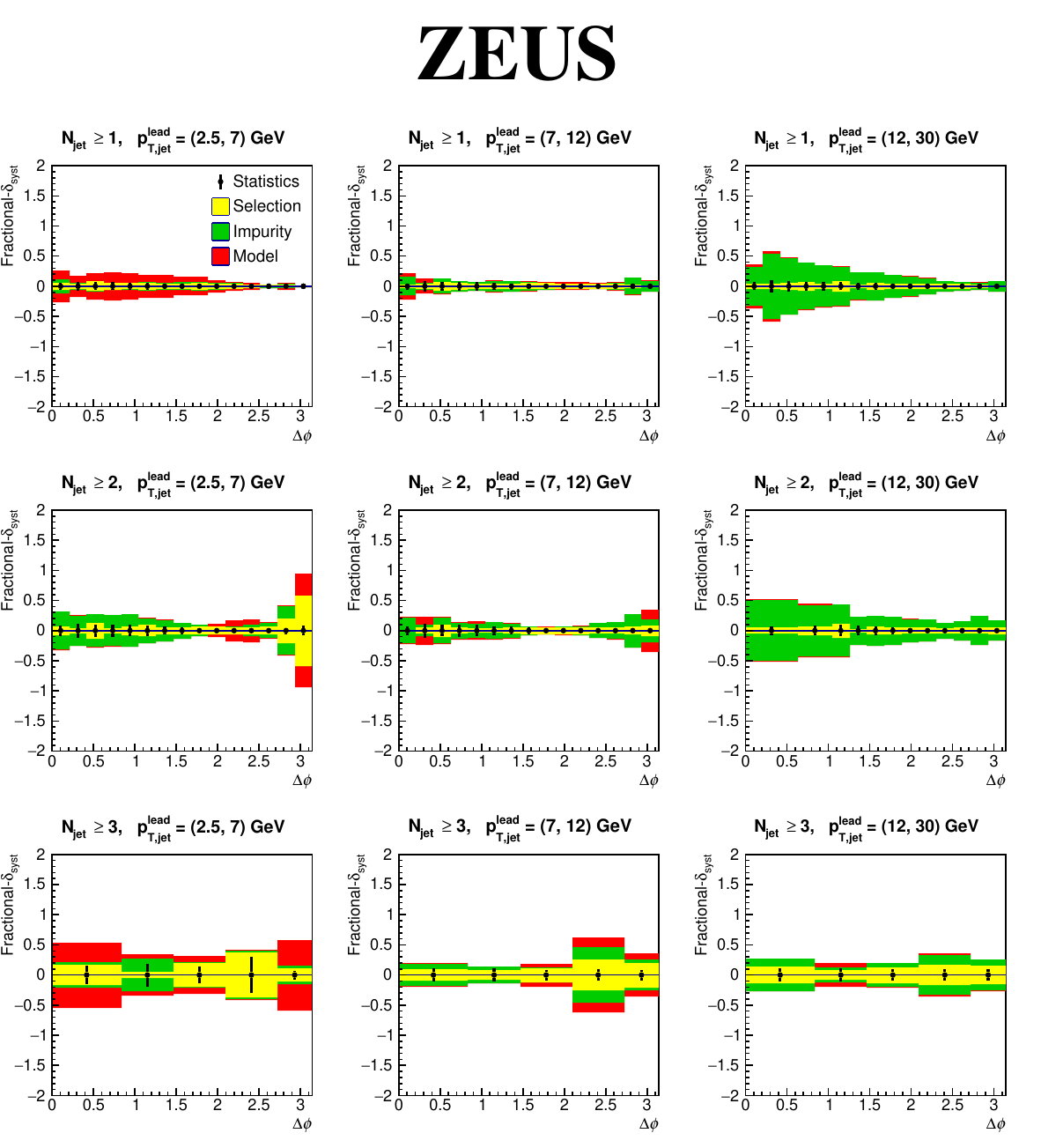}
\caption{ 
Breakdown of the systematic uncertainty in various ranges of $N_\mathrm{jet}$ and $p_\mathrm{T,jet}$ compared to the statistical uncertainty.
Other details are as in the caption to Fig.~\ref{syst_0}.
}
\label{syst_1}
\end{figure}

\begin{figure}[h]
\centering
\includegraphics[width=\textwidth]{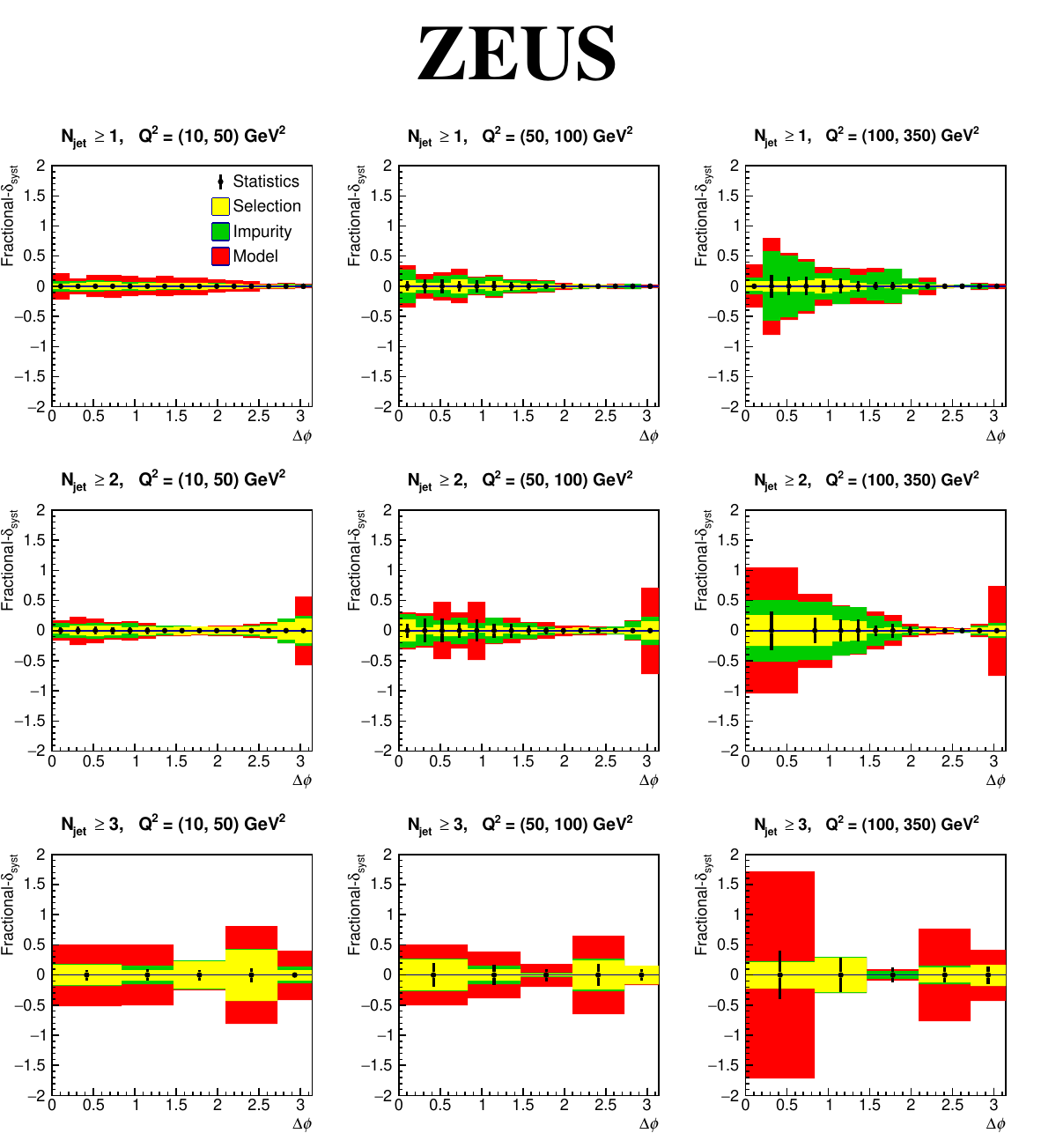}
\caption{ 
Breakdown of the systematic uncertainty in various ranges of $N_\mathrm{jet}$ and $Q^{2}$ compared to the statistical uncertainty.
Other details are as in the caption to Fig.~\ref{syst_0}.
}
\label{syst_2}
\end{figure}

\clearpage
%%%%%%%%%%%%%%%%%%%%%%%%%%%%%%%%%%%%%%%%%%%%%
% HadCor
%%%%%%%%%%%%%%%%%%%%%%%%%%%%%%%%%%%%%%%%%%%%%
\section{Hadronisation correction}
\label{app_hadcor}

Corrections were applied to the parton-level calculations to account for hadronisation effects.
Parton-level jets were found from the collision information provided by the ARIADNE simulation with the $k_\mathrm{T}$-clustering algorithm with $R = 1$ in the massless mode.
This jet definition is the same as used in the perturbative calculations.
A two-dimensional distribution of hadron-level correlation angle, $\Delta\phi_\mathrm{had}$, versus parton-level angle, $\Delta\phi_\mathrm{par}$, was formed for each $p_\mathrm{T,jet}$, $Q^2$, and $N_\mathrm{jet}$ range.
This distribution was normalised along the $\Delta\phi_\mathrm{had}$ axis, ensuring that the sum of each column, $\sum_i m_{ij}$, was equal to unity.
Thus, each element in the matrix represents the probability of a parton-level pair in $\Delta\phi$ bin $j$ contributing to the hadron-level yield in bin $i$.
These distributions are shown in Figs.~\ref{hadcor1}, ~\ref{hadcor2}, and  ~\ref{hadcor3}.
In order to correct for the migration of the kinematic quantities that define the fiducial region of the measurement, two additional correction factors, denoted as $c_1$ and $c_2$, were computed using the MC simulation.
The $c_1$ factor is the fraction of parton-level lepton--leading-jet pairs with a matching hadron-level pair over the total parton-level yield.
On the other hand, $c_2$ is the fraction of hadron-level pairs with a matching parton-level pair over the total hadron-level yield.
Hadron-level predictions were obtained from the perturbative calculations at parton level by using the following expression:

\begin{equation}
\left( \frac{d\sigma}{d\Delta\phi} \right)_{\mathrm{had},i}
= \frac{1}{c_{2,i}}
\times \sum_{j} m_{ij} \, c_{1,j} \, \left(\frac{d\sigma}{d\Delta\phi} \right)_{\mathrm{par},j}.
\end{equation}

In this expression, $\left( \frac{d\sigma}{d\Delta\phi} \right)_{\mathrm{had},i}$ represents the hadron-level differential cross section in bin $i$ to be directly compared to the measurement, and $\left(\frac{d\sigma}{d\Delta\phi} \right)_{\mathrm{par},j}$ represents the parton-level differential cross section in bin $j$ obtained with the P2B method.

The dependence on model-specific assumptions made in the ARIADNE simulation was tested by repeating the hadronisation correction with the LEPTO simulation.
The difference in the result derived using the two models was found to be $\sim 5\%$ at maximum.
This was taken as an additional source of uncertainty in the predictions and added in quadrature to the uncertainty arising from scale variations.

\begin{figure}[h]
\centering
\includegraphics[width=\textwidth]{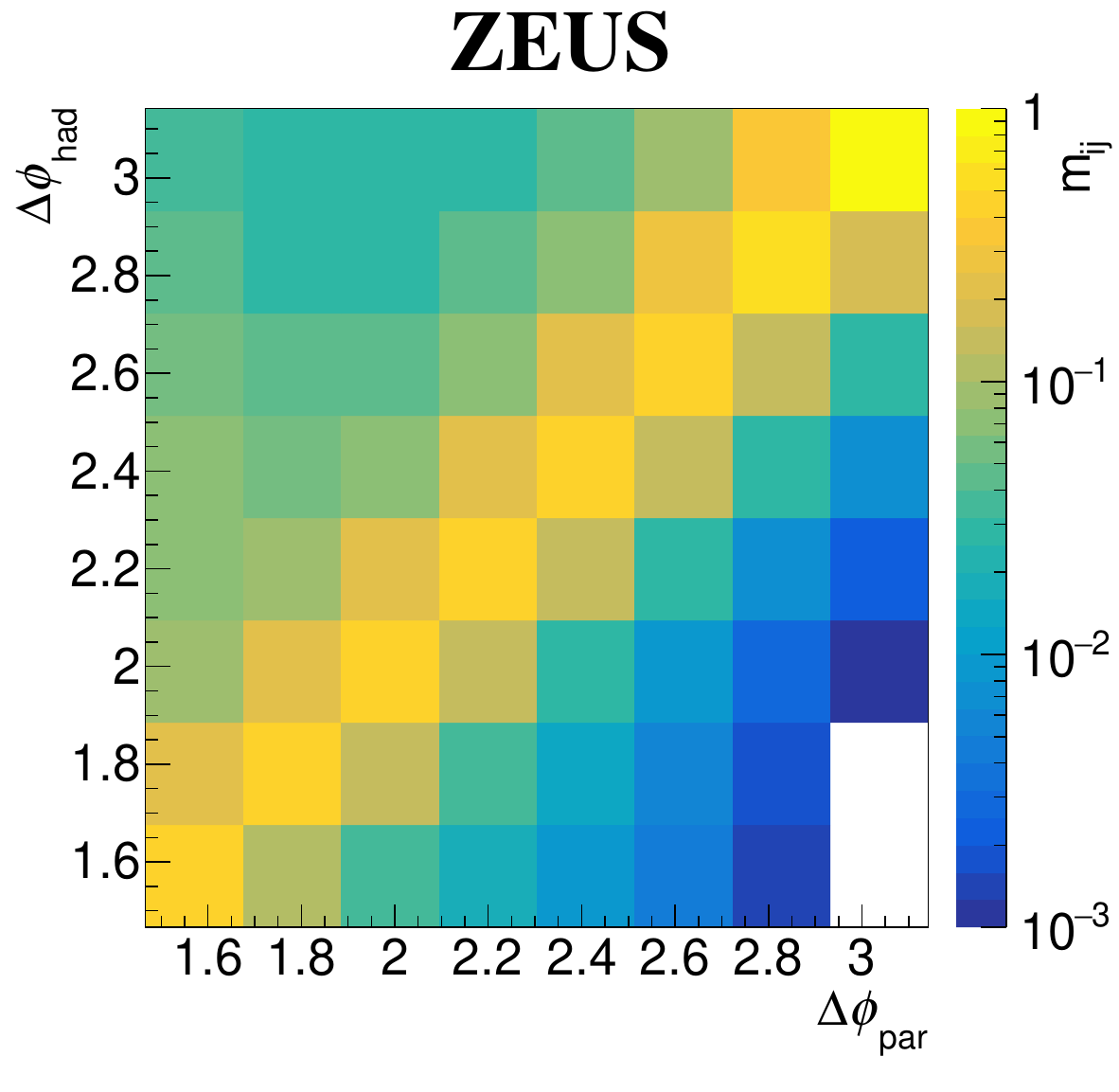}
\caption{ 
The migration matrix as input for the hadronisation correction of the perturbative calculations of the inclusive differential cross section.
Each element represents the probability of a parton-level lepton--leading-jet pair with an azimuthal correlation angle of $\Delta\phi_\mathrm{par}$ to give rise to a pair with a hadron-level angle $\Delta\phi_\mathrm{had}$.
}
\label{hadcor1}
\end{figure}

\begin{figure}[h]
\centering
\includegraphics[width=\textwidth]{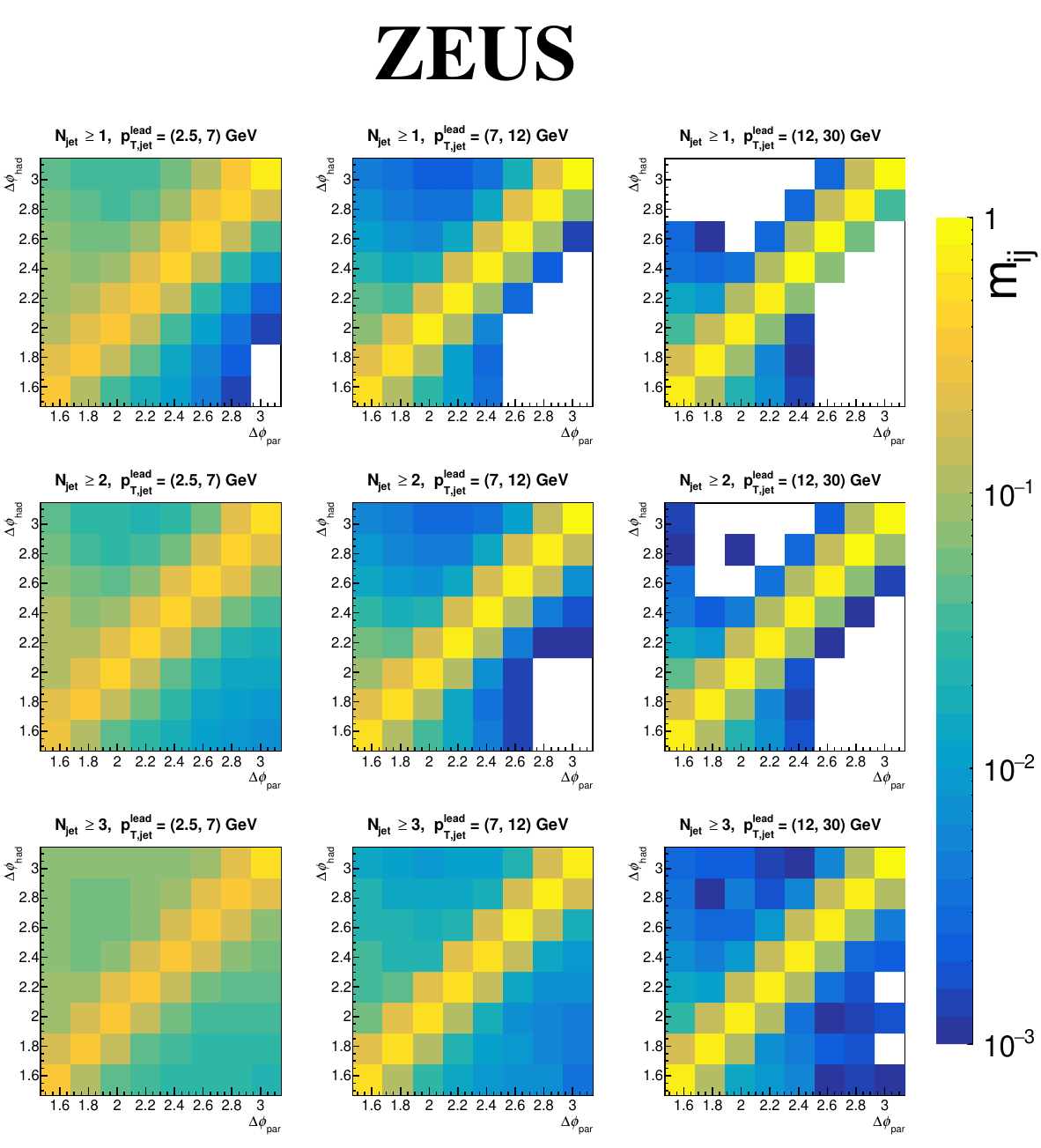}
\caption{ 
The migration matrices as input for the hadronisation correction of the perturbative calculations of the differential cross sections in varying ranges of jet multiplicity and $p_\mathrm{T,jet}$.
Each element represents the probability of a parton-level lepton--leading-jet pair with an azimuthal correlation angle of $\Delta\phi_\mathrm{par}$ to give rise to a pair with a hadron-level angle $\Delta\phi_\mathrm{had}$.
}
\label{hadcor2}
\end{figure}

\begin{figure}[h]
\centering
\includegraphics[width=\textwidth]{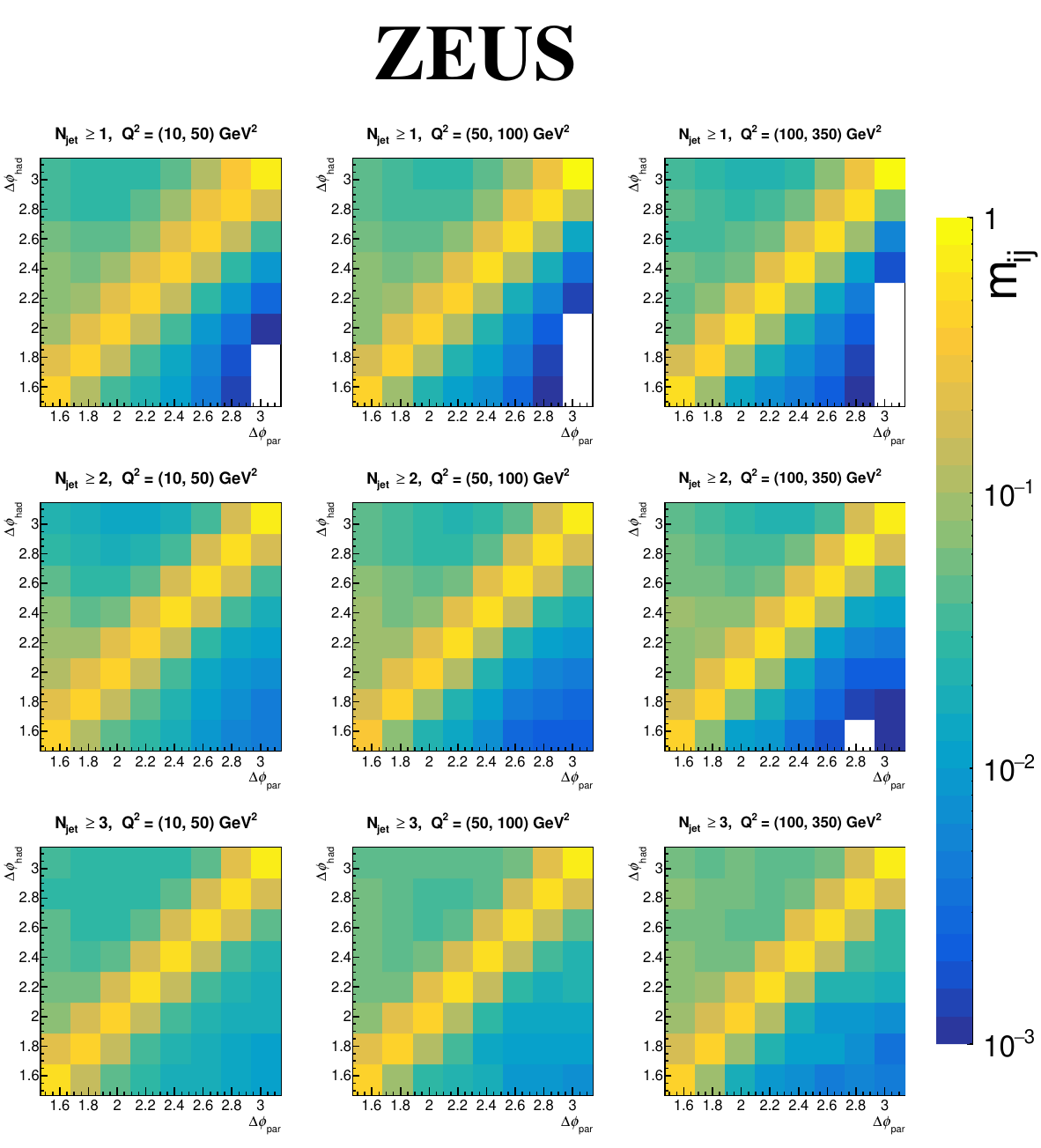}
\caption{ 
The migration matrices as input for the hadronisation correction of the perturbative calculations of the differential cross sections in varying ranges of jet multiplicity and $Q^{2}$.
Each element represents the probability of a parton-level lepton--leading-jet pair with an azimuthal correlation angle of $\Delta\phi_\mathrm{par}$ to give rise to a pair with a hadron-level angle $\Delta\phi_\mathrm{had}$.
}
\label{hadcor3}
\end{figure}

\clearpage
%%%%%%%%%%%%%%%%%%%%%%%%%%%%%%%%%%%%%%%%%%%%%
% QEDCor
%%%%%%%%%%%%%%%%%%%%%%%%%%%%%%%%%%%%%%%%%%%%%
\section{QED radiation correction}
\label{app_QED}

The resulting cross sections were corrected to QED Born-level, which is defined by the absence of QED-radiative effects, while including the scale dependence of the electromagnetic coupling.
Corresponding MC samples were generated using the RAPGAP 3.308 event generator~\cite{Jung:1993gf} with QED radiation simulated by HERACLES~\cite{Kwiatkowski:1990es}.
Bin-wise correction factors were determined by comparing the cross sections derived from
these samples to those from the nominal MC samples.
The resulting correction factors, $c_\mathrm{QED}$, are listed in Table~\ref{TAB_QEDcor_0} for the inclusive cross section, Table~\ref{TAB_QEDcor_1} for the $p_\mathrm{T,jet} \otimes N_\mathrm{jet}$-dependent cross sections, and Table~\ref{TAB_QEDcor_2} for the $Q^2 \otimes N_\mathrm{jet}$-dependent cross sections.

\begin{table}[h]
\begin{center}
\resizebox{0.4\textwidth}{!}{
\begin{tabular}{ c | c c c }
\hline
 & \multicolumn{3}{c}{Inclusive}\\
\cline{2-4}
 & $\Delta\phi^\mathrm{low}$ & $\Delta\phi^\mathrm{up}$ & $c_\mathrm{QED}$\\
\hline
\multirow{15}{*}{$N_\mathrm{jet} \geq 1$}  & 0.000 & 0.209 & 0.766 \\
 & 0.209 & 0.419 & 0.949 \\
 & 0.419 & 0.628 & 0.979 \\
 & 0.628 & 0.838 & 0.987 \\
 & 0.838 & 1.047 & 1.010 \\
 & 1.047 & 1.257 & 0.992 \\
 & 1.257 & 1.466 & 1.007 \\
 & 1.466 & 1.676 & 1.003 \\
 & 1.676 & 1.885 & 1.004 \\
 & 1.885 & 2.094 & 1.016 \\
 & 2.094 & 2.304 & 1.019 \\
 & 2.304 & 2.513 & 1.012 \\
 & 2.513 & 2.723 & 1.014 \\
 & 2.723 & 2.932 & 1.018 \\
 & 2.932 & 3.142 & 1.009 \\
\hline
\end{tabular}
}
\end{center}
\caption{
QED correction factors for inclusive measurement, as estimated with RAPGAP.
}
\label{TAB_QEDcor_0}
\end{table}

\begin{table}
\begin{center}
\resizebox{\textwidth}{!}{
\begin{tabular}{ c | c c c | c c c | c c c }
\hline
 & \multicolumn{3}{c|}{$2.5 \;\mathrm{GeV} < p_\mathrm{T,jet}^\mathrm{lead} < 7 \;\mathrm{GeV}$} & \multicolumn{3}{c|}{$7 \;\mathrm{GeV} < p_\mathrm{T,jet}^\mathrm{lead} < 12 \;\mathrm{GeV}$} & \multicolumn{3}{c}{$12 \;\mathrm{GeV} < p_\mathrm{T,jet}^\mathrm{lead} < 30 \;\mathrm{GeV}$}\\
\cline{2-10}
 & $\Delta\phi^\mathrm{low}$ & $\Delta\phi^\mathrm{up}$ & $c_\mathrm{QED}$ & $\Delta\phi^\mathrm{low}$ & $\Delta\phi^\mathrm{up}$ & $c_\mathrm{QED}$ & $\Delta\phi^\mathrm{low}$ & $\Delta\phi^\mathrm{up}$ & $c_\mathrm{QED}$\\
\hline
\multirow{15}{*}{$N_\mathrm{jet} \geq 1$}  & 0.000 & 0.209 & 0.814 & 0.000 & 0.209 & 0.548 & 0.000 & 0.209 & 0.483 \\
 & 0.209 & 0.419 & 0.953 & 0.209 & 0.419 & 0.927 & 0.209 & 0.419 & 0.856 \\
 & 0.419 & 0.628 & 0.979 & 0.419 & 0.628 & 0.980 & 0.419 & 0.628 & 0.989 \\
 & 0.628 & 0.838 & 0.990 & 0.628 & 0.838 & 0.982 & 0.628 & 0.838 & 0.897 \\
 & 0.838 & 1.047 & 1.008 & 0.838 & 1.047 & 1.061 & 0.838 & 1.047 & 0.884 \\
 & 1.047 & 1.257 & 0.990 & 1.047 & 1.257 & 1.019 & 1.047 & 1.257 & 0.977 \\
 & 1.257 & 1.466 & 1.006 & 1.257 & 1.466 & 0.981 & 1.257 & 1.466 & 1.154 \\
 & 1.466 & 1.676 & 1.007 & 1.466 & 1.676 & 0.975 & 1.466 & 1.676 & 0.979 \\
 & 1.676 & 1.885 & 1.004 & 1.676 & 1.885 & 1.012 & 1.676 & 1.885 & 0.936 \\
 & 1.885 & 2.094 & 1.016 & 1.885 & 2.094 & 1.024 & 1.885 & 2.094 & 1.005 \\
 & 2.094 & 2.304 & 1.019 & 2.094 & 2.304 & 1.016 & 2.094 & 2.304 & 0.984 \\
 & 2.304 & 2.513 & 1.012 & 2.304 & 2.513 & 1.012 & 2.304 & 2.513 & 0.982 \\
 & 2.513 & 2.723 & 1.018 & 2.513 & 2.723 & 0.989 & 2.513 & 2.723 & 0.907 \\
 & 2.723 & 2.932 & 1.026 & 2.723 & 2.932 & 0.970 & 2.723 & 2.932 & 0.877 \\
 & 2.932 & 3.142 & 1.030 & 2.932 & 3.142 & 0.942 & 2.932 & 3.142 & 0.828 \\
\hline
\multirow{15}{*}{$N_\mathrm{jet} \geq 2$}  & 0.000 & 0.209 & 0.926 & 0.000 & 0.209 & 0.728 & \multirow{3}{*}{0.000} & \multirow{3}{*}{0.628} & \multirow{3}{*}{0.855} \\
 & 0.209 & 0.419 & 0.966 & 0.209 & 0.419 & 0.955 &  &  &  \\
 & 0.419 & 0.628 & 0.971 & 0.419 & 0.628 & 0.998 &  &  &  \\
 & 0.628 & 0.838 & 0.974 & 0.628 & 0.838 & 0.995 & \multirow{2}{*}{0.628} & \multirow{2}{*}{1.047} & \multirow{2}{*}{0.908} \\
 & 0.838 & 1.047 & 0.997 & 0.838 & 1.047 & 1.078 &  &  &  \\
 & 1.047 & 1.257 & 0.978 & 1.047 & 1.257 & 1.005 & 1.047 & 1.257 & 0.987 \\
 & 1.257 & 1.466 & 0.983 & 1.257 & 1.466 & 1.003 & 1.257 & 1.466 & 1.171 \\
 & 1.466 & 1.676 & 1.032 & 1.466 & 1.676 & 0.981 & 1.466 & 1.676 & 0.992 \\
 & 1.676 & 1.885 & 1.022 & 1.676 & 1.885 & 1.006 & 1.676 & 1.885 & 0.924 \\
 & 1.885 & 2.094 & 1.044 & 1.885 & 2.094 & 1.023 & 1.885 & 2.094 & 1.019 \\
 & 2.094 & 2.304 & 1.027 & 2.094 & 2.304 & 1.026 & 2.094 & 2.304 & 0.987 \\
 & 2.304 & 2.513 & 1.014 & 2.304 & 2.513 & 1.009 & 2.304 & 2.513 & 0.991 \\
 & 2.513 & 2.723 & 0.999 & 2.513 & 2.723 & 0.982 & 2.513 & 2.723 & 0.910 \\
 & 2.723 & 2.932 & 1.011 & 2.723 & 2.932 & 0.961 & 2.723 & 2.932 & 0.869 \\
 & 2.932 & 3.142 & 1.003 & 2.932 & 3.142 & 0.930 & 2.932 & 3.142 & 0.654 \\
\hline
\multirow{5}{*}{$N_\mathrm{jet} \geq 3$}  & 0.000 & 0.838 & 0.969 & 0.000 & 0.838 & 0.847 & 0.000 & 0.838 & 0.884 \\
 & 0.838 & 1.466 & 0.984 & 0.838 & 1.466 & 0.998 & 0.838 & 1.466 & 0.939 \\
 & 1.466 & 2.094 & 1.064 & 1.466 & 2.094 & 1.009 & 1.466 & 2.094 & 0.994 \\
 & 2.094 & 2.723 & 1.017 & 2.094 & 2.723 & 0.966 & 2.094 & 2.723 & 0.934 \\
 & 2.723 & 3.142 & 0.983 & 2.723 & 3.142 & 0.928 & 2.723 & 3.142 & 0.764 \\
\hline
\end{tabular}
}
\end{center}
\caption{
QED correction factors for various $p_\mathrm{T,jet}^\mathrm{lead}$ and $N_\mathrm{jet}$ ranges, as estimated with RAPGAP.
}
\label{TAB_QEDcor_1}
\end{table}

\begin{table}
\begin{center}
\resizebox{\textwidth}{!}{
\begin{tabular}{ c | c c c | c c c | c c c }
\hline
 & \multicolumn{3}{c|}{$10 \;\mathrm{GeV}^{2} < Q^{2} < 50 \;\mathrm{GeV}^{2}$} & \multicolumn{3}{c|}{$50 \;\mathrm{GeV}^{2} < Q^{2} < 100 \;\mathrm{GeV}^{2}$} & \multicolumn{3}{c}{$100 \;\mathrm{GeV}^{2} < Q^{2} < 350 \;\mathrm{GeV}^{2}$}\\
\cline{2-10}
 & $\Delta\phi^\mathrm{low}$ & $\Delta\phi^\mathrm{up}$ & $c_\mathrm{QED}$ & $\Delta\phi^\mathrm{low}$ & $\Delta\phi^\mathrm{up}$ & $c_\mathrm{QED}$ & $\Delta\phi^\mathrm{low}$ & $\Delta\phi^\mathrm{up}$ & $c_\mathrm{QED}$\\
\hline
\multirow{15}{*}{$N_\mathrm{jet} \geq 1$}  & 0.000 & 0.209 & 0.930 & 0.000 & 0.209 & 0.486 & 0.000 & 0.209 & 0.118 \\
 & 0.209 & 0.419 & 0.969 & 0.209 & 0.419 & 0.886 & 0.209 & 0.419 & 0.577 \\
 & 0.419 & 0.628 & 0.994 & 0.419 & 0.628 & 0.941 & 0.419 & 0.628 & 0.655 \\
 & 0.628 & 0.838 & 0.998 & 0.628 & 0.838 & 0.966 & 0.628 & 0.838 & 0.710 \\
 & 0.838 & 1.047 & 1.023 & 0.838 & 1.047 & 0.971 & 0.838 & 1.047 & 0.715 \\
 & 1.047 & 1.257 & 1.003 & 1.047 & 1.257 & 0.964 & 1.047 & 1.257 & 0.727 \\
 & 1.257 & 1.466 & 1.019 & 1.257 & 1.466 & 0.959 & 1.257 & 1.466 & 0.742 \\
 & 1.466 & 1.676 & 1.013 & 1.466 & 1.676 & 0.990 & 1.466 & 1.676 & 0.740 \\
 & 1.676 & 1.885 & 1.014 & 1.676 & 1.885 & 0.987 & 1.676 & 1.885 & 0.747 \\
 & 1.885 & 2.094 & 1.028 & 1.885 & 2.094 & 0.994 & 1.885 & 2.094 & 0.748 \\
 & 2.094 & 2.304 & 1.031 & 2.094 & 2.304 & 1.000 & 2.094 & 2.304 & 0.757 \\
 & 2.304 & 2.513 & 1.025 & 2.304 & 2.513 & 1.007 & 2.304 & 2.513 & 0.760 \\
 & 2.513 & 2.723 & 1.028 & 2.513 & 2.723 & 1.015 & 2.513 & 2.723 & 0.767 \\
 & 2.723 & 2.932 & 1.037 & 2.723 & 2.932 & 1.024 & 2.723 & 2.932 & 0.776 \\
 & 2.932 & 3.142 & 1.039 & 2.932 & 3.142 & 1.037 & 2.932 & 3.142 & 0.796 \\
\hline
\multirow{15}{*}{$N_\mathrm{jet} \geq 2$}  & 0.000 & 0.209 & 0.943 & 0.000 & 0.209 & 0.558 & \multirow{3}{*}{0.000} & \multirow{3}{*}{0.628} & \multirow{3}{*}{0.451} \\
 & 0.209 & 0.419 & 0.973 & 0.209 & 0.419 & 0.919 &  &  &  \\
 & 0.419 & 0.628 & 0.994 & 0.419 & 0.628 & 0.960 &  &  &  \\
 & 0.628 & 0.838 & 0.986 & 0.628 & 0.838 & 0.975 & \multirow{2}{*}{0.628} & \multirow{2}{*}{1.047} & \multirow{2}{*}{0.751} \\
 & 0.838 & 1.047 & 1.015 & 0.838 & 1.047 & 1.000 &  &  &  \\
 & 1.047 & 1.257 & 0.992 & 1.047 & 1.257 & 0.977 & 1.047 & 1.257 & 0.753 \\
 & 1.257 & 1.466 & 1.004 & 1.257 & 1.466 & 0.969 & 1.257 & 1.466 & 0.765 \\
 & 1.466 & 1.676 & 1.033 & 1.466 & 1.676 & 1.010 & 1.466 & 1.676 & 0.753 \\
 & 1.676 & 1.885 & 1.027 & 1.676 & 1.885 & 1.012 & 1.676 & 1.885 & 0.760 \\
 & 1.885 & 2.094 & 1.057 & 1.885 & 2.094 & 1.015 & 1.885 & 2.094 & 0.765 \\
 & 2.094 & 2.304 & 1.043 & 2.094 & 2.304 & 1.032 & 2.094 & 2.304 & 0.778 \\
 & 2.304 & 2.513 & 1.033 & 2.304 & 2.513 & 1.037 & 2.304 & 2.513 & 0.784 \\
 & 2.513 & 2.723 & 1.015 & 2.513 & 2.723 & 1.042 & 2.513 & 2.723 & 0.793 \\
 & 2.723 & 2.932 & 1.030 & 2.723 & 2.932 & 1.031 & 2.723 & 2.932 & 0.785 \\
 & 2.932 & 3.142 & 1.026 & 2.932 & 3.142 & 0.974 & 2.932 & 3.142 & 0.688 \\
\hline
\multirow{5}{*}{$N_\mathrm{jet} \geq 3$}  & 0.000 & 0.838 & 0.973 & 0.000 & 0.838 & 0.740 & 0.000 & 0.838 & 0.531 \\
 & 0.838 & 1.466 & 1.003 & 0.838 & 1.466 & 1.010 & 0.838 & 1.466 & 0.774 \\
 & 1.466 & 2.094 & 1.078 & 1.466 & 2.094 & 1.020 & 1.466 & 2.094 & 0.771 \\
 & 2.094 & 2.723 & 1.054 & 2.094 & 2.723 & 0.998 & 2.094 & 2.723 & 0.744 \\
 & 2.723 & 3.142 & 1.026 & 2.723 & 3.142 & 0.961 & 2.723 & 3.142 & 0.688 \\
\hline
\end{tabular}
}
\end{center}
\caption{
QED correction factors for various $Q^{2}$ and $N_\mathrm{jet}$ ranges, as estimated with RAPGAP.
}
\label{TAB_QEDcor_2}
\end{table}

\end{appendices}

\end{document}